\documentclass[onecolumn]{article}
\usepackage{amsmath,amssymb}

\usepackage{hyperref}
\begin{document}
\centerline{\bf{Existence of Negative Gravity Material}}
\centerline{\bf{Identification of Dark Energy}}
\centerline{\bf{Dark Matter and Galactic Rotation}}
\vskip0.5cm
\centerline{James G. Gilson\quad  j.g.gilson@qmul.ac.uk}
\centerline{School of Mathematical Sciences}
\centerline{Queen Mary University of London}
\centerline{Mile End Road London E14NS}
\vskip0.2cm
\centerline{March 27, 2006}
\section{Abstract}
A solution to Einstein's field equations via the Friedman equations is shown to produce a cosmological model that is in exact agreement with the measurements made by the dark energy astronomers. All the essential physical parameters are obtained as epoch dependent functions all in closed form. The equations of state are obtained for a non-dark energy density, a dark energy density and a total gravitationally weighted mass density. An interpretation of the structure involving a dark energy mass distribution that is twice the usual value is shown to clarify greatly the physical significance of the mathematics. It is asserted that the astronomer's measurements together with the mathematical model proves that the universe is permeated uniformly with a positive mass density that carries a negative gravitational constant, -G, characteristic. This mass component is identified with the dark energy content of the universe that has been postulated to explain the observed acceleration. Another result implied by the model is that there is twice the amount of dark energy that is usually considered to be present. This last point is analysed in more detail in appendix 1 using Einstein's field equations. Five additional appendices, 2, 3, 4, 5 and 6  in which isothermal gravitational dark matter equilibrium and the galactic rotations curve flatness problem are examined in detail. Appendix 5 is concerned with mass clumping and expressing gravitational isothermal equilibrium constraints using a cosmological Schr\"odinger equation to demonstrate the existence of a new quantum force involved with galactic stability. Appendix 6 is concerned with gravitational quantization.  Each appendix has its own abstract.
\vskip0.3cm  
\centerline{Keywords: Dust Universe, Dark Energy, Dark Matter, Friedman Equations}
\centerline{Isothermal Gravitational Equilibrium, Cosmological Schr\"odinger Equation}
\vskip0.3cm  
\centerline{PACS Nos.: 98.80.-k, 98.80.Es, 98.80.Jk, 98.80.Qc}
\section{Introduction}
\setcounter{equation}{0}
\label{sec-intr}
The work to be described in this paper is a detailed analysis of a conclusion hinted at in the paper {\it A Dust Universe Solution to the Dark Energy Problem \/}\cite{45:gil}. The conclusion arrived at there was that the dark energy {\it substance\/} is physical material with a positive density, as is usual, but with a negative gravity, -G,  characteristic.  References to equations in that paper will be prefaced with the letter $A$. The work in $A$ and its extension here has it origins in the studies of Einstein's general relativity in the Friedman equations context to be found in references (\cite{03:rind},\cite{43:nar},\cite{42:gil},\cite{41:gil},\cite{40:gil},\cite{39:gil},\cite{04:gil},\cite{45:gil}) and similarly motivated work in references (\cite{10:kil},\cite{09:bas},\cite{08:kil},\cite{07:edd},\cite{05:gil}) and 
(\cite{19:gil},\cite{28:dir},\cite{32:gil},\cite{33:mcp},\cite{07:edd}).
Other useful sources of information have been references (\cite{3:mis},\cite{44:berr}) with the measurement essentials coming from references (\cite{01:kmo},\cite{02:rie},\cite{18:moh}).

After the astronomical observations some eight years ago implying that the expansion of the universe is accelerating there has been overwhelming research into trying to find an explanation of what is happening and how it all does or does not fit into general relativity theory. The acceleration phenomenon is widely described as very mysterious. This is particularly in connection with the material called {\it dark energy\/} which is thought to produce a {\it negative pressure\/} that causes the acceleration. As a positive pressure in the general relativity generated Friedman equations is thought to be due to gravitational attraction, the measured acceleration is seen {\it vaguely\/} as being caused by dark energy being gravitationally repulsive. However, it has  seemed to be difficult to fit the observations into general relativity theory. I think the basis of all the theoretical difficulties lies with what Einstein called his greatest mistake, the introduction of his cosmological constant $\Lambda$ to stabilise his model which without $\Lambda$ would itself suffer negative acceleration. I think the mistake was in fact {\it not\/} the introduction of $\Lambda$ which actually brilliantly achieved his objective but rather the association of $\Lambda$ with the mass density function $\rho_\Lambda $ which was assumed to represent the mystery repulsive dark energy material.
As this concept was obviously effective in stabilising Einstein's model it has for the last hundred years been regarded as a physically correct assumption and so it always appears in $\Lambda$ orientated discussions of the Friedman equations. In $A$, I showed that if the density function for dark energy is rather defined as $\rho^\dagger _\Lambda =\Lambda c^2/(4 \pi G)= 2 \rho _\Lambda $ then, dark energy becomes a distribution of positive mass that carries a negatively valued gravitational constant, $-G$. The dark energy model for the universe introduced in $A$, then accurately describes the dynamics and kinematics of the actual measured universe parameters and most of the mystery vanishes.  It all turns on recognising that the universe contains a uniform and minutely small density distribution of negative gravity characterised material. I give a strong case for these ideas in the following pages by showing that the model from $A$ has a {\it complete\/} solution giving for example the equations of state for the total {\it gravitationally positive\/} mass distribution density, $\rho (t)$, the dark energy material distribution, $\rho _\Lambda$, and the density a distribution, $\rho _G(t)$, which expresses the positive gravitational mass density relative to the dark energy mass density. 
\section{Summary of Mathematical Structure of Model}
\setcounter{equation}{0}
\label{sec-sms}
The main theoretical basis for the work to be discussed here are the two Friedman equations that derive from general relativity,
\begin{eqnarray}
8\pi G \rho (t) r^2/3 & = & {\dot r}^2 +(k - \Lambda r^2/3) c^2\label{B1}\\
-8 \pi GP(t) r/c^2 & = & 2 \ddot r + {\dot r}^2/r +(k/r -\Lambda r) c^2 \label{B2}
\end{eqnarray}
in the case $k=0$ and a positively valued $\Lambda$ we have,
\begin{eqnarray}
8\pi G \rho (t) r^2/3 & = &  {\dot r}^2 - |\Lambda | r^2 c^2/3 \label{B3}\\
-8\pi GP(t) r/c^2 & = & 2 \ddot r + {\dot r}^2/r -|\Lambda | r c^2.\label{B4}
\end{eqnarray}
The main results obtained from the dark energy model for ease of reference are listed next,
\begin{eqnarray}
r(t) & = & (R_\Lambda /c)^{2/3} C^{1/3} \sinh^{2/3} ( \pm 3ct/(2R_\Lambda)) \label{B77}\\
b & = &  (R_\Lambda /c)^{2/3} C^{1/3}\label{B77.1}\\
C &=& 8\pi G \rho (t) r^3/3\label{B6}\\
R_{\Lambda} &=& |3/\Lambda|^{1/2} \label{B5}\\
\theta _\pm(t) &=& \pm 3ct/(2R_\Lambda) \label{B77.3}\\
r(t) & = &  b\sinh^{2/3}(\theta_\pm (t) )\label{B77.2}\\
v(t)&=& \pm (bc/R_\Lambda)\sinh ^{-1/3} (\theta _\pm (t))\cosh ( \theta _\pm(t))\label{B80.1}\\
a(t)&=& b(c/(R_\Lambda) )^2\sinh ^{2/3} (\theta _\pm (t))(3 - \coth ^2 (\theta _\pm(t)))/2 \label{B80.2}\\
H(t)&=& (c/R_\Lambda ) \coth(\pm 3ct/(2R_\Lambda)) \label{B81.1}
\end{eqnarray}
\begin{eqnarray}
P(t)&=&(-c^2/(8\pi G))(2\ddot r(t)/r(t) + H^2(t) -3(c/R_\Lambda)^2) \label{B81.12}\\
P(t)&\equiv& 0 \ \forall\  t\label{B81.13}\\
P_\Lambda &=& (-3c^2/(8\pi G))(c/R_\Lambda)^2\label{B811.13}\\
\rho _\Lambda & =& (3/(8 \pi G))(c/R_\Lambda)^2 \label{B81.2}\\
\rho ^\dagger _\Lambda & =& (3/(4 \pi G))(c/R_\Lambda)^2\ =\  2\rho _\Lambda. \label{B81.3}
\end{eqnarray}
The function $r(t)$ is the radius or scale factor\footnote{After writing $A$, I found mention of this scale factor in Michael Berry's Book, p. 129.} for this model(\cite{44:berr}).
The acceleration $a(t)$ can also be expressed as
\begin{eqnarray}
\ddot r(t)&=& 4\pi r G(\rho^\dagger_\Lambda -\rho (t))/3\ =\ a(t) \label{B99.1}\\
& = &4\pi r^3 G(\rho^\dagger_\Lambda -\rho (t))/(3 r^2)\label{B99}\\
& = &{M^\dagger}_\Lambda G/r^2 - M_UG/r^2\label{B100}\\
M^\dagger_\Lambda & = & 4\pi r^3 \rho^\dagger_\Lambda/3\label{B101}\\
M_U & = & 4\pi r^3 \rho (t)/3\label{B102}
\end{eqnarray}
where $M^\dagger_\Lambda$ is the total dark energy mass within the universe and $M_U$ is the total non-dark energy mass within the universe.
From equation(A2.19), we have
\begin{eqnarray}
8\pi G \rho (t) r^2/3 + \Lambda r^2 c^2/3 &=&  {\dot r}^2\label{B1.0}\\
8\pi G \rho (t) + \Lambda c^2 &=&  3H^2(t)\label{B1.1}\\
8\pi G \rho (t) &=&  3H^2(t) -\Lambda c^2 \label{B1.2}\\
\rho (t) &=&(3/(8\pi G))(H^2(t)- (c/R_\Lambda)^2)\label{B1.3}\\
&=&(3/(8\pi G))(c/R_\Lambda)^2(\coth^2(\theta _+(t))- 1)\label{B1.4}\\
&=&(3/(8\pi G))(c/R_\Lambda)^2(\sinh^{-2}(3ct/(2R_\Lambda))\label{B1.5}\\
&=&(3/(8\pi G))C/r^3(t)\label{B1.6}\\
&=&3 M_U/(4\pi r^3(t))= M_U/V_U(t).\label{B1.7}
\end{eqnarray}
Equation (\ref{B1.7}) confirms the mass density of the universe character of $\rho (t) $.
\section{The Pseudo Mass Density}
\setcounter{equation}{0}
\label{sec-pmd}
The mass densities introduced in this theoretical construction, so far, are
$\rho _\Lambda $, $\rho ^\dagger_\Lambda$ and $\rho (t)$. The first of these is what Einstein would have regarded as the repulsive material density  within the universe to be associated with his $\Lambda$ term added to overcome the intrinsic expansion process and give his static universe. The second one, is numerically twice the first which I have introduced for reasons about to be explained. The third density, $\rho(t)$, is the total {\it gravitational positive\/} mass density for the model being discussed. These are three legitimate mass densities in particular conforming to the usually accepted requirement that mass should always be {\it positive\/}. I now introduce another density of great importance which will be denoted by $\rho_G(t)$ and defined by
 \begin{eqnarray}
\rho _G (t)= \rho (t) - \rho ^\dagger _\Lambda.\label{B1.8}
\end{eqnarray}
This is clearly {\it not\/} a legitimate mass density as it can become negative if $\rho (t) < \rho ^\dagger _\Lambda $. However, if we rewrite it as
\begin{eqnarray}
\rho _G(t) &=& (G_+  \rho (t)  + G_- \rho ^\dagger _\Lambda)/G\label{B1.9}\\
G_+&=& +G\label{B1.10}\\
G_-&=& -G,\label{B1.11}
\end{eqnarray}
the quantity $\rho _G(t) $ can then be regarded as a legitimate gravitationally {\it weighted mass density\/} and the equation (\ref{B99.1}) becomes
\begin{eqnarray}
\ddot r(t) &=& -4\pi r(t)G \rho_G (t) /3\label{B1.12}\\
2\ddot r(t)/r(t) &=& -8\pi G \rho_G(t) /3.\label{B1.13}
\end{eqnarray}
There is another way in which the quantity $\rho _G(t)$ can be legitimised. This is by calling it a {\it relative mass density\/}. Then possible negative values can be explained by arguing that it falls below the reference density $\rho ^\dagger _\Lambda $ for negative values. In that situation the uniform constant dark energy value is simply acting as a standard reference value throughout all space. Equation (\ref{B1.12}) or equation (\ref{B1.13}) is exactly the classical Newtonian result for the acceleration at the boundary of a spherical distribution  positive G mass density.   
\section{Component Masses}
\setcounter{equation}{0}
\label{sec-cm}
Using equations (\ref{B81.12}) and (\ref{B1.13}) the identically zero total pressure $P(t)$ can be expressed in the form
\begin{eqnarray}
P(t) &=& c^2 \rho _G(t)/3 + (c^2/(8\pi G))(3(c/R_\Lambda)^2 - H^2(t))\ \equiv\ 0.\label{B1.14}
\end{eqnarray}
In $A$,  $P_M(t) $, the positive component of the pressure associated with ordinary non-dark energy mass, was defined as $P(t)  - P_\Lambda$. Here the notation has been changed so that the $M$ subscript is replaced with the $G$ subscript, $P_G(t) = P_M(t)$, an improved designation for the pressure produced by the non-dark energy material component of the universe.  
It follows from equation (\ref{B1.14}) with the total pressure $P(t) = P_G(t) +P_\Lambda$ that
\begin{eqnarray}
\rho _G(t) &=& (3/(8\pi G))(H^2(t) - 3(c/R_\Lambda)^2) \label{B1.15}\\
P_G (t) &=& P(t) - P_\Lambda\label{B1.16}\\
&=& c^2\rho _G(t)/3 + (-c^2/(8\pi G))H^2(t) +(6c^2/(8\pi G))(c/R_\Lambda)^2\label{B1.17}
\end{eqnarray}
using equation (\ref{B811.13}).
Then as we have both $P_G (t) $ and $\rho _G(t)$ for this component, we can write down the equation of state for this component as
\begin{eqnarray}
P_G (t) /(c^2\rho _G(t)) &=& 1/3 -(c^2/(8\pi G))/(3 c^2/(8\pi G)) \nonumber\\
& +& (c/R_\Lambda)^2/(H^2(t)-3(c/R_\Lambda)^2) \label{B1.18}\\
&=& 1/3 - 1/3 +1/(\coth^2(3ct/(2R_\Lambda))-3) \label{B1.19}\\
&=& 1/(\coth^2(3ct/(2R_\Lambda))-3)\ =\ \omega_G(t). \label{B1.20}
\end{eqnarray}
The equation of state for the density, $\rho (t)$, is given by equation (\ref{B1.14}) in either of the forms,
\begin{eqnarray}
P(t) &=& 0\label{B1.21}\\
\rho _G(t) &=& (3/(8\pi G))( H^2(t)-3(c/R_\Lambda)^2) \label{B1.22}
\end{eqnarray}
and because of the identically zero value of $P(t)$ an $\omega$ value is not defined. The equation of state for the dark energy component is given by equation $A$(4.50) as
\begin{eqnarray}
\omega _\Lambda &=& P_\Lambda /(c^2 \rho _\Lambda)= -( 3c^2/(8 \pi G))(c/R_\Lambda)^2 /(c^2\rho _\Lambda) \label{B1.23}\\
&=&  -1.\label{B1.24}
\end{eqnarray}
\section{Conclusions}
The cosmological model developed in $A$ and further amplified here is {\it rigorously\/} a solution to the full set of Einstein's field equations of general relativity via the Friedman equations.  The model also satisfies {\it exactly\/} the recent measurements by the astronomical dark energy workers. The model is mathematically complete as all the essential physical functions for, scale, velocity, acceleration, and Hubble's {\it not so\/} constant are given in closed form. The equations of state are obtained for the total {\it positive G\/} or non-dark energy density, $\rho (t)$; the dark energy density, $\rho _\Lambda$, used originally by Einstein and a total gravitationally weighted mass density, $\rho _G(t)$, are  given in closed form. One difference from the usual physical interpretation is that the dark energy material is associated with a density, $\rho _\Lambda ^\dagger$, twice the usual density that Einstein used. Thus if this density is used to replace the Einstein density in the equation of state (\ref{B1.24}) for dark energy, the $-1$ would be replaced with $-1/2$. This difference has radical consequences for the understanding  of the whole physical structure of the theory. In particular, it implies that there is twice a much dark energy material present as hitherto has been thought to be present. Another major interpretational consequence is that dark energy material is {\it positive\/} mass that carries a negative valued, -G, characteristic property. With regard to the often suggested {\it mystery\/} that the enormous amount of dark energy present cannot be seen, a calculation of the density value of $\rho ^\dagger _\Lambda$ shows it to be about the equivalent of less than nine protons or hydrogen atoms per cubic meter. No wonder it cannot be seen. It is also the case that if all the observable mass where un-clumped from being galaxies and spread uniformly throughout the universe it also would be unobservable. The flatness of the model means that the expanding universe can be regarded simply as an expanding 3-sphere enveloping more dark energy as it increases in radius. Thus as this density is an absolute constant this increase must come from the regions outside the universe's spherical boundary. Thus dark energy is more than just a characteristic of the interior of the universe, it extends uniformly out into the three dimensional enveloping space into which the universe is expanding. It is hyper-universal. It is my contention that the astronomical measurements\cite{01:kmo} together with the inevitably following general relativity theoretical model derived in $A$\cite{45:gil} and further developed in this paper constitute {\it proof\/} that there exists throughout the universe and beyond extending into all hyperspace {\it positive\/} mass with a {\it negative\/} gravitational characteristic, $-G$, additional to the rest of the universe of swarming galaxies largely controlled {\it locally\/} by positive $G$.
\section{Appendix 1}
\centerline{\Large {\bf Newtonian Limit of}}
\centerline{\Large {\bf Einstein's Field Equations} }
\centerline{\Large {\bf Dark Matter and Dark Energy}}
\centerline{\Large {\bf with Einstein's Lambda}}
\vskip0.5cm
\centerline{December 9 2009}
\vskip 0.5cm  
\section{Appendix 1 Abstract}  
The stress energy momentum tensor with its invariant contracted scalar from
Einstein's field equations are used to show that the pressure term they involve is responsible for inducing {\it additional\/} mass density above that which is historically thought to be present. This has significant consequences for both the dark matter and dark energy problems. The additional material in the case of normally gravitating material might reasonably be taken to be the missing dark matter required for the stability of galaxies. The negative pressure of the dark energy is shown to induce just the right amount of extra negatively gravitating material to account for its usually assumed equation of state and also confirming the physical dark energy density I have deduced as being present in
astrospace in earlier papers.
\vskip0.2cm 
\section{Appendix 1 Introduction}
\setcounter{equation}{0}
\label{sec-a1intr}
The work to be described in this paper is an application of the cosmological model introduced in the papers {\it A Dust Universe Solution to the Dark Energy Problem\/} \cite{45:gil}, {\it Existence of Negative Gravity Material. Identification of Dark Energy\/} \cite{46:gil} and {\it Thermodynamics of a Dust Universe\/} \cite{56:gil}. All of this work and its applications has its origin in the studies of Einstein's general relativity in the Friedman equations context to be found in references (\cite{03:rind},\cite{43:nar},\cite{42:gil},\cite{41:gil},\cite{40:gil},\cite{39:gil},\cite{04:gil},\cite{45:gil}) and similarly motivated work in references (\cite{10:kil},\cite{09:bas},\cite{08:kil},\cite{07:edd},\cite{05:gil}) and 
(\cite{19:gil},\cite{28:dir},\cite{32:gil},\cite{33:mcp},\cite{07:edd},\cite{47:lem},\cite{44:berr}). The applications can be found in 
(\cite{45:gil},\cite{46:gil},\cite{56:gil},\cite{60:gil},\cite{58:gil}\cite{64:gil}). Other useful sources of information are (\cite{3:mis},\cite{44:berr},\cite{53:pap},\cite{49:man},\cite{52:ham},\cite{54:riz}) with the measurement essentials coming from references (\cite{01:kmo},\cite{02:rie},\cite{18:moh},\cite{61:free}).  Further references will be mentioned as necessary.
\section{Einstein's Lambda}
\setcounter{equation}{0}
\label{sec-elam}
Einstein presented his field equation for general relativity in $1915$,
ten years after the publication of his {\it special\/} theory of relativity which, as its name implies, was all about observers views from frames of reference in relative motion and this replaced the precursor {\it Newton's Theory of dynamics}. The big jump from the special theory to the general theory was that gravitation was incorporated into the simple space time geometrical structure of the special theory to produce a much more involved space time geometry. Thus general relativity replaced {\it Newton's Theory of Gravity\/}. However, you look at these developments one thing is certain, the evolution of theory must involve the more advanced forms reducing to the less advanced forms under some approximation conditions. Only if theory develops under such a constraint, can we be sure that the evolution process is consistent and built on sound foundations. There are other important constraints on theory evolution, the most important of which is theoretical agreement with measurement. Here we consider a case of the first mentioned, {\it limiting consistency with earlier theory\/}. In particular, I shall examine an important aspect of the extensively studied Newtonian limit of general relativity. This limit is discussed in detail in all the serious textbooks on general relativity with correct mathematical and physical analysis. However, it seems to me that they all miss important consequences that can be seen from a careful and possibly alternative interpretation of the physics of that limit. One of these missed consequences involves the Lambda term and its involved pressure and the other missed consequence is about  the normal mass density contribution and its involvement with pressure. The first issue is the question of the theoretical {\it amount\/} of physical dark energy density that is present as the vacuum state of three space. This aspect is studied in this paper using the Einstein field equations together with the Friedman equations rather than as in earlier work when I used the Friedman equations exclusively. The second missed consequence could supply a general relativistic explanation for the missing {\it dark matter\/} problem and essentially falls out from the first $\Lambda$ issue.

Einstein's fundamental and brilliant idea in writing down his field equations was to make a connection between physical conditions in space time as given by the stress energy momentum tensor, $T_{\mu\nu}$, with a geometrical interpretation of these conditions as given by the geometrical structure terms on the left hand side of the equation. See reference (\ref{1}) . Thus the right hand side is essentially input with the left hand side describing an output geometrical pattern explanation of the input physical structure on the right. Clearly, however, the introduction of the Lambda term, (\ref{2}),  does complicate this simple input-output relationship idea because mathematically the Lambda term could be transposed to the right hand side and then be regarded as actually a contribution to the physical input rather than to the geometry.
Either side will do for this term in its raw form as $ \pm g_{\mu\nu}\Lambda $ with, of course, the appropriate sign. However, if it is put on the right hand side
there is a temptation to interpret it as being the consequence of a physical space density additional to the physical input tensor $ -\kappa T_{\mu\nu}$ and of the same  form such as $-\kappa T_{0,\mu\nu}$ but with the extra $0$ subscript to distinguish it. $\Lambda$  would then acquire a simple physical interpretation in terms of an additional energy density $\rho_0$, the (44) component of $T_{0,\mu\nu}$. Consequently we would get
\begin{eqnarray}
-g_{\mu\nu}\Lambda = -\kappa T_{0,\mu\nu}.\label{3} 
\end{eqnarray}
The (44) term of $ T_{\mu\nu}$ is $c^2\rho$ and the (44) term of $g_{\mu\nu}$ is $=1$ so that (\ref{3}) would imply
\begin{eqnarray}
\Lambda = \kappa T_{0,44}=\kappa c^2\rho_0\label{4} 
\end{eqnarray}
and with $\kappa$ as usual identified as
\begin{eqnarray}
\kappa =8\pi G/c^4\label{5} 
\end{eqnarray}
the conclusion is that
\begin{eqnarray}
\rho_0 =\frac{\Lambda c^2}{8\pi G}.\label{6} 
\end{eqnarray}
Readers familiar with this area of study will recognise (\ref{6}) as giving the value of the constant physical density that Einstein actually chose to account for the addition of his Lambda term in his field equations. I think it is important to understand that the above argument leading to the result (\ref{6}) is completely correct if the original question to be answered is taken to be the following. What value of the (44) term of a stress energy momentum tensor on the right is required to correctly represent the newly introduced Lambda term on the left? I expect that Einstein's derivation of (\ref{6}) was finding the answer to the above question in his original reasoning which led to his result.  However, a different answer is obtained from a closer study of the field equations 
if the following alternative question is asked. What is the value of physical energy density in three space implied by the addition of the Einstein Lambda term on the left? This last question will be answered in the following pages.
\section{Einstein's Field Equations}
\setcounter{equation}{0}
\label{sec-eife}
The first tensor equation that Einstein proposed in $1915$ embodying the general theory was,
\begin{eqnarray}
R_{\mu\nu} -\frac{1}{2}g_{\mu\nu}R = -\kappa T_{\mu\nu}. \label{1}
\end{eqnarray}
Later, $1917$, he modified this with the addition of the so called {\it Lambda\/}, $\Lambda$, term so that his equation became
\begin{eqnarray}
R_{\mu\nu} -\frac{1}{2}g_{\mu\nu}R +g_{\mu\nu}\Lambda = -\kappa T_{\mu\nu}.\label{2}
\end{eqnarray}
The right hand side of this equation is the stress energy momentum tensor,
$T_{\mu\nu}$ multiplied by a negative constant $-\kappa $ which can be evaluated on the basis of the assumption that Newtonian gravitation theory is a limiting consequence of these equations. The term $R_{\mu\nu}$ is the Ricci curvature tensor and the term $R$ is a scalar invariant obtained from the Ricci tensor. The term $g_{\mu\nu}$ is the metric tensor which in special relativity has the form and components which can be taken to be given by
\begin{eqnarray}
g_{\mu\nu}=diag(-1,-1,-1,+1).\label{2.2}
\end{eqnarray}
The stress energy momentum tensor, describing as it does the local conditions of a physical continuous medium of some sort or other, is in general a very complicated mathematical physical object and only in the cases of a limited number of actual physical structures has it been used as a source term in Einstein's field equations to yield a reasonable geometric structure describable by the left hand side of the field equations. One such physical situation that is viable is what is called a perfect fluid which using special relativity and the assumption that such a fluid will have no internal stresses other than just pressure, $P$, $ T_{\mu\nu}$ can be expressed in the form
\begin{eqnarray}
T_{\mu\nu}=g_{\mu\sigma}g_{\nu\zeta}\left( \rho +\frac{P}{c^2}\right)\frac{dx^\sigma}{ds}\frac{dx^\zeta}{ds} - Pg_{\mu\nu}. \label{2.3}
\end{eqnarray}
If the local material described by the $T_{\mu\nu}$ has no $3$-dimensionsl macroscopic movement, then the $4$-dimensional velocity vectors in the expression (\ref{2.3}) can be replaced by velocity vectors of the form,
\begin{eqnarray}
\frac{dx^\sigma}{ds}=(0,0,0,c).\label{2.31}
\end{eqnarray}
to give
\begin{eqnarray}
T_{\mu\nu}=g_{\mu\sigma}g_{\nu\zeta}\left( \rho +\frac{P}{c^2}\right)c^2- Pg_{\mu\nu}. \label{2.32}
\end{eqnarray}
Thus we can read off from this that all component with $\nu\ne\mu$ are given by the first entry below as being zero. The elements with $\nu =\mu\ne 4$ are given by the second, third and fourth entries below. The (44) component is give as $c^2\rho$ at the fifth entry below.
\begin{eqnarray}
T_{\mu\nu}&=&0,\ \nu\ne\mu \label{2.33}\\
T_{11}&=&P\label{2.34}\\
T_{22}&=&P\label{2.35}\\
T_{33}&=&P\label{2.36}\\
T_{44}&=& c^2\rho .\label{2.37}
\end{eqnarray}
Thus the stress energy momentum tensor in the limiting situation chosen can be represented by the diagonal matrix
\begin{eqnarray}
T_{\mu\nu}=diag(P,P,P, c^2\rho) \label{2.38}
\end{eqnarray}
and the trace of this matrix called T which is the sum of these diagonal elements is the invariant scalar or zero order tensor, 
\begin{eqnarray}
T=3P + c^2\rho =c^2(3P/c^2+\rho ).\label{2.39}
\end{eqnarray}
From this we see that in general $T$ involves adding to the energy density an extra energy density $3P$ or equivalently adding to the mass density an extra mass density $3P/c^2$. It is well known that in relativity cosmological pressure is an additional contribution to mass density that modifies the mass density and so makes an extra relativistic contribution to Newton's law of gravitation. Thus an effective physical mass density is generated of amount $\rho_{eff}= 3P/c^2+\rho$. This is most easily seen from an equation that can be obtained from the Friedman equation for the acceleration field due to density. This equation is
\begin{eqnarray}
\frac{\ddot r }{ r }  = \frac{\Lambda c^2 }{3} - \frac{4\pi G}{3} \left( \rho + \frac{3 P}{c^2} \right) \label{2.40}
\end{eqnarray}
and includes a contribution from the Lambda term, $\Lambda c^2 $.
This is a very important equation in relation to the acceleration due to gravity at radius $r$ and how that depends on the mass density term $\rho $. Clearly, the pressure term $3P/c^2$ adds to the mass density to produce an effective or physical  mass density $3P/c^2+\rho $, the same addition we noted above in the invariant tensor $T$ derived from the full $ T_{\mu\nu}$ by contraction. This makes an extra contribution to the acceleration caused by gravity due to the mass density that occurs in Newtonian theory and indeed is a relativistic contribution that has been known about for years. This extra term makes very good sense in the density context and can be explained as follows. Suppose we wish to determine the magnitude of the acceleration $ a_g $ due to gravity at some distance from a {\it galaxy\/} of total mass $m_g$ using the Newtonian inverse square law formula,
\begin{eqnarray}
a_g = -\frac{m_g G}{r^2}.\label{2.41}
\end{eqnarray}
The minus sign in the above incorporates the fact that the acceleration is normally always towards the {\it source\/} of gravitational force. This formula will only be useful if we can find a value to give to the mass symbol $m_g$ of the galaxy.  However, galaxies are greatly diverse objects and their masses are obviously not just the sum of the masses of their individual component stars and planets. All the components of a galaxy are held together in relative motions or at rest by forces in the very simplest picture. All such forces and motions will modify the simple sum of mass components.  A very simplified type picture of a galaxy is a gas or a fluid with interacting particles so that the stress energy tensor used above does fit the bill and reduces our options in general relativity to that form. However, it should be clear that the stress energy momentum tensor can only be an hyper idealisation of the structure it represents, essentially smoothing out all the complications into the two entities density and pressure. Thus we have little choice but to describe a galaxy using a density function and a pressure term. Also we need to recognise that a galaxy will need to have some sort of bounded spherical volume $v_g$, say,  within which its mass density will account for all its mass and outside which the galaxy will have no mass. The frame of reference in which we wish to study the acceleration determines the frame of reference in which the source will be taken to be at rest and I think it is clear that in this frame the mass $m_g=\rho_{eff}v_g$ should be called the {\it rest mass\/} of the galaxy and as with rest mass in the case of elementary particles, in this frame, the galaxy will have no total overall motion but it can have the local type of random motions that are responsible for pressure as in the gas analogy, the contribution to the pressure term being strictly internal. Clearly the frame invariant $T$ term is just what is needed to emulate the frame invariant rest mass of an elementary particle.

Returning to the equation for acceleration (\ref{2.40}) we can see that the negative acceleration is reinforced by the additional {\it positive\/} pressure term $3P/c^2$. That is to say it helps holds the system together. 
This aspect prompts me to speculate that such an addition to the mass of a galaxy could be just what is needed to explain  {\it Dark Matter\/}. This is the extra mass on top of the usually assumed amount of visible mass due to average density calculations or mass density that is required to balance the dynamical bookkeeping. However, as we have seen, it would require a non zero pressure which as it stands is not a property of my dust universe model. However, I think having spatially extended regions, galaxies, scattered throughout the universe in which pressure is not zero whilst in the massively larger regions between them in which pressure is zero would be allowable with the overall effect being a dust universe. This is because the term involved in the acceleration formula is that due to normally gravitating material in which extra energy due to pressure would clump together with the density by which it was induced. Clearly this argument does not apply to the $\Lambda$ term because that is self repulsive material that does not clump and so the extra density due to pressure would disperse uniformly throughout space as is its origin density. As far as I know, this explanation for the existence of {\it dark matter\/} has not appeared before in spite of the mathematical physical structure concerned in general relativity having been known about for many years, see Rindler on pressure page $395$(\cite{03:rind}).     

It is possible to further simplify $T_{\mu\nu}$ by introducing the concept of equation of state for the fluid motion which is an equation expressing the pressure linearly as a function density similar to an equation used in thermodynamics, $P=RT/V$, taking  $\rho$ to have the mass density $RT/(V c^2\omega)$. The standard form for this equation of state as used in cosmology is
\begin{eqnarray}
\frac{P}{c^2}= \rho\omega,\label{2.4}
\end{eqnarray}
where $\omega$ is a dimensionless quantity with a numerical value that is determined by the physical character of the system, it can be negative, zero or positive. If the fluid particles are photons it has the value $1/3$. If the fluid particles are what is called dust that is to say they exert no fluid pressure it has the value $0$. If the fluid particles exerts a negative pressure it has to have a negative value, $\omega_\Lambda=-1\implies P= c^2\omega_\Lambda\rho_\lambda$, assuming that density is always positive.
Thus in these three cases explained above the pressure $P$ can be replaced in favour of $\rho$ so that the stress energy momentum tensor  invariant scalar, T,  at (\ref{2.39}) can be put into the forms listed below. The original  formula (\ref{2.39}) given first then the case with substitution for the pressure in the general case, $P=c^2\rho w$. Next the case with the pressure substitution for the photon case, $\omega =1/3$. Next with the pressure substitution for the dust case, $P=0$ and finally the case with pressure substitution for the negative pressure case, $\omega =-1$. 
\begin{eqnarray}
T&=& c^2(3P/c^2+\rho ),\ original formula \label{2.5}\\
T_g&=& c^2\rho(3\omega +1),\ general \label{2.6}\\
T_p&=& 2c^2\rho,\ photons \label{2.7}\\
T_d&=& c^2\rho ,\ dust \label{2.8}\\
T_n&=& -2 c^2\rho ,\ negative\ pressure.\label{2.9}
\end{eqnarray}
Using the original formula at (\ref{2.5}), equation (\ref{2.40}) can be expressed as
\begin{eqnarray}
\frac{\ddot r }{ r }  = \frac{\Lambda c^2 }{3} - \frac{4\pi G T}{3c^2}. \label{2.91}
\end{eqnarray}
From (\ref{2.9}) it can be seen that we can account for the $\Lambda$ term by making an identification with the appropriate, $T_n$, as exists with the $T$ term as follows
\begin{eqnarray}
\Lambda c^2&=& - \frac{4\pi G T_n}{c^2}\label{2.92}\\
&=& - \frac{4\pi G (-2 c^2)\rho }{c^2} \label{2.93}\\
&=& 8\pi G \rho.  \label{2.94}
\end{eqnarray}
Clearly $\rho$ above is Einstein's, $ \rho_{\Lambda,E}=\Lambda c^2/(8\pi G)$, but in the acceleration formula the Lambda term should be represented as 
\begin{eqnarray}
\frac{\Lambda c^2 }{3}=\frac{4\pi G \times2\rho }{3}, \label{2.95}
\end{eqnarray}
indicating that twice Einstein's dark energy, $\rho_\Lambda^\dagger = 2\rho_{\Lambda,E}$, is the physical energy density that causes the acceleration.
In reference (\cite{45:gil}), I showed that the general relativistic version for acceleration due to gravity for a mass $M_U$ embedded in and permeated by an infinitely extended uniformly field of $\Lambda$ dark energy corresponds exactly to the Newtonian inverse square law form provided, the Lambda density is taken to be $\rho_\lambda^\dagger$. This is easily derivable from (\ref{2.91}).
\begin{eqnarray}
\frac{\ddot r}{r}& = &4\pi r^3 G(\rho^\dagger_\Lambda -\rho)/(3 r^2)\label{99}\\
& = &{M^\dagger}_\Lambda G/r^2 - M_UG/r^2\label{100}\\
M^\dagger_\Lambda & = & 4\pi r^3 \rho^\dagger_\Lambda/3\label{101}\\
M_U & = & 4\pi r^3 \rho/3.\label{102}
\end{eqnarray}
\section{Appendix 1 Conclusions}
Using the physical information source term, $T_{\mu\nu}$,  from Einstein's field equations for general relativity and its invariant scalar contraction, $T$, I have reappraised the significance of the pressure variable $P$ that occurs therein.  I make use of well known and well established formulae which shows that the cosmological pressure $P$ measures {\it induced\/}  mass density additional to the mass density that is described by the $44$ component of $T_{\mu\nu}$. This is achieved using the well known details of the Newtonian limit of general relativity. In the case of Einstien's Lambda term, I use the same structured argument to show that the negative pressure associated with the lambda term similarly induces an additional density contribution so doubling up to the amount of dark energy to give for the {\it physically\/} present amount $\rho_\Lambda^\dagger=2\Lambda c^2/(8\pi G)$ or twice Einstein's value. This is equal to the value I have used for dark energy density in the papers in the sequence preceding this paper, there deduced by exclusively using the Friedman equations. I suggest that the increased amount of density implied for the density of normally gravitating material induced by its pressure could account for the missing dark matter an issue that greatly exercises the astronomical community. Both of these additional {\it dark\/} mass energy contributions could well have some bearing on the {\it Pioneer\/} anomaly problem.   
\vskip 0.5cm
\section{Appendix 2}
\centerline{\Large {\bf Physical Applicability of Self Gravitating}}
\centerline{\Large {\bf Isothermal Sphere Equilibrium Theory to}}
\centerline{\Large {\bf Dark Matter Galactic Halos}}
\vskip 0.35cm  
\centerline{February 8, 2011}
\vskip 0.75cm  
\section{Appendix 2 Abstract}
In this paper, a derivation of the basic gas self-gravity equilibrium equation is carried through in detail using a totally Newtonian macroscopic view of the problem. The standard derivation involves a mixed local microscopic and macroscopic approach. Consequently the final equation obtained here differs subtly but significantly from the usual formula. This is achieved by formulating the {\it self gravitating action\/} in Newtonian terms with both the gravitating source part and gravitated object part of the system both described macroscopically. This approach has the effect of giving thrust precedence over local pressure and thus taking into account the geometrically extended character of a spherical distribution. The equilibrium condition is obtained carefully taking into account that the source gravitation field will effectively act at the mass centroid of any object spherical shell. The mathematics of this is worked out in detail and the new equilibrium equation is obtained. The steps in the usual standard derivation are given to arrive at the standard equation. Solutions of the standard and new equation are obtained and comparisons are made. It is suggested that solutions of the new version of the equilibrium equation may be more appropriate for modelling galactic halos than were the standard solutions. It is also suggested that, if the new version of gas gravity equilibrium theory is accepted, then the Lane-Emden solutions that arise in this context from the standard approach will need to be reappraised in terms of the new derivation and its solutions.       

\vskip 0.2cm 
\centerline{Keywords: Cosmology, Dust Universe, Dark Energy, Dark matter}
\centerline{Cosmological Constant, Friedman Equations, Galactic Halo}
\centerline{Isothermal Sphere, Newton's Gravitation constant}
\vskip 0.3cm

\centerline{PACS Nos.: 98.80.-k, 98.80.Es, 98.80.Jk, 98.80.Qc}
\section{Appendix 2 Introduction}
\setcounter{equation}{0}
\label{sec-a2intro}
In a recent paper, appendix 2 of (\cite{58:gil}), I have made the case that the pressure term in Einstein's general relativity can account for the very large amount of matter that seems to missing when current astrophysics theory is used to study galactic kinematics and dynamics. It is generally assumed that the missing matter resides in the halos associated with galaxies or groups of galaxies. It is also generally agreed that the material of which these halos are formed is invisible in the sense that it does not emit or reflect light and so cannot be seen in the usual optical telescopes. The exact physical identity of this material is consequently considered to be a big mystery. A large range of macroscopic and microscopic  objects have been suggested as possible candidates to represent this missing material but so far no consensus has emerged. Although I am claiming that Einstein's pressure term can account for the {\it amount\/} of missing mass, I have made no identification of the actual physical nature or particle type that may be involved. The work in this paper was originally aimed at an identification of the possible dark mater halos as composed of a gas like substance spread out over large distances and held in place by its own positively self gravitating character. However, when I looked into the basic physics and mathematics \cite{70:wei} which is usually used to describe self gravitation equilibrium in astro-physics, I found a possible significant and interesting possible deviation from the standard theory. This deviation implies that all the work in astrophysics on the isothermal sphere problem and its offshoots such as the Lane-Emden equation and its polytropic solutions may need some substantial interpretational revision. 
\section{Galactic Gas Halo}
Consider a spherical volume of astronomical 3-space within which any external gravitational influence is negligible but which does contain only some specific gas like material under its own gravitational influence. The first question that arises from this idea is, can such a system be in static equilibrium? This issue I shall now investigate using Newtonian gravitation theory with the simple gas dynamics theory of the ideal gas equation. I shall also assume the gas is compressible and is described by a mass density function, $\varrho (r)$, only dependent on radial distance $r$ from the centre of the sphere, with a temperature which remains constant throughout the whole volume of the sphere. The density function I shall be using, $ \varrho (r)$, mostly in this work is mass density per unit radius and I shall call it the {\it one dimensional density\/}. Its relation to the usual mass density function, $ \varrho_\prime ({\bf r})$, mass per unit volume which I shall call the {\it three dimensional density\/}, distinguished here by the subscripted prime, is  $ \varrho (r) = 4\pi r^2\varrho_\prime (r)$, valid when the density is purely spherically symmetric. This papers develops the gas self gravity problem along different lines from that which is usually employed with rather subtle but significant differences from the usual theory emerging. This different point of view of the problem is fully {\it macroscopically\/} orientated rather than partly {\it local microscopic\/} as is the usual point of view. This difference which will be explained in more detail later is partly implemented by my use of the one dimensional density, $\varrho$, rather than the three dimension density, $\varrho_\prime$, and will become clearer as the theory is developed. Thus to avoid confusion with the standard theory I shall use the {\it var}-$\rho$ version, $\varrho (r)$, for density when discussion the new theory and revert to the usual form $\rho (r)$ when discussing the standard theory.
The assumption confining the mass density to spherical symmetry is simply to reduce the complexity of the problem. Because of the space variable, $r$, involved the system implied by the description is of a {\it inhomogeneous\/} type. The system to be studied here is also extremely non-linear. Such systems are best described by pure non-dimensioned parameters and can be very involved mathematically. Thus to avoid one of dimensionality problems that can arise, I shall settle the way functions of the usual {\it dimensioned\/} space variable, $r$, are defined at the start as explained next.
Suppose that $r$ is the usual length dimensioned position variable that is often used to represent radial distance from the origin as it would appear in the three dimensioned density $\varrho_\prime (r)$. The $r$ parameter can be replaced with a dimensionless radius variable $\acute{r}  =r/r_0$ where $r_0$ is any suitable dimensioned length with $ \varrho_\prime $ replaced with $ \varrho_{\prime,0} $, where $0$ is a temporary subscript, zero, representing a change in functional form,  as below at (\ref{g01}) so that
\begin{eqnarray}
\varrho_\prime (r)&=&\varrho_\prime (\acute{r}
r_0)=  \varrho_{\prime,0} (\acute{r})= \varrho_{\prime,0} (r/r_0) \label{g01}\\
\frac{\partial\varrho_\prime (r)}{\partial r}&=& \frac{\partial\varrho_{\prime,0} (r/r_0) }{\partial r}= \frac{\partial\varrho_{\prime,0}(r/r_0) }{\partial (r/r_0) r_0}=\frac{\partial\varrho_{\prime,0}(r/r_0) }{\partial r} \label{g011}\\
\int_0^{r_d}\varrho_\prime (r)dr&=& \int_0^{r_d}\varrho_{\prime,0} (\acute r r_0)d(\acute{r}r_0)= \int_0^{r_d}\varrho_{\prime,0} (r^\prime)dr^\prime ,\label{g012}
\end{eqnarray}
where after the last equality above $r^\prime$ is a dummy variable as is the $r$ in the first entry at the line above.
The simple ideal gas equation written below is usually not directly applicable when inhomogeneous systems are involved as normally the gas density, $\varrho_\prime $, will not depend on position within the gas container.
\begin{eqnarray}
P&=& \frac{\varrho_\prime RT}{M}\label{g0}\\
&=& \varrho_\prime c^2\omega=\sigma^2\varrho_\prime \label{0g}\\
\omega &\equiv& \frac{RT}{Mc^2}.\label{1g}
\end{eqnarray}
Equation (\ref{0g}) gives the usual cosmological version of the relation between pressure and density and equation (\ref{1g}) identifies $\omega=(\sigma/c)^2$ when the system obeys the ideal gas equation law.
However, we can bypass the non-applicability problem by the procedures discussed in the next few pages.

 Consider a concentric sphere with radius $r$ within the extent of a spherical gas distribution. The surface area of this sphere will be $4 \pi r^2$. Assuming there is a definite density function, $\varrho (r)$, mass per unit radius giving the mass density at all radial positions $r$, we can write down the exact amount of mass $M(r_d, r_u)$ between two concentric spheres of radii $r_d$ and $r_u$ with {\it r-up\/} greater than {\it r-down\/}, $r_u> r_d $, as follows 
\begin{eqnarray}
M(r_d, r_u)= \int_{r_d}^{r_u} \varrho (r) dr.\label{g1}
\end{eqnarray}
Using this formula, we can also write down the total quantity of mass within the sphere of radius r-down as follows
\begin{eqnarray}
M(0, r_d)= \int_{0}^{r_d} \varrho (r) dr.\label{g2}
\end{eqnarray}
Both of these quantities of mass involve mass over a great extent of the entire sphere and so I shall designate them as macroscopic quantities rather than local quantities in contrast with local quantities would only involve mass in the vicinity of a definite point.
The mass within the annular region between $r_d$ and $r_u$ will have radial mass centroid, at a radius $r_c$ say, with $r_d<r_c<r_u$ determined  by  the mass density function density $\varrho(r)$. Thus there will be an effective  Newtonian gravitational square law force, $F_G(r_c)$, of attraction acting effectively at the centroidal position $r_c$ on the annular region between $r_d$ and $r_u$ from all the mass within the $r_d$ sphere of magnitude 
\begin{eqnarray}
F_G(r_c)= -\frac{M(r_d, r_u)M(0, r_d) G}{r_c^2}.\label{g3}
\end{eqnarray}
If the mass quantity, $M(r_d, r_u)$, subject to this gravitational pull were not otherwise constrained it would experience, according to Newtonian theory, a radial inwards acceleration due to gravity of amount given by $\ddot r_c$ below
\begin{eqnarray}
F_G(r_c)= M(r_d, r_u)\ddot r_c.\label{g4}
\end{eqnarray}
However, the gaseous material enclosed by the $r_d,r_u$ radii is subject also to a force thrust $F_P(r_d)$ at the $r_d$ surface due to pressure, $P(r_d)$, away from the origin from material below and a force thrust due to pressure, $P(r_u)$, towards the origin from material above the annular region. These thrusts also an aspect of a macroscopic view  will be determined by the surface areas of the bounding spheres times the local pressures as
\begin{eqnarray}
F_P(r_d)&=&+4 \pi r_d^2P(r_d)\label{g5}\\
F_P(r_u)&=&-4 \pi r_u^2P(r_u).\label{g6}
\end{eqnarray}
It follows that there can be taken to be a resultant thrust from pressure away from the origin, $F_P(r_c)$, due to pressure on the annulus $r_d,r_u$ acting at a mean or centroidal radius $r_c$  given by
\begin{eqnarray}
F_P(r_c)=F_P(r_d)+ F_P(r_u)>0\label{g7}
\end{eqnarray}
and as indicated this needs to be positive, away from the origin, if equilibrium between the gravity effect, assumed acting negatively or towards the origin,  and the pressure effect is to occur.
$F_P(r_c)$ can be expressed exactly as the integral below
\begin{eqnarray}
F_P(r_c)=F(r_d)+ F(r_u)= -\int_{r_d}^{r_u}\frac{\partial (4\pi r^{ 2}P(r))}{\partial r}dr\label{g8}
\end{eqnarray}
because $r_u>r_d$.
Thus for a static system to pertain the upward thrust due to pressure must balance the downward force due to gravity implying 
\begin{eqnarray}
F_P(r_c)+ F_G(r_c)=0\label{g9}
\end{eqnarray}
and this is an exact statement from classical Newtonian theory with the usual understanding of the meaning of the term pressure. Writing out this equilibrium equation in detail using (\ref{g2}), (\ref{g3}) and (\ref{g8}),we obtain
\begin{eqnarray}
\int_{r_d}^{r_u}  \frac{\partial (4\pi r^{ 2}P(r))}{\partial r}dr  +\frac{ G \int_{r_d}^{r_u} \varrho (r) dr \int_{0}^{r_d} \varrho (r) dr }{r_c^2} =0\label{g10}
\end{eqnarray}
whereas in the non equilibrium case we obtain, the exact form
\begin{eqnarray}
\int_{r_d}^{r_u}  \frac{\partial (4\pi r^{2}P(r))}{\partial r}dr  +\frac{ G \int_{r_d}^{r_u} \varrho (r) dr \int_{0}^{r_d} \varrho (r) dr }{r_c^2} =+\ddot r \int_{r_d}^{r_u} \varrho (r) dr.\label{g11}
\end{eqnarray}
I shall from now on confine discussion to the equilibrium case (\ref{g10}). An inspection of that equation reveals a dependence on two independent radius variables $r_u$ and $r_d$ and on a further radial variable $r_c$ which is the radial density mean of the first two and so can be expressed as
\begin{eqnarray}
r_c(r_d,r_u)\int_{r_d}^{r_u}\varrho (r)dr&=&\int_{r_d}^{r_u}r\varrho (r)dr\label{g12}\\
\frac{\partial r_c(r_d,r_u)}{\partial r_u}\int_{r_d}^{r_u}\varrho (r)dr+ r_c(r_d,r_u)\varrho (r_u)&=& r_u\varrho (r_u)\label{g12b}\\
\frac{\partial r_c(r_d,r_u)}{\partial r_d}\int_{r_d}^{r_u}\varrho (r)dr- r_c(r_d,r_u)\varrho (r_d)&=&- r_d\varrho (r_d).\label{g12bb}
\end{eqnarray}
The objective is to find the differential equation for $\varrho(r)$ implied by the integral equation (\ref{g10}). This will involve differentiating through the equation by either $r_u$ or $r_d$ to undo the integrations. Thus if in proceeding on that course we encounter terms involving functions of the two variables other than as integral limits such as the $r_c^{-2}(r_u,r_d)$ in (\ref{g11}), they will have to be differentiated taking into account the formula (\ref{g12}).
In anticipation of this complication suppose we encounter a differentiable function $f(r_c)$. If we differentiate this with respect to $r_u$ for example, we get
\begin{eqnarray}
\frac{\partial f(r_c)}{\partial r_u}&=& \dot f(r_c)\frac{\partial r_c(r_d,r_u)}{\partial r_u} \label{g13}\\
&=& \dot f(r_c) \frac{(r_u -r_c(r_d,r_u))\varrho (r_u)}{\int_{r_d}^{r_u}\varrho (r)dr} \label{g14}\\
\lim_{r_d\rightarrow r_u}\frac{\partial f(r_c)}{\partial r_u}&\rightarrow&\dot f(r_u)\frac{(r_u -r_c(r_d,r_u))\varrho (r_u)}{(r_u-r_d)\varrho (r_c)} \label{g15}\\
&\rightarrow&\dot f(r_u) \label{g16}\\
E_m(r_u)= \frac{\partial r_c(r_d,r_u)}{\partial r_u} &=&\frac{ (r_u -r_c(r_d,r_u))\varrho (r_u)}{\int_{r_d}^{r_u}\varrho (r)dr} \label{g17}\\
\lim_{r_d\rightarrow r_u}\frac{\partial r_c(r_d,r_u)}{\partial r_u} &\rightarrow&1\label{g17a}
\end{eqnarray}
The last two lines above arise from taking the limit $r_d$ approaching $r_u$ when effectively $r_c$ approaches $r_u$ from below and $r_d$ from above.
Thus the prescription for dealing with functions such as $f(r_c)$ is differentiate with respect to its argument $r_c$ to get the result (\ref{g14}) and in the final result after all differentiations take the limit $r_d\rightarrow r_u$ to replace $r_c$ with $r_u$ in $\dot f(r_u)$  and remove the extraneous multiplier of $\dot f(r_u)$ at (\ref{g15}).
The extraneous multiplier $E_m(r_u)$ for the first differentiation with respect to $r_u$ has been recorded at {\ref{g17}} for future reference.
 The result of a first differentiation through of (\ref{g10}) by $r_u$ before taking any limits is 
\begin{eqnarray}
4\pi \frac{\partial (r_u^2P(r_u))}{\partial r_u}+\frac{G\varrho (r_u)\int_0^{r_d}\varrho(r)dr}{r_c^2(r_d,r_u)}&-&2\frac{G\int_{r_d}^{r_u}\varrho(r)dr\int_0^{r_d}\varrho (r)dr E_m(r_u)}{r_c^3(r_d,r_u)}.\nonumber\\
&=&0\label{g18}
 \end{eqnarray}
Looking back at the definition of $E_m(r_u)$ at (\ref{g17}), we see that this last equation can be simplified through the following equation and then further to (\ref{g19c})  
\begin{eqnarray}
4\pi \frac{\partial (r_u^2P(r_u))}{\partial r_u}+\frac{G\varrho (r_u)\int_0^{r_d}\varrho(r)dr}{r_c^2(r_d,r_u)}&-&2\frac{G (r_u -r_c(r_d,r_u))\varrho (r_u) \int_0^{r_d}\varrho (r)dr }{r_c^3(r_d,r_u)}\nonumber\\
&=& 0\label{g19b}
\end{eqnarray}
\begin{eqnarray}
4\pi \frac{ \partial (r_u^2P(r_u))}{\partial r_u}&-&\left(\frac{2r_u}{r_c^3(r_d,r_u)}-\frac{3}{ r_c^2(r_d,r_u)}\right)G\varrho (r_u) \int_0^{r_d}\varrho (r)dr \nonumber\\
&=& 0.\label{g19c}
\end{eqnarray}
Inspection of this last equation shows that if the limit $r_d\rightarrow r_u$ is taken which as we have seen involves $r_c\rightarrow r_u$ from below, we obtain the equation
\begin{eqnarray}
4\pi \frac{\partial (r_u^2P(r_u))}{\partial r_u}&+&\left(\frac{ G\varrho (r_u)}{ r_u^2}\right) \int_0^{r_u}\varrho (r)dr=0.\label{g19d}
\end{eqnarray}
 Replacing $P(r_u)$ with the ideal gas equation form, $P(r_u)= c^2 \varrho_\prime (r_u) \omega (r_u)$, and then expanding, we have 
\begin{eqnarray}
2 r_u \varrho_\prime (r_u)\omega (r_u) + r_u^2\frac{\partial \varrho_\prime (r_u)}{\partial r_u}\omega (r_u)+ r_u^2\varrho_\prime (r_u) \frac{\partial \omega (r_u)}{\partial r_u}&+&\frac{ G\varrho (r_u)}{4\pi c^2  r_u^2} \int_0^{r_u}\varrho (r)dr\nonumber\\
=0.\label{g19e}
\end{eqnarray}
Replacing the $\varrho_\prime (r_u)= \varrho (r_u) /(4\pi r_u^2)$ density with its form in terms of $\varrho (r_u)$ we get
\begin{eqnarray}
2 r_u \frac{\varrho(r_u)}{4\pi r_u^2}
\omega (r_u) &+& \left(\frac{\partial \varrho (r_u)}{\partial r_u}-\frac{2 \varrho (r_u)}{r_u}\right)\frac{\omega (r_u)}{4\pi}\nonumber\\
&+& r_u^2 \frac{\varrho(r_u)}{4\pi r_u^2}
\frac{\partial \omega (r_u)}{\partial r_u}+\frac{ G\varrho (r_u)}{4\pi c^2  r_u^2} \int_0^{r_u}\varrho (r)dr=0.\label{g19e1}
\end{eqnarray}
This last equation can be expressed as
\begin{eqnarray}
2\frac{\varrho(r_u)}{ r_u}
\omega (r_u) &+& \left(\frac{\partial \varrho (r_u)}{\partial r_u}-\frac{2 \varrho (r_u)}{r_u}\right)\omega (r_u)\nonumber\\
&+&\varrho(r_u)
\frac{\partial \omega (r_u)}{\partial r_u}+\frac{ G\varrho (r_u)}{ c^2  r_u^2} \int_0^{r_u}\varrho (r)dr=0.\label{g19e2}
\end{eqnarray}
The equation above arises from the  assumption that the form of the equation of state for the system involves the dimensionless quantity  $\omega$ depending on the variable $r_u$. I shall here make the next step here by using the more restrictive assumption that $\omega$ is a constant which is effectively the same as taking the temperature as being constant throughout the gas mass distribution. I shall possibly return to the more general case at a later date. Under this restriction and having multiplied through by $r_u^2$ the equation takes the form
\begin{eqnarray}
+ \left(\frac{\partial \ln (\varrho (r_u))}{\partial r_u} \right) =-\frac{ G}{\omega r_u^2c^2 } \int_0^{r_u}\varrho (r)dr\label{g19e3}
\end{eqnarray}
which I shall refer to as the fundamental $\varrho$ self gravity equilibrium gas density equation.
This can be differentiated with respect to $r_u$ to give
\begin{eqnarray}
 2r_u\frac{\partial \ln (\varrho (r_u))}{\partial r_u}
+ r_u^2\frac{\partial^2 \ln (\varrho (r_u))}{\partial r_u^2} = -\frac{ G}{ c^2\omega}\varrho (r_u) = - \acute{\varrho}(r_u).\label{g19h}
\end{eqnarray}
It looks as though it might be more easily integrated if $ \ln \acute{\varrho} (r_u)$ which differs from $\ln (\varrho(r_u))$ by a constant as can be seen from (\ref{g19h}) is denoted by the dimensionless quantity $ X(r_u)= \ln \acute{\varrho} (r_u)$ and then the equation is written in the form
\begin{eqnarray}
r_u^2\frac{\partial^2 X(r_u)}{\partial r_u^2} +2r_u\frac{\partial X(r_u)}{\partial r_u}
&=&-\exp (X(r_u)). \label{g19j}
\end{eqnarray}
In the step from equation (\ref{g19h}) to equation (\ref{g19j}) we have avoided obtaining an equation that is dimensionally dubious by introduce a dimensionless density function denoted by the acute accent sign indicating that it is dimensionless and also by denoting $\ln \acute{\varrho} (r_u)$ by $X(r_u)$ as follows
\begin{eqnarray}
\acute{\varrho} (r_u)&=& \frac{ G}{ c^2\omega}\varrho (r_u) \label{g19j1}\\
X(r_u)=\ln\acute{\varrho} (r_u) &=&\ln \varrho (r_u)+\ln (\frac{G}{ c^2\omega})\label{g19j2}\\
\frac{\partial X(r_u)}{\partial r_u}= \frac{ \partial \ln \varrho(r_u)}{\partial r_u}
&=&\frac{\partial \ln\acute{\varrho} (r_u)}{\partial r_u}\label{g19j3}
\end{eqnarray}
 with the last two equations a consequence of the first so that now equation (\ref{g19j}) can be written as follows
\begin{eqnarray}
r_u^2\frac{\partial^2 X(r_u)}{\partial r_u^2} +2r_u\frac{ \partial X(r_u)}{\partial r_u}
=-\acute{\varrho} (r_u) =-\exp (X(r_u)). \label{g19k}
\end{eqnarray}
Only $r_u$ appears in this equation so that the subscript $u$ can from now on be dropped and at the same time the $r_u$ variable can be replaced with a dimensionless radius variable $\acute{r}=r_u/r_0$ where $r_0$ is any suitable dimensioned length with also $X(r_u)$ and $ \acute{\varrho} (r_u)$ replaced with $X(\acute{r})$ and $\acute{\varrho} (\acute{r})$ respectively as discussed earlier so that 
\begin{eqnarray}
X(r_u)=X(\acute{r} r_0) &\rightarrow & X(\acute{r})  \label{g19k0}\\
\acute{\varrho} (r_u)=\acute{\varrho} (\acute{r} r_0)&\rightarrow & \acute{\varrho} (\acute{r})\label{g19k1}
\end{eqnarray}
and then the equation simplifies through (\ref{g20k1}) to (\ref{g20k1.1}). 
\begin{eqnarray}
r^2\frac{\partial^2 X(r)}{\partial r^2} +2r\frac{ \partial X(r)}{\partial r}
&=&-\acute{\varrho} (r) =-\exp (X(r)). \label{g20k1}\\
\frac{\partial}{\partial r}\left(r^2\frac{ \partial ln\acute{\varrho}(r)}{\partial r}\right)
&=& - \acute{\varrho} (r)=-\frac{4\pi r_0^2G}{c^2\omega}r^2\varrho_\prime (r), \label{g20k1.1}
\end{eqnarray}
where the acute sign on the $\acute{r}$ has been dropped on the understanding that $r$ from now on will represent a dimensionless parameter.
Equation (\ref{g20k1}) or (\ref{g20k1.1}) is now dimensionally unambiguous in all respects. We should note for future reference, the structure of equation (\ref{g20k1.1}) because this replaces the equation in the standard theory that is used by transforming variables to deduce the Lane-Emden equation and so claim that the Lane-Emden equation solutions also satisfy the gravitational thermal equilibrium equation solutions.  This structure can be described as follows.
Equation (\ref{g20k1.1}) is the basic thermal equilibrium equation that arises from the alternative fully macroscopic derivation used in the paper. It should be noticed that the non-linear differential equation (\ref{g20k1.1}) involves $r^2$ both within the operand of first differentiation with respect to $r$ on the far left and also on the far right hand side  as a multiplier outside the differentiated zone, with $\acute{\varrho}$ appearing within the differentiation zone and also outside the differentiation zone after the first equality. Importantly, $r^2$ and the {\it same\/} density function do not appear together inside the differentiation zone and outside the differentiation zone which ever equality is used to express the relation. These remarks should be compared with the remarks after equation (\ref{g20j14.1}).   

The issue now is the mathematical problem of finding the solution of the non linear differential equation (\ref{g20k1}) or with its rearranged form with $X(r)= \ln\acute{\varrho} (r)$ at (\ref{g20k1.1}).
It can be seen that an immediate formal integral is given by
\begin{eqnarray}
\frac{\partial \ln\acute{\varrho} (r)}{\partial r}
&=& -\frac{1}{r^2}\int_0^r\acute{\varrho} (r^\prime) d r^\prime =-\frac{G}{c^2\omega r^2}\int_0^r\varrho (r^\prime) d r^\prime. \label{g20k22}
\end{eqnarray}
Equation (\ref{g20k22}) is not a surprising result as it is just one of the originating equations  for the alternative structure, the fundamental self gravity equilibrium gas
density equation. This last equation can be taken back further towards physical basics by writing it as 
\begin{eqnarray}
\acute{\varrho} (r)&=&\acute{\varrho} (0) \exp \left(-\int_0^r \frac{GM(r^\prime) d r^\prime }{ c^2\omega r^{\prime 2}}\right)
\label{g20k23}\\
M (r)&=&\int_0^r\varrho (r^\prime)d r^\prime \label{g20k23.1}
\end{eqnarray}
and if we define the gravitational potential $\varPhi (r)$ simultaneously restoring its dimensionality with the multiplied $\sigma^2$  as 
\begin{eqnarray}
\varPhi (r) = \sigma^2\int_0^r \frac{GM (r^\prime) d r^\prime }{c^2\omega r^{\prime 2}}
\label{g20k24}
\end{eqnarray}
then the dimensionless density $ \acute{\varrho} $ of the alternative structure becomes
\begin{eqnarray}
\acute{\varrho} (r)= \acute{\varrho} (0) \exp \left(-\frac{\varPhi (r) }{\sigma^2}\right).
\label{g20k25}
\end{eqnarray}
Thus the {\it one\/} dimensional density, $\acute{\varrho} (r)$, of this derivation is of the same {it form\/} as the three dimensional density  of the standard solution, $\rho_\prime (r)$, at (\ref{g20k6}) except for the different gravitational potential functions.
Let us next look in more detail at how these steps and this equation relates to known solutions of the gas gravity equilibrium problem. The usual differential equation for the spherical gravity equilibrium problem is derived as follows.
\section{The Standard Equations}
\setcounter{equation}{0}
\label{sec-tse}
The standard equations used to study the gas gravity equilibrium problem are very well established and derive entirely from continuum classical fluid dynamics. The starting point in the derivation process can be taken to be two equations, Euler's equation and the equation that defines the gravitational potential $\Phi$ in the fluid context, the first two equations displayed consecutively below. Equation (\ref{g20j8.1}) defines the fluid vorticity ${\bf \Omega}$. 
\begin{eqnarray}
\frac{\partial {\bf v} }{\partial t} +\frac{1}{2} {\bf\nabla}(v^2) - {\bf v} \wedge {\bf \Omega} &=& -\frac{{\bf\nabla}P}{\varrho_\prime}- {\bf\nabla} \Phi \label{g20j6.1}\\
{\bf\nabla}^2\Phi&=&4\pi G\varrho_\prime \label{g20j7.1}\\
{\bf \Omega}&=& {\bf\nabla}\wedge {\bf v}.\label{g20j8.1} 
\end{eqnarray}
It follows that if we confine attention to the completely {\it static\/} equilibrium situation such that ${\bf v}=0$ then the Euler equation simplifies to the line below and this is followed by the result of further applying the divergence operator $ {\bf\nabla\cdot}$ at (\ref{g20j10.1}).
\begin{eqnarray}
0&=& - \frac{{\bf\nabla}P}{\varrho_\prime }- {\bf\nabla}\Phi \label{g20j9.1}\\
{\bf\nabla\cdot}\left( \frac{ {\bf\nabla}P}{\varrho_\prime } \right)&=&-{\bf\nabla}^2 \Phi. \label{g20j10.1}
\end{eqnarray}
If we now further restrict the discussion to the pure spherically symmetric situation when the density function and the pressure function will only depend on the radius variable $r$, the gradient operator and the Laplacian operator become respectively 
\begin{eqnarray}
{\bf\nabla}& \rightarrow& \hat r\frac{\partial}{\partial r} \label{g20j11.1}\\
{\bf\nabla}^2&\rightarrow & \frac{\partial r^2\partial }{r^2\partial r\partial r }\label{g20j12.1}\\
\frac{\partial P }{\rho_\prime \partial r} &=& - \frac{\partial\Phi}{\partial r}=\frac{\partial (\sigma^2\rho_\prime (r)) }{\rho_\prime \partial r}  \label{g20j13.12}\\
{\bf\nabla\cdot}\left( \frac{ {\bf\nabla}P}{\rho_\prime } \right)&=&-{\bf\nabla}^2 \Phi= \frac{\partial r^2\partial \Phi }{r^2\partial r\partial r } =\frac{\partial r^2\partial P }{r^2\partial r\rho_\prime \partial r }= \frac{\sigma^2\partial r^2\partial \rho_\prime }{r^2\partial r\rho_\prime \partial r } \label{g20j13.1}\\
-4\pi G\rho_\prime(r)  &=& \frac{\sigma^2\partial}{r^2\partial r}\left(r^2\frac{ \partial \ln\rho_\prime(r)}{\partial r}\right)\label{g20j14.1}\\
-\frac{G}{\sigma^2}\rho (r) &=& \frac{\partial}{\partial r}\left(r^2\frac{ \partial \ln\acute{\rho}(r)}{\partial r}\right)
=-\acute{\rho} (r) \label{g20j14.2}
\end{eqnarray}
again using
 the acute sign to denote a dimensionless function at the last equality above.
Equation (\ref{g20j14.1}) is the basic thermal equilibrium equation that supplies the standard argument for the transformation of variables that generates the Lane-Emden equation and so associates the polytropic solutions of the Lane Emden equation with the self gravitating equilibrium equation. It should be noticed that the non-linear differential equation (\ref{g20j14.1}) involves $r^2$ both within the operand of first differentiation with respect to $r$ on the right hand side of the first equality and also as a divisor outside the differentiation under the $\sigma^2$ with the same function, $\rho_\prime(r)$, appearing within the differentiation zone and outside the differentiation zone on the far left. Importantly, $r^2$ and the {\it same\/} density function, $\rho_\prime$, do appear together inside the differentiation zone and outside the differentiation zone. This is the main distinguishing feature of the standard equilibrium from the alternative deduced in this paper.
These remarks should be compared with the remarks after equation (\ref{g20k1.1}).
Having got to the differential equation (\ref{g20j14.1}), we can now identify Newtonian analogues of the fluid structure objects by firstly integrating once after multiplying through by $r^2$ as follows
\begin{eqnarray}
- \int_0^r 4\pi G\rho_\prime (r^\prime)r^{ \prime 2}d r^\prime&=& \frac{\sigma^2 r^2\partial\ln\rho_\prime }{\partial r }. \label{g20j15.1}
\end{eqnarray}
Thus by noting that  $M(r)=\int_0^r 4\pi \rho_\prime (r^\prime)r^{ \prime 2}d r^\prime $, we find the analogue of Newton's equation immediately below followed by Newton's equation for acceleration in terms of the radial component of force, $f_r$, {\it per unit mass\/} of influenced particle for comparison. 
\begin{eqnarray}
- \frac{ G M(r)}{ r^2}&=& \frac{\sigma^2 \partial\ln\rho_\prime (r) }{\partial r }=\frac{\partial P (r)}{\rho_\prime (r)\partial r}=-\frac{\partial \Phi(r)}{\partial r}\label{g20j16.1}\\
- \frac{ G M(r)}{ r^2}&=&f_r=-\frac{\partial \Phi(r)}{\partial r}. \label{g20j16.12}
\end{eqnarray}
Thus the fluid gravitational potential $\Phi$ should be indentified as the potential for force per unit mass or the local {\it acceleration potential\/}. Clearly this is the principal of equivalence under fluid conditions.
The definition of the total mass within the radial limits $(0,r)$, $M(r)$, is clearly the macroscopic quantity of mass which is responsible for attracting any mass outside that region towards the spherical centre at $r=0$ and it takes the same form in the alternative discussion used in this paper. This constitutes the gravitational active mass of the Newtonian picture so that with respect to that mass both arguments are macroscopic. However, the standard theory effectively uses {\it local density\/} to represent the attracted massive Newtonian object. The alternative theory given above uses a transversally fully extended thick shell of mass to represent the attracted Newtonian mass so that the alternative theory is fully macroscopic in Newtonian terms in contrast with the standard theory as being semi-macroscopic. 
Collecting all the results together for the standard theory, we have
\begin{eqnarray}
\frac{\partial P(r)}{\partial r}&=&\sigma^2\frac{\partial \rho_\prime(r)}{\partial r}=-\rho_\prime (r)\frac{G M(r)}{r^2} =-\rho_\prime (r)\frac{\partial \Phi(r)}{\partial r} \label{g20k2}\\
\Phi (r)&=&\int_0^r \frac{G M (r^\prime) dr^\prime}{r^{\prime 2}}
 \label{g20k4ab2}\\
&=&-\frac{G( M(r)+ \tilde M(r))}{r}\label{g20k4ab}\\
M(r) &=& \int_0^r4\pi r^{\prime 2}\rho_\prime (r^\prime)dr^\prime\label{g20k3}\\
P(r)&=&\rho_\prime (r)\sigma^2\label{g20k4}\\
\tilde M(r)&=&r\int_0^r 4\pi r^\prime \rho_\prime (r^\prime)dr^\prime\label{g20k4ac}\\
\frac{\partial\ln \rho_\prime (r)}{\partial r}&=&- \frac{G}{ \sigma^2r^2}\int_0^r4\pi r^{\prime 2}\rho_\prime (r^\prime)dr^\prime =-\frac{\partial \Phi(r)/\sigma^2}{ \partial r} \label{g20k5}\\
\rho_\prime (r)&=& \rho_\prime (0)\exp \left(-\frac{\Phi (r)}{ \sigma^2}\right)  \label{g20k6}\\
\rho_{\prime ,sis} (r)&=& \frac{\sigma^2}{2\pi G r^2}\rightarrow \rho_{sis} (r)=4\pi r^2\rho_{\prime ,sis} (r)= \frac{2 \sigma^2}{G}.
 \label{g20k6.1}
\end{eqnarray}
The subscript prime here being used to draw attention to the fact that the usual analysis involves the three dimensional density function $\rho_\prime (r)$ in contrast with the one dimensional density, $\varrho (r)$, density that I have mostly used earlier in this paper. The density, $\rho_{\prime ,sis}$, at (\ref{g20k6.1}) is a well known and rather simple standard theory solution. This solution is easily obtained assuming it is of the form $\rho_\prime (r)=C/r^2$ where $C$ is a constant, using the standard equation as follows with the result appearing at (\ref{g20k10a})
\begin{eqnarray}
\frac{\partial\ln \rho_\prime (r)}{\partial r}&=& - \frac{G}{ \sigma^2r^2}\int_0^r4\pi r^{\prime 2}\rho_\prime (r^\prime)dr^\prime \label{g20k7a}\\
0-\frac{2}{r}&=& - \frac{G}{ \sigma^2r^2}\int_0^r4\pi r^{\prime 2}\rho_\prime (r^\prime)dr^\prime \label{g20k8a}\\
-2r&=& - \frac{G}{ \sigma^2}\int_0^r4\pi r^{\prime 2}\rho_\prime (r^\prime)dr^\prime \label{g20k9a}\\
2&=& \frac{G}{\sigma^2}4\pi r^2\rho_\prime (r)= \frac{G}{\sigma^2}4\pi r^2\rho_{\prime ,sis} (r). \label{g20k10a}
\end{eqnarray}
The equation given by the first equality at (\ref{g20k5}) is repeated below followed by the equivalent $\varrho$ equation from (\ref{g19e3}).
\begin{eqnarray}
\frac{\partial\ln \rho_\prime (r)}{\partial r}&=& - \frac{G}{ \sigma^2r^2}\int_0^r4\pi r^{\prime 2}\rho_\prime (r^\prime)dr^\prime \label{g20k7}\\
\frac{\partial \ln \varrho (r_u)}{\partial r_u} &=&-\frac{ G}{ \omega c^2 r_u^2 } \int_0^{r_u}\varrho (r)dr =- \frac{G}{ \omega c^2 r_u^2 }\int_0^{r_u}4\pi r^{\prime 2}\varrho_\prime (r^\prime)dr^\prime \label{g20k8}\\
\varrho (r) &=& 4\pi r^2\varrho_\prime \label{g20k9}\\
\sigma^2&=&\omega c^2. \label{g20k10}
\end{eqnarray}
Apart from a minor difference in notation and taking into account the last two equations above, a first impression is that these two equations contain the same physical information about the gravity system but express it in terms of the two different densities. However, this is a false impression as can be seen by expressing the second equation in terms of $\varrho_\prime$ a result given next and simplified at the following line.
\begin{eqnarray}
\frac{\partial (\ln \varrho_\prime (r_u)+\ln(r_u^2)+\ln(4\pi))}{\partial r_u} &=&-\frac{ G}{ \omega c^2 r_u^2 } \int_0^{r_u} 4\pi r^2\varrho_\prime (r) dr\label{g20k11}\\
\frac{\partial \ln \varrho_\prime (r_u)}{\partial r_u}+ \frac{2}{r_u} &=&-\frac{ G}{ \omega c^2 r_u^2 } \int_0^{r_u} 4\pi r^2\varrho_\prime (r) dr. \label{g20k12}
\end{eqnarray}
If we now compare (\ref{g20k12}) with (\ref{g20k7}) we see the major difference between their descriptions of the {\it three\/} dimensional densities $\rho_\prime$ or $\varrho_\prime$ alone is the additional term $2/r_u$ with its strong divergence for small $r_u$ in the $\varrho$ system. There is also a superficial appearance  from these two equations that they give the {\it same\/} solution, (\ref{g20k6}), for the two {\it distinctly\/} different quantities the one dimensional density from the $\varrho$ system and the three dimensional density from the standard system $\rho_\prime$ and this appears to occur because the same gravitation potential derivative, occurs on their right hand sides (\ref{g20k5}). However, a closer inspection of these two {\it different} gravitational potentials $\Phi$ and $\varPhi$ exposes the superficiality of this appearance.  
Also from (\ref{g20k8})
\begin{eqnarray}
r_u^2\frac{\partial \ln \varrho(r_u)}{\partial r_u} &=&-\frac{ G}{ \omega c^2} \int_0^{r_u}\varrho (r) dr \label{g20k12.1}\\
\frac{\partial}{\partial r_u}\left(\frac{r_u^2\partial \ln \varrho (r_u)}{\partial r_u}\right)&=&-\frac{G\varrho (r_u)}{ \omega c^2}  \label{g20k12.2}\\
0&=&-\frac{G\varrho (r_u)}{\omega c^2}. \label{g20k12.3}
\end{eqnarray}
The last three equations above take the alternative equation (\ref{g20k8}) and under the assumption that there is a solution of the form $\varrho (r_u)$ which is constant under $r$ variation  which then implies that the differentiated bracketed form on the left reduces to the value $0$ and so the only constant value for the density on the left at (\ref{g20k12.3}) is zero. In other words, there is no constant valued density solution to this equation in the $\varrho$ system. This apparently trivial fact is important theoretically because it implies that what is called the $\rho_{\prime ,sis} (r)$ solution in the standard system has no analogue solution in  the alternative system. Thus if the alternative theory is correct there is no simple physical isothermal sphere solution. 
It follows that the two fundamental self gravitation equilibrium equations (\ref{g20k7}) and (\ref{g20k8}) give inconsistent mathematical descriptions for the character of both pairs of densities, the $\rho$ and $\rho_\prime$ and the $\varrho$ and $\varrho_\prime$ solutions.  Thus the question arises which description is correct, the orthodox version (\ref{g20k7}) or the alternative version, (\ref{g20k8}), derived in this paper? This uncertain situation can be summarised as follows. The standard theory implies that three different density functions $\rho$, $\rho_\prime$ and $\rho_{\prime ,sis}$ describe equilibrium situations and are given by 
\begin{eqnarray}
\rho (r) &=&4\pi r^2\rho_{\prime ,0} \exp \left(-\frac{\Phi (r)}{ \sigma^2}\right)\label{g20k13}\\
\rho_\prime (r)&=& \rho_{\prime ,0}(r)\exp \left(-\frac{\Phi (r)}{ \sigma^2}\right) \label{g20k14}\\
\rho_{\prime ,sis} (r)&=& \frac{\sigma^2}{2\pi G r^2}\label{g20k14.1}
\end{eqnarray}
and the alternative equation for gravitational equilibrium derived in this paper implies that only the first and second densities above $\rho$ and $\rho_\prime$ above have equilibrium analogues $\varrho$ and $\varrho_\prime$ given by 
\begin{eqnarray}
\varrho (r)&=& \varrho_0\exp \left(-\frac{\varPhi (r)}{ \sigma^2}\right)\label{g20k15}\\
\varrho_\prime (r)&=&\frac{\varrho_0}{4\pi r^2}\exp \left(-\frac{ \varPhi (r)}{ \sigma^2}\right).\label{g20k16}
\end{eqnarray}
Both versions cannot be simultaneously correct.
Clearly, the gravitation potential, $\Phi (r)$ or the var-Phi version  $\varPhi (r)$, plays a central role in getting at the solutions above whatever the truth about their physical validity happens to turn out to be. In the next section, the differences between standard theory and the alternative developed in this paper are examined.  
\section{Comparing Standard with Macroscopic Approach}
\setcounter{equation}{0}
\label{sec-wsff}
The basic standard formula for spherical gas gravitational equilibrium from (\ref{g20k2}), repeated below, is
\begin{eqnarray}
\frac{\partial \ln\rho_\prime(r)}{\partial r}&=&- \frac{G \int_0^{r}4\pi r^{\prime 2}\rho_\prime (r^\prime)dr^\prime }{\omega c^2 r^2}. \label{g20k18}  
\end{eqnarray}
My derived gas equilibrium equation (\ref{g20k8}) is
\begin{eqnarray}
\frac{\partial \ln \varrho (r_u)}{\partial r_u} &=& -\frac{G\int_0^{r_u}\varrho (r)dr }{\omega c^2 r_u^2 }=-\frac{G\int_0^{r_u}4\pi r^2\varrho_\prime (r)dr }{\omega c^2 r_u^2 }. \label{g20k19}
\end{eqnarray}
The apparent  similarity of these two formulae is deceptive, particularly as their right hand sides appear to be essentially the same as indicated at (\ref{g20k19}). The important difference is between the densities, $\rho_\prime$, on the left hand side of the first equation and the density, $\varrho$, on the left hand side of the second equation.  
The first formula's structure is based on the well known and tested case of the hydrostatic structure of the earth's atmosphere, of course, above the earth's surface. It may not be so well known that the formula for that case is an approximation anyway. This is because the formula is derived as shown above from classical local fluid theory which is essentially about the local microscopic conditions. It is an approximation in that it does not take into account the earth's curvature which is the macroscopic arena of the general gravity pressure equilibrium problem. However, in the earth's atmosphere context this approximation is so good that correct results are given by its use. If on the other hand, the earth was a minute planet with an enormous extended atmosphere, the formula would fail miserably. The required formula in the case of planetary solar or galactic greatly extended atmospheres needs to take into account just how Newton's dynamics needs to be expressed in the gas with gravity context in order to make gas pressure formalism and Newton's dynamics formalism consistent.
The linking characteristic between these two different points of view is the macroscopic concept of {\it thrust\/} rather than the local concept of {\it pressure\/}. Thrust can be taken to be the dynamical face of gas pressure. The analogues of subjected object and source force in that context are a {\it finite\/} thickness spherical shell of gas and the sphere of gravitating gas within its smaller surface respectively. It is not just pressure change between inner and outer surface of the enveloping gas shell that determines the dynamics of the whole shell but rather thrust change between these two surfaces. Only in a non-curvilinear situation would pressure become sufficient as would become the case for the region near the surface of a very large radius spherical volume. 
 My formula at (\ref{g20k19}) takes these aspects into account and also involves a careful use of the gravitational radial centroid associated with the spherical shell object which is subjected to the gravity of all the mass included within its {\it smaller\/} surface. The standard argument used to derive (\ref{g20k18}) is almost entirely {\it local\/} and does not take account of the macroscopic dynamical and geometrical character of the system  but rather is based on what might be called a local flat earth point of view when locally pressure is the controlling parameter.
\section{Appendix 2 Conclusions}
The work in this paper was motivated by a search for a {\it good physical\/} model for galactic halos constituted from the dark matter material which is nowadays postulated to exist. Here the possibility that this material is present under in a self gravitating gas like equilibrium condition has been mathematically analysed. It has been shown that if the self gravitation gas equilibrium physical structure is generated and analysed more from the macroscopic point of view than from the semi-microscopic point of view that is usually the case, the basic equilibrium equation that arise is {\it superficially\/} like the usual standard gas gravity equilibrium equation but has serious differences from that equation, particularly near the radial distance, $r$, origin where it diverges markedly. It can be inferred, though it has not been shown, that as a consequence of this difference the use of a specific type of transformation that takes the isotropic equilibrium equation into the form of the Lane-Emden equation will not work in the same way. This has the drastic consequence that the important isotropic solutions of the Lane Emden equation are not immediately available as a theoretical support for many astrophysical applications. The divergence near the origin of the macroscopically derived thermal equilibrium equation does render it more interesting in application to the galactic halo problem because of the well known fact that the physical galactic structures do usually have a greatly increased density or even a singularity towards their centres. There is thus finally the firm conclusion that if the new theory given in this paper for isotropic equilibrium is accepted then a reappraisal of thermal equilibrium and the Lane Emden solutions in cosmology needs to be set in train. Some of these issues will be examined in a following paper.    
\section{Appendix 3}
\centerline{\Large {\bf Physical Applicability of Self Gravitating}}
\centerline{\Large {\bf Isothermal Sphere Equilibrium Theory II}}
\centerline{\Large {\bf  Effect of New Theory on Consequently Modified}}
\centerline{\Large {\bf Lane Emden Equation Theory and its Solutions}}
\vskip 0.35cm  
\centerline{March 28, 2011}
\vskip 0.75cm  
\section{Appendix 3 Abstract}
In this paper, limitations of the usual Lane Emden theory of the polytropic gas gravity  equilibrium states is analysed and discussed. A modified form of the Lane Emden type equation is derived from a new gas gravity equilibrium equation obtained earlier. The new version of the Lane Emden equation is shown to have a continuous infinity and a discrete set of available equilibrium states and so is suitable for evolutionary galactic modelling. The new state structure is worked out in mathematical detail and the covering formula is obtained for all possible available density distributions within the system.  
\vskip 0.2cm 
\centerline{Keywords: Cosmology, Dust Universe, Dark Energy, Dark matter}
\centerline{Newton's Gravitation Constant, Galactic Halo}
\centerline{Isothermal Gravitational Equilibrium, Lane Emden Equation Modification}
\vskip 0.3cm

\centerline{PACS Nos.: 98.80.-k, 98.80.Es, 98.80.Jk, 98.80.Qc}
\section{Appendix 3 Introduction}
\setcounter{equation}{0}
\label{sec-a3intro}
This paper is a follow up to a paper, \cite{71:gil}, of similar title on the problem of formulating the equation that describes the equilibrium of a gaseous material in a self gravitational equilibrium  condition in the galaxy modelling context \cite{70:wei}, see also, appendix 2 of (\cite{58:gil}). The equation found although looking much like the standard equation for self gravitating thermal equilibrium is significantly different from the standard equation and this has a strong influence on the form attributed to polytropic gas theory which can be regarded as an offshoot of the standard theory. This paper considers basic changes to polytropic theory in the Lane Emden context that are implied by the new theory. I start this account by displaying the new equation in some rearrangements using the symbol $\varrho$ for mass density followed by some of the standard equations using the symbol $\rho$ for density. 
Consider the form of the gravitational thermal equilibrium equation below obtained in the paper preceding this one and which was displayed as equation (3.30) in the previous paper but here appears at (\ref{h01}) and with various manipulations down to (\ref{h011111}). The equations from (\ref{h01111b}) to (\ref{h04}) are the standard forms for the equivalent thermal equilibrium equation for comparison and identified with $\rho$ rather than with $\varrho$. The subscripted prime on $\varrho_\prime$ indicates the usual density per unit volume where as plain $\varrho$ is density per unit radius. The acute accent on $\rho$ or $\varrho$ indicates a dimensionless quantity underneath.  
\begin{eqnarray}
4\pi \frac{\partial (r^2P(r))}{\partial r}&+&\left(\frac{ G\varrho (r)}{ r^2}\right) \int_0^{r}\varrho (r^\prime)dr^\prime =0\label{h01}\\
\frac{\partial (r^2P(r))}{\partial r}&+&G\varrho_\prime (r) \int_0^{r}\varrho (r^\prime)dr^\prime =0\label{h011}\\
\frac{\partial (r^2P(r))}{ \varrho_\prime (r)\partial r}&=&-G  \int_0^{r}\varrho (r^\prime)dr^\prime \label{h0111}\\
\frac{\partial}{\partial r}\left(\frac{\partial (r^2P(r))}{ \varrho_\prime (r)\partial r}\right)&=&-G \varrho (r) = -4\pi r^2 G\varrho_\prime(r)\label{h01111}\\
\frac{\sigma^2\partial}{\partial r}\left(\frac{r^2\partial (\ln\varrho (r))}{\partial r}\right) &=&-G \varrho (r) = -4\pi r^2 G\varrho_\prime(r)\label{h011111}\\
\frac{\partial}{r^2\partial r}\left( \frac{ r^2 \partial P (r) }{\rho_\prime (r)\partial r}\right) &=&-4\pi G\rho_\prime(r)\label{h01111b}\\
\frac{\sigma^2\partial}{r^2\partial r}\left(\frac{ r^2 \partial \ln\rho_\prime(r)}{\partial r}\right) &=& -4\pi G\rho_\prime(r)  \label{h02}\\
-\frac{G}{\sigma^2}\rho (r) &=& \frac{\partial}{\partial r}\left(r^2\frac{ \partial \ln\acute{\rho}(r)}{\partial r}\right)
=-\acute{\rho} (r) \label{ho3}\\
P=P_{\rho_\prime}&=& K\rho_\prime^\gamma . \label{h04}\\
=P_{\varrho_\prime}&=& K\varrho_\prime^\gamma. \label{h05}
\end{eqnarray}
The last two equations above give a class of possible pressure dependencies on density that will be used in the following work. In principle, any value of the numerical index parameter $\gamma$ could be considered. However, it should be noted that this $\gamma$ {\it is not\/} the adiabatic index of gas theory. \section{Standard Lane Emden Equation and its Solutions}
\setcounter{equation}{0}
\label{sec-sle}
The Lane Emden equation is often derived from equation (\ref{h01111b}) using the following function variable transformation from ($\rho (r),r$) to ($\theta (\xi),\xi$) expressed as,
\begin{eqnarray}
\rho_\prime (r) &\rightarrow& \rho_{\prime,c} \theta^n(\xi)\label{h1}\\
r&=& \alpha\xi\label{h2}\\
\rho_{\prime,c}&=&\rho_\prime (0), \label{h3}
\end{eqnarray}
where the symbol $n$ is the polytropic gas index. 
Using the second and third equation above, the first equation can be written as
\begin{eqnarray}
\theta^n(\xi) &\leftarrow& \frac{\rho_\prime (\alpha \xi)}{\rho_\prime (0) } \label{h4}
\end{eqnarray}
and this implies from (\ref{h2}) that when $r=0$,  $\xi=0$ and from $\ref{h4}$
\begin{eqnarray}
\theta (0)= (\rho_\prime (0)/\rho_\prime (0))^{1/n}=1^{1/n}=1,\ \forall n\label{h5}
\end{eqnarray}
when $\xi =0$. Thus, if  $\theta (0)$ is not to turn out to be indeterminate, then the density function $\rho_\prime (r)$ has to assume a  {\it finite\/} value, $\rho_{\prime,c}$, at the $r$ origin. 
Let us now using (\ref{h04}) transform equation (\ref{h01111b}) by substituting for the ($\rho_\prime (r),r$) variables in terms of the ($\theta (\xi),\xi$) variables as below    
\begin{eqnarray}
\frac{\partial}{r^2\partial r}\left( \frac{ r^2 \partial P (r) }{\rho_\prime (r)\partial r}\right) &=&-4\pi G\rho_\prime(r) \label{h666}\\
\frac{\partial}{(\alpha\xi )^2\partial (\alpha\xi) }\left( \frac{ (\alpha\xi )^2 \partial (K\rho_{\prime,c}^\gamma\theta^{n\gamma} (\xi)) }{ \rho_\prime \theta^n(\xi)\partial (\alpha\xi ) }\right) &=&-4\pi G \rho_{\prime,c} \theta^n(\xi) \label{h6}\\
\frac{ K\rho_c^{\gamma -2} \partial}{4\pi G \alpha^2\xi^2\partial \xi }\left( \frac{ \xi^2 \partial (\theta^{n\gamma} (\xi)) }{ \theta^n(\xi)\partial (\xi ) }\right) &=&- \theta^n(\xi). \label{h6.1}
\end{eqnarray}
It seems to me that the game at this stage is to simplify the expression (\ref{h6.1}) sufficiently to be able to recognise a possibilities of finding solutions to this {\it nonlinear\/} differential  equation. There are five free valued constants, $K$, $\rho_{\prime,c}$, $\gamma$, $\alpha$, $n$, to play with and one obvious relation that could be introduced between these constants to render the differential expression within the large brackets on the left significantly and usefully simplified. Consider that expression below with brackets retained and having been subjected to the consequence of the relation between constants at (\ref{h6.3}).
\begin{eqnarray}
\left( \frac{\xi^2 \partial (\theta^{n\gamma} (\xi))}{ \theta^n(\xi)\partial \xi }\right)&\rightarrow&\left( \frac{\xi^2 \partial (\theta^{n+1} (\xi))}{ \theta^n(\xi)\partial \xi }\right)= (n+1) \left(\frac{\xi^2 \partial \theta(\xi)}{\partial \xi }\right) \label{h6.2}\\
\gamma &=&1 +1/n.\label{h6.3}
\end{eqnarray}
If this changed expression is now substituted for its original in equation (\ref{h6.1}) after imposing a second condition on constants at (\ref{h6.5}) the result is the Lane Emden equation at (\ref{h6.4}),
\begin{eqnarray}
\frac{1}{\xi^2}\frac{\partial }{\partial \xi }\left(\frac{\xi^2 \partial \theta(\xi)}{\partial \xi }\right)&=&- \theta^n(\xi) \label{h6.4}\\
\frac{(n+1) K\rho_{\prime,c}^{\gamma -2} }{4\pi G \alpha^2}&=&1 \implies\label{h6.5}\\
\alpha&=&\left(\frac{(n+1) K\rho_{\prime,c}^{(1-n)/n} }{4\pi G}\right)^{1/2}, \label{h6.6}
\end{eqnarray}
where in the last entry above the condition (\ref{h6.3}) has been used.
For ease of reference, I now give what are referred to as  the three, (n=0,1,5), {\it analytic\/} solutions of the Lane Emden equation (\ref{h6.4}),
\begin{eqnarray}
n=0, \quad \theta_0(\xi)&=&1-\xi^2/6\label{h06.7}\\
n=1, \quad \theta_1(\xi)&=&\sin(\xi)/\xi\label{h06.8}\\
n=5, \quad \theta_5(\xi)&=&(1+\xi^2/3)^{-1/2}.\label{h06.9}
\end{eqnarray}
The solution $n=0$ implies a spherical mass distribution of constant density and so could be used as a {\it very\/} rough approximation in some astrophysical situations. In fact, it is not of much use at all. The solution $n=1$ has been used to model planetary interiors and brown dwarfs. The case $n=5$ is the only non-linear solution and consequently it is the only important solution in the context of galactic modelling and has been used in that context but for various reasons it is greatly limited. This aspect will be discussed after the first two solution cases are briefly reviewed.
Consider the standard Lane Emden equation,
\begin{eqnarray}
\frac{1}{\xi^2}\frac{\partial }{\partial \xi }\left(\frac{\xi^2 \partial \theta(\xi)}{\partial \xi }\right)&=&- \theta^n(\xi) \label{h6.64}\\
+ \frac{ \partial^2 \theta(\xi)}{\partial \xi^2}+\frac{2 \partial \theta(\xi)}{\xi\partial \xi } &=&- \theta^n(\xi) \label{h6.65}
\end{eqnarray}
we proceed as follows by transforming  
into a {\it normal\/} form, only second and zeroth order differentiations of $\phi $, with the transformation of $\theta (\xi) $ to $\phi (\xi) $ below
\begin{eqnarray}
\theta (\xi) &=& \phi (\xi) \exp \left(-\frac{1}{2}\int \frac{2 d\xi }{\xi}\right) = \xi^{-1}\phi (\xi) \label{h6.66}\\
\frac{\partial\theta (\xi)}{\partial\xi}&=& -\frac{\xi^{-2}\phi (\xi)}{1} + \frac{\xi^{-1}\partial\phi (\xi)}{\partial \xi}\label{h6.67}\\
\frac{\partial^2\theta (\xi)}{\partial\xi^2} &=& +\frac{2\xi^{-3}\phi (\xi)}{1} -\frac{\xi^{-2}\partial\phi (\xi)}{\partial \xi}  \label{h6.68}\\
&=&- \frac{\xi^{-2}\partial\phi (\xi)}{\partial \xi} + \frac{\xi^{-1}\partial^2 \phi (\xi)}{\partial \xi^2}.\label{h6.69}
\end{eqnarray}
After inserting the above parts into the equation (\ref{h6.65}), the result is
\begin{eqnarray}
&\ & +\frac{2\xi^{-3}\phi (\xi)}{1} -\frac{2\xi^{-2}\partial\phi (\xi)}{\partial \xi}   \nonumber\\
&+&\frac{\xi^{-1}\partial^2 \phi (\xi)}{\partial \xi^2} -\frac{2\xi^{-3}\phi (\xi)}{1} + \frac{2\xi^{-2}\partial\phi (\xi)}{\partial \xi} \nonumber\\
&+& \xi^{-n}\phi (\xi)^n  = 0 . \label{h6.70}\\
&\ & \frac{\xi^{-1}\partial^2 \phi (\xi)}{\partial \xi^2} + \xi^{-n}\phi (\xi)^n  = 0 . \label{h6.71}\\
&\ & \frac{\partial^2 \phi (\xi)}{\partial \xi^2} + \xi^{1-n}\phi (\xi)^n  = 0. \label{h6.72}
\end{eqnarray}
A simple and obvious general solution of this equation occurs at $n=1$ and is
\begin{eqnarray}
\phi (\xi)=A\cos (\xi) +B\sin (\xi),\label{h6.73}
\end{eqnarray}
where A and B are arbitrary constants. Consequently the $\theta$ solution that goes along with this is
\begin{eqnarray}
\theta (\xi)= \xi ^{-1}(A\cos (\xi) +B\sin (\xi)),\label{h6.74}
\end{eqnarray}
using (\ref{h6.66}).
The second term here $ B\xi ^{-1}\sin (\xi)=\theta_1 (\xi)$ which is the well known solution of the Lane Emden equation for $n=1$ and importantly for the boundary condition $\theta (+0)\rightarrow 1$. However, the first term $ A \xi ^{-1} \cos (\xi) $ would be rejected with this boundary condition and for survival and used alone requires $\theta (\xi_b)=1$, when $\xi_b=A\cos (\xi_b)$ according to the value of $A$ chosen and the outer boundary radius of the spherical mass distribution could be at the value of $\xi$ which gives the first zero of $\cos (\xi)$  which, to my mind, is as good astronomically  as is the other and usually used case. In fact as it diverges towards $\xi=0$, it could be physically more realistic. Of course, these two functions could also be used together with a different boundary condition. Another simple and obvious solution to equation (\ref{h6.72}) occurs at (n=0) and with two arbitrary constants is 
\begin{eqnarray}
\phi (\xi)=-\xi^3/6 +A\xi +B \label{h6.75}
\end{eqnarray}
and leads to a $\theta$ version,
\begin{eqnarray}
\theta (\xi)=  -\xi^2/6 + A +B/\xi. \label{h6.76}
\end{eqnarray}
The usual boundary condition $ \theta (+0)\rightarrow 1$ involves $B=0$ and $A=1$ but this discounts the $B/\xi$ term  which however could be brought into use, not alone in this case, with the boundary condition $\theta (\xi_b)=1$ with $\xi_b$ satisfying
\begin{eqnarray}
\theta (\xi_b)=1=  -\xi_b^2/6 + A +B/\xi_b. \label{h6.77}
\end{eqnarray}
 according to the values of $A$ and $B$ chosen and again we get a divergent at the origin possible mass distribution and which could be regarded as having a limiting {\it outer\/} boundary given by a solution of (\ref{h6.78}) for $\xi_o$.
\begin{eqnarray}
-\xi_o^2/6 + A +B/\xi_o=0. \label{h6.78}
\end{eqnarray}
 The last of the well known solutions (\ref{h06.9})
 is probably the most important of  the three. To explain why this is the case, let us use $n$,  the power  of $\theta$, in equation (\ref{h6.64}) to represent the degree of non-linearity of the equation. As we have seen the cases $n=0$ and $n=1$ produce linear equations for which general solutions with two arbitrary constants are easily found which leaves us with scope for choosing boundary conditions. However, for $n\ge 2$ the equation becomes, in the usual sense, non-linear and only the one solution $\theta_5(\xi)$ has been found. This seems to me to be a strange situation in the sense that the theory is only yielding up the one non-linear solution for the gravity gas equilibrium problem which is clearly a very fundamental and physically significant cosmological problem. Significant because something like gravity gas equilibrium must play a central role in galactic dynamics. The gas gravity equilibrium equation describes a physical system with a remarkable and unique intrinsic non-linear character. Further, if thermal gas gravity equilibrium is so important, then the time evolution of such a system must also be important which implies equilibrium distributions of mass will have to change with time and so evolve through many phases of quasi static equilibrium continuously. This in turn suggests that many static equilibrium continuously related solutions must exist physically. Mathematically, it is possible to show that there are no other solutions similar to  $\theta_5(\xi)$ but with different numerical constants involved. However, this does not exclude possible solutions structurally different from the $\theta_5(\xi)$ solution but, it seems to me that such additional solutions are unlikely to exist. There is also the aesthetic aspect that that the whole ethos of the gravity gas equilibrium equations is that it is essentially non-linear so that its important solutions will also be non-linear. The solutions for n=0 and n=1 above are {\it hidden\/} extreme approximations that are possible because of the way $n$ is introduced in the function variable transformation (\ref{h1}) in which basically $n$ should have an arbitrary value taken from all the real numbers if the transformation is not to impose the hidden restrictions involve in giving $n$ integer or any value less that than arbitrary real. Taking $n$ to be integral or otherwise restricted kills off the essential, fundament and continuous non-linearity of the gas gravity system for n=0 and n=1 and above that for $n>1$ and integral great damage is done to the continuity aspect.   The situation with regard to the modified thermal equilibrium equation and its consequent modified Lane Emden equation version and its solutions is very different as will be shown in the next section. 
\section{Modified Lane Emden Equation and its Solutions}
\setcounter{equation}{0}
\label{sec-mle}
Let us now using (\ref{h05}) transform the new thermal equilibrium equation equation (\ref{h01111}) by substituting for the ($\varrho_\prime (r),r$) variables in terms of the ($\vartheta (\xi),\xi$) variables as below
\begin{eqnarray}
 \frac{\partial}{r^2\partial r}\left(\frac{\partial (r^2P(r))}{ \varrho_\prime (r)\partial r}\right)&=& -4\pi G\varrho_\prime(r) \label{h6.7}\\
\frac{ K\varrho^{\gamma -2}_b \partial}{4\pi G \alpha^2\xi^2\partial \xi }\left(\frac{\partial ( \xi^2 \vartheta^{n\gamma}(\xi)) }{ \vartheta^n(\xi)\partial \xi }\right)&=& - \vartheta^n(\xi)\label{h6.8}
\end{eqnarray}
The transformation here involves the three dimensional density, radius pair ($\varrho_\prime (r),r$) to the ($\vartheta (\xi),\xi$) pair expressed in as,
\begin{eqnarray}
\varrho_\prime (r) &\rightarrow& \varrho_{\prime,b} \vartheta^n(\xi)\label{h66.8}\\
r&=& \alpha\xi\label{h66.9}\\
\varrho_{\prime,b}&=&\varrho_\prime (r_b),\label{h66.10}
\end{eqnarray}
where the $\rho_c$ in the standard case transformation has been replaced by the constant $\varrho_{\prime,b}$ defined as follows
\begin{eqnarray}
\varrho_{\prime,b}&=&\varrho_\prime (r_b)=\varrho_\prime(\alpha\xi_b) \label{h6.81}\\
\vartheta(\xi_b)&=&1\label{h6.82}
\end{eqnarray}
in anticipation that some solutions at least of the modified Lane Emden equation will diverge at the $\xi$ origin.
The steps from equation (\ref{h666}) to the standard Lane Emden equation at (\ref{h6.4}) also take the rather changed form in the new $\vartheta$ system to give the modified Lane Emden equation as below at (\ref{h6.12}) with expanded forms following from (\ref{h6.13}) to (\ref{h6.15})
\begin{eqnarray}
\left( \frac{ \partial (\xi^2 \vartheta^{n\gamma} (\xi))}{ \vartheta^n(\xi)\partial \xi }\right)&\rightarrow&\left(2\xi\vartheta(\xi)+ \frac{\xi^2 \partial (\vartheta^{n+1} (\xi))}{ \vartheta^n(\xi)\partial \xi }\right)=\nonumber\\
&\ & \left( 2\xi\vartheta(\xi)+  (n+1) \frac{\xi^2 \partial \vartheta(\xi)}{\partial \xi }\right) \label{h6.9}\\
\gamma &=&1 +1/n\label{h6.10}
\end{eqnarray}
\begin{eqnarray}
\frac{ K\varrho^{\gamma -2}_b \partial}{4\pi G \alpha^2\xi^2\partial \xi }\left(2\xi\vartheta(\xi)+  (n+1) \frac{\xi^2 \partial \vartheta(\xi)}{\partial \xi }\right)&=& - \vartheta^n(\xi)\label{h6.11}\\
\frac{\partial}{\xi^2\partial \xi }\left(\frac{2\xi\vartheta(\xi)}{n+1}+ \frac{\xi^2 \partial \vartheta(\xi)}{\partial \xi }\right)&=& - \vartheta^n(\xi)\label{h6.12}
\end{eqnarray}
\begin{eqnarray}
\frac{(n+1) K\varrho_b^{\gamma -2} }{4\pi G \alpha^2}&=&1\implies \label{h6.124}\\
\alpha&=&\left(\frac{(n+1) K\varrho_b^{(1-n)/n} }{4\pi G}\right)^{1/2} \label{h6.125}
\end{eqnarray}
\begin{eqnarray}
\frac{2\vartheta(\xi)}{(n+1)\xi^2}+ \frac{2 \partial \vartheta(\xi)}{ (n+1)\xi  \partial \xi }+ \frac{\partial}{\xi^2\partial \xi }\left(\frac{\xi^2 \partial \vartheta(\xi)}{\partial \xi }\right)&=& - \vartheta^n(\xi)\label{h6.13}\\
\frac{2\vartheta(\xi)}{(n+1)\xi^2}+ \frac{2 \partial \vartheta(\xi)}{ (n+1)\xi  \partial \xi }+\frac{2 \partial \vartheta(\xi)}{\xi\partial \xi}+\frac{\partial^2 \vartheta(\xi)}{\partial \xi^2 }&=&- \vartheta^n(\xi) \label{h6.14}\\
\frac{2\vartheta(\xi)}{(n+1)\xi^2}+ \frac{2(n+2) \partial \vartheta(\xi)}{ (n+1)\xi  \partial \xi } +\frac{\partial^2 \vartheta(\xi)}{\partial \xi^2 }&=&- \vartheta^n(\xi) . \label{h6.15}
\end{eqnarray}
Consider the special case of (\ref{h6.15}) with (n=0) and having multiplied through with $\xi^2$,
\begin{eqnarray}
2\vartheta(\xi)+ \frac{4 \xi  \partial \vartheta(\xi)}{ \partial \xi } +\frac{\xi^2 \partial^2 \vartheta(\xi)}{\partial\xi^2 }&=&- \xi^2 \label{h7}
\end{eqnarray}
The general solution to this homogeneous linear equation is 
\begin{eqnarray}
\vartheta(\xi)&=& \frac{A}{\xi} +\frac{B}{\xi^2}-\frac{\xi^2}{12},\label{h7.1}\\
\frac{\partial\vartheta(\xi)}{\partial \xi} &=& -\frac{A}{\xi^2}-\frac{2B}{\xi^3}-\frac{\xi}{6},\label{h7.2}\\
\frac{\partial^2\vartheta(\xi)}{\partial \xi^2} &=&+ \frac{2A}{\xi^3} +\frac{6B}{\xi^4}-\frac{1}{6},\label{h7.3}
\end{eqnarray}
where $A$ and $B$ are arbitrary constants.
We can confirm this solution by considering the arithmetical difference of the standard, (\ref{h6.4}) or (\ref{h6.16}), and new versions of the Lane Emden equation,(\ref{h6.12}) or (\ref{h6.17}), as below at (\ref{h6.18})
\begin{eqnarray}
\frac{1}{\xi^2}\frac{\partial }{\partial \xi }\left( \frac{\xi^2 \partial \theta(\xi)}{\partial \xi }\right)&=&- \theta^n(\xi)\label{h6.16}\\
\frac{\partial}{\xi^2\partial \xi }\left(\frac{2\xi\vartheta(\xi)}{n+1}+ \frac{\xi^2 \partial \vartheta(\xi)}{\partial \xi }\right)&=& - \vartheta^n(\xi)\label{h6.17}\\
\frac{\partial}{\xi^2\partial \xi }\left(\frac{2\xi\vartheta(\xi)}{n+1}+ \frac{\xi^2 \partial \vartheta(\xi)}{\partial \xi}-\frac{\xi^2 \partial \theta(\xi)}{\partial \xi }\right)&=& - \vartheta^n(\xi)+ \theta^n(\xi) \label{h6.18}\\
\frac{\partial}{\xi^2\partial \xi }\left(\frac{2\xi\vartheta(\xi)}{n+1}+ \frac{\xi^2 \partial (\vartheta(\xi)- \theta(\xi))}{\partial \xi}\right) &=& - \vartheta^n(\xi)+ \theta^n(\xi) \label{h6.19}
\end{eqnarray}
and in the case $n=0$ this last equation reduces to
\begin{eqnarray}
\frac{\partial}{\xi^2\partial \xi }\left(2\xi\vartheta(\xi)+ \frac{\xi^2 \partial (\vartheta(\xi)- (1-\xi^2/6))}{\partial \xi}\right) &=& - 1+1=0\label{h6.20}\\
2\xi\vartheta(\xi)+ \frac{\xi^2 \partial (\vartheta(\xi)- (1-\xi^2/6))}{\partial \xi} &=& C_1\label{h6.21}\\
\vartheta(\xi)+ \frac{\xi\partial \vartheta(\xi)}{2\partial \xi}   &=& \frac{C_1\xi^{-1}}{2} -\xi^2/6\label{h6.22}
\end{eqnarray}
having substituted for $\theta (\xi)$ the known solution, $\theta_0 (\xi) $, at line (\ref{h6.20}) for $n=0$ of the Lane Emden equation at (\ref{h06.7}). Using (\ref{h7.2}) in (\ref{h6.22}) we get
\begin{eqnarray}
\vartheta(\xi)+ \frac{\xi}{2}\left(-\frac{A}{\xi^2}-\frac{2B}{\xi^3}-\frac{\xi}{6}\right)   &=& \frac{C_1\xi^{-1}}{2} -\xi^2/6\label{h6.23}\\
\vartheta(\xi)+ \left(-\frac{A}{2\xi}-\frac{B}{\xi^2}-\frac{\xi^2}{12}\right)   &=& \frac{C_1\xi^{-1}}{2} -\xi^2/6\label{h6.24}\\
\vartheta(\xi)  &=&\frac{A+ C_1}{2\xi}+\frac{B}{\xi^2}-\frac{\xi^2}{12}. \label{h6.25}
\end{eqnarray}
As there can only be two arbitrary constants and for agreement with (\ref{h7.1})
$C_1=A$. To write down the full $n=0$ solution, the coefficient $\varrho_{\prime,b}$,  $\ref{h6.27}$, is required and is obtained by solving the equation from (\ref{h7.1}) for $\xi_b $ as below
\begin{eqnarray}
1=\vartheta(\xi_b)&=& \frac{A}{\xi_b } +\frac{B}{ \xi_b ^2}-\frac{ \xi_b ^2}{12},\label{h6.26}\\
\varrho_{\prime,b}&=& \varrho_\prime(\xi_b) \label{h6.27}\\
\varrho_{\prime,0} (r) &=&\lim \varrho_{\prime,b}\vartheta^n (\xi)\rightarrow \varrho_{\prime,b},\ n\rightarrow 0 \label{h6.28}\\
P_{\varrho_0}(r) & =& \lim K\varrho_{\prime,0}^{\gamma (n)} (r) = \lim K\varrho_{\prime,0}^{1+1/n}(r)\rightarrow \infty, \ n\rightarrow 0 \label{h6.29}
\end{eqnarray}
and the two free arbitrary constants would be determined by the physical characteristics thought to apply in the case under study. However, the last two entries above indicating that the density function is constant and the pressure function is infinite for this solution undermines its possible physical applicability.
Consider the special case of (\ref{h6.15}) or (\ref{temp}) with (n=1),
 \begin{eqnarray}
\frac{2\vartheta(\xi)}{(n+1)\xi^2}+ \frac{2(n+2) \partial \vartheta(\xi)}{ (n+1)\xi  \partial \xi } +\frac{\partial^2 \vartheta(\xi)}{\partial \xi^2 }&=&- \vartheta^n(\xi)\label{temp}\\ 
\frac{\vartheta(\xi)}{\xi^2}+ \frac{3 \partial \vartheta(\xi)}{\xi  \partial \xi } +\frac{\partial^2 \vartheta(\xi)}{\partial \xi^2 }&=&- \vartheta(\xi)\label{h6.30}\\
\frac{\partial^2 \vartheta(\xi)}{\partial\xi^2 } + \frac{3 \partial \vartheta(\xi)}{ \xi  \partial \xi}
+ \left(1+\frac{1}{\xi^2}\right)\vartheta(\xi) &=& 0 . \label{h6.31}
\end{eqnarray}
The differential equation, (\ref{h6.31}) above, looks like the  first order Bessel equation following at (\ref{h6.31.1})
\begin{eqnarray}
\frac{\partial^2 \vartheta(\xi)}{\partial\xi^2 } + \frac{\partial \vartheta(\xi)}{ \xi  \partial \xi}
+ \left(1-\frac{p^2}{\xi^2}\right)\vartheta(\xi) &=& 0  \label{h6.31.1}
\end{eqnarray}
apart from the obvious differences that in it there is no $3$ coefficient in the second term and a minus sign is in the place of the plus sign before the $\xi^{-2}$ term. However, there does seem to be some relation between these two equations.
I will continue with this section after making a digression to obtain a general result a special case of which I shall use when I return from the digression.
\section{The General Case}
\setcounter{equation}{0}
\label{sec-tgc1}
Let us now consider the general non-linear case in which the $n$ parameter can initially, at least, be allowed to take any real number value in the range $-\infty\rightarrow +\infty$.
We can clarify this situation by getting rid of the first derivative term in equations (\ref{h55.bb}) by a well known trick of a variable transformation. We proceed as follows by transforming  
into a {\it normal\/} form, only second and zeroth order differentiations of $\varphi $, with the transformation of $\vartheta (\xi) $ to $\varphi (\xi) $ below
\begin{eqnarray}
&\ &\frac{\partial^2 \vartheta(\xi)}{\partial \xi^2} + \frac{2(n+2) \partial \vartheta(\xi)}{ (n+1)\xi  \partial \xi } + \frac{2\vartheta(\xi)}{(n+1)\xi^2}=- \vartheta^n(\xi)\label{h55.bb}\\
 \vartheta (\xi) &=& \varphi (\xi) \exp \left(- \frac{n+2}{n+1}\int \frac{d\xi }{\xi}\right) = \xi^{- \frac{n+2}{n+1}}\varphi (\xi) \label{h6.56b}\\
\frac{\partial\vartheta (\xi)}{\partial\xi}&=& -\frac{(n+2)\xi^{-
\frac{2n+3}{n+1}}\varphi (\xi)}{n+1} + \frac{\xi^{-\frac{n+2}{n+1}}\partial\varphi (\xi)}{\partial \xi}\label{h6.57b}\\
\frac{\partial^2\vartheta (\xi)}{\partial\xi^2} &=& +\frac{(n+2)(2n+3)\xi^{-\frac{3n+4}{n+1}}\varphi (\xi)}{(n+1)^2} -\frac{(n+2)\xi^{-\frac{2n+3}{n+1}}\partial\varphi (\xi)}{(n+1)\partial \xi}  \label{h6.58b}\\
&=&- \frac{(n+2)\xi^{-\frac{2n+3}{n+1}}\partial\varphi (\xi)}{(n+1)\partial \xi} + \frac{\xi^{-\frac{n+2}{n+1}}\partial^2 \varphi (\xi)}{\partial \xi^2}.\label{h6.59b}
\end{eqnarray}
After inserting the above parts into the equation (\ref{h55.bb}), the result is
\begin{eqnarray}
&\ & +\frac{(n+2)(2n+3)\xi^{-\frac{3n+4}{n+1}}\varphi (\xi)}{(n+1)^2} -2\frac{(n+2)\xi^{-\frac{2n+3}{n+1}}\partial\varphi (\xi)}{(n+1)\partial \xi}  \nonumber\\
&\ & +\frac{\xi^{-\frac{n+2}{n+1}}\partial^2 \varphi (\xi)}{\partial \xi^2}
 -\frac{2(n+2)^2\xi^{-\frac{3n+4}{n+1}}\varphi (\xi)}{(n+1)^2} + \frac{2(n+2)\xi^{-\frac{2n+3}{n+1}}\partial\varphi (\xi)}{(n+1)\partial \xi} \nonumber\\
&\ & -\frac{(n+2)\xi^{-\frac{3n+4}{(n+1)}}\varphi (\xi)}{(n+1)^2} + \frac{2\xi^{-\frac{3n+4}{n+1}}\varphi (\xi) }{(n+1)}+\xi^{- \frac{n(n+2)}{n+1}}\varphi^n (\xi)   = 0 . \label{h6.60b}\\
&\ & \frac{\xi^{-\frac{n+2}{n+1}}\partial^2 \varphi (\xi)}{\partial \xi^2}
 -\frac{(n+2)\xi^{-\frac{3n+4}{(n+1)}}\varphi (\xi)}{(n+1)^2}\nonumber\\
 &+& \frac{2\xi^{- \frac{3n+4}{n+1}}\varphi (\xi) }{(n+1)}+\xi^{- \frac{n(n+2)}{n+1}}\varphi^n (\xi)= 0. \label{h6.60bb}\\
&\ & \frac{\partial^2 \varphi (\xi)}{\partial \xi^2}  -\frac{(n+2)\xi^{-2}\varphi (\xi)}{(n+1)^2} + \frac{2\xi^{- 2}\varphi (\xi) }{(n+1)} =- \xi^{-\frac{n^2 +n -2}{n+1}}\varphi^n (\xi) . \label{h6.62b}\\
&\ & \frac{\partial^2 \varphi (\xi)}{\partial \xi^2}  +\frac{n\xi^{-2}\varphi (\xi)}{(n+1)^2}  =- \xi^{-\frac{n^2 +n -2}{n+1}}\varphi^n (\xi) . \label{h6.62bbc}
\end{eqnarray}
It follows that if we can find a solution for $\varphi$ from the formula (\ref{h6.62bbc}) then via (\ref{h6.56b}) a solution for $\vartheta$ of the modified Lane Emden equation will follow immediately. I now return from the digression to continue with the $n=1$ case.
For $n=1$, we note that the equation (\ref{h6.62bbc})reduces to
\begin{eqnarray}
&\ & \frac{\partial^2 \varphi (\xi)}{\partial \xi^2}  +\left(1+\xi^{-2}/4\right) \varphi (\xi)  = 0 . \label{h6.62bbcd}
\end{eqnarray}
The Bessel equation (\ref{h6.31.1}) written below with the subscript $B$ everywhere necessary to avoid confusion
\begin{eqnarray}
\frac{\partial^2 \vartheta_B(\xi)}{\partial\xi^2 } + \frac{\partial \vartheta_B (\xi)}{ \xi  \partial \xi}
+ \left(1-\frac{ p^2}{\xi^2}\right)\vartheta_B (\xi) &=& 0  \label{h6.39.1}
\end{eqnarray}
can be transformed into a {\it normal\/} form, only second and zeroth order differentiations of $\varphi_B $, with the transformation of $\vartheta_B (\xi) $ to $\varphi_B (\xi) $ below
\begin{eqnarray}
\vartheta_B (\xi) &=& \varphi_B (\xi) \exp \left(-\frac{1}{2}\int \frac{ d\xi }{\xi}\right) = \xi^{-1/2}\varphi_B (\xi) \label{h6.40}\\
\frac{\partial\vartheta_B (\xi)}{\partial\xi}&=& -\frac{\xi^{-3/2}\varphi_B (\xi)}{2} + \frac{\xi^{-1/2}\partial\varphi_B (\xi)}{\partial \xi}\label{h6.41}\\
\frac{\partial^2\vartheta_B (\xi)}{\partial\xi^2} &=& +\frac{3\xi^{-5/2}\varphi_B (\xi)}{4} -\frac{\xi^{-3/2}\partial\varphi_B (\xi)}{2\partial \xi}  \label{h6.42}\\
&=&- \frac{\xi^{-3/2}\partial\varphi_B (\xi)}{2\partial \xi} + \frac{\xi^{-1/2}\partial^2 \varphi_B (\xi)}{\partial \xi^2}.\label{h6.43}
\end{eqnarray}
After inserting the above parts into the equation (\ref{h6.39.1}), the result is
\begin{eqnarray}
&\ &\frac{3\xi^{-5/2}\varphi_B (\xi)}{4} -\frac{\xi^{-3/2}\partial\varphi_B (\xi)}{\partial \xi}  \nonumber\\
&+&\frac{\xi^{-1/2}\partial^2 \varphi_B (\xi)}{\partial \xi^2}-\frac{\xi^{-5/2}\varphi_B (\xi)}{2} + \frac{\xi^{-3/2}\partial\varphi_B (\xi)}{\partial \xi} \nonumber\\
&+& \left(\xi^{-1/2}-\xi^{-5/2} p^2\right) \varphi_B (\xi)  = 0 . \label{h6.44}\\
&\ & \frac{\xi^{-1/2}\partial^2 \varphi_B (\xi)}{\partial \xi^2}  +\left(\xi^{1/2}-\xi^{-5/2} (p^2+1)/4\right) \varphi_B (\xi)  = 0 . \label{h6.45}\\
&\ & \frac{\partial^2 \varphi_B (\xi)}{\partial \xi^2}  +\left(1-\xi^{-2}(p^2-1)/4\right) \varphi_B (\xi)  = 0 . \label{h6.46}\\
&\ & \frac{\partial^2 \varphi (\xi)}{\partial \xi^2}  +\left(1+\xi^{-2}/4\right) \varphi (\xi)  = 0 . \label{h6.47}
\end{eqnarray}
The new version of the $\varphi$ thermal-equilibrium Lane-Emden equivalent equation equation, (\ref{h6.62bbcd}), written again at (\ref{h6.47}) can now be compared with the well known Bessel equivalent equation obtained at (\ref{h6.46}) to make the firm identification $p=0$. Thus we can conclude  that a $\varphi$ solution version of the new theory for $n=1$ is also a $\varphi_B$ solution version of a Bessel equation for $p=0$
\begin{eqnarray}
\varphi_1 (\xi)= \varphi_{0,B}(\xi).\label{h6.48}
\end{eqnarray}
We can get the $\vartheta$ version of equation (\ref{h6.48}) by using the transformations involved at (\ref{h6.56b}) and (\ref{h6.40}) and redisplayed below
\begin{eqnarray}
\vartheta (\xi) &=& \xi^{-3/2}\varphi (\xi) \label{h6.49}\\
\vartheta_B (\xi) &=&  \xi^{-1/2}\varphi_B (\xi). \label{h6.50}
\end{eqnarray}
Thus
\begin{eqnarray}
\xi^{3/2}\vartheta_1 (\xi)&=&\xi^{1/2}\vartheta_{0,B}(\xi)\label{h6.51}\\
\vartheta_1 (\xi)&=& \xi^{-1} \vartheta_{0,B}(\xi)= \xi^{-1}(J_0(\xi)A_1-N_0 (\xi)A_2), \label{h6.52}
\end{eqnarray}
where finally $\vartheta_1$ is given by identifying $\vartheta_{0,B}(\xi)$ as being the $p=0$ general solution of the Bessel equation involving also the $p=0$ Neumann function, $N_0 (\xi)$.

Returning now to the general case for all allowable $n$, let us look for simple solutions of the form, 
\begin{eqnarray}
\varphi (\xi)=  A \xi^p, \label{h6.62d}
\end{eqnarray}
where $A$ and  $p$ are to be determined. Substituting this possible solution into the $\varphi (\xi)$ equation at (\ref{h6.62bbc}), we get
\begin{eqnarray}
p(p-1)A\xi^{p-2} +\frac{n\xi^{p-2} A }{(n+1)^2}+\xi^{-\frac{n^2 +n -2}{n+1}+pn} A^n =0.\label{h6.63d}
\end{eqnarray}
It follows that
\begin{eqnarray}
p(p-1)A +\frac{n A }{(n+1)^2}+ A^n =0\label{h6.64d}\\
\left(p(p-1) +\frac{n}{(n+1)^2}\right) =- A^{n-1} \label{h6.64.1d}\\
p-2= -\frac{n^2 +n -2}{n+1}+pn \label{h6.65d}\\
p= \frac{n^2 -n-4}{n^2-1}\label{h6.66d}\\
p-1= \frac{n^2 -n-4-(n^2 -1)}{n^2-1}= \frac{ -n-3}{n^2-1}\label{h6.67d}\\
p(p-1)= \frac{n^2 -n-4}{n^2-1}\times \frac{ -n-3}{n^2-1}\label{h6.69d}\\
\left(\frac{n^2 -n-4}{n^2-1}\times \frac{ -n-3}{n^2-1} +\frac{n}{(n+1)^2}\right) =- A^{n-1} =B(n),\  say.\label{h6.70.1d}\\
\frac{1}{(n+1)^2(n-1)^2}\left(n(n-1)^2-(n^2 -n-4)(n+3)\right) =B(n)\label{h6.71.1d}\\
\frac{1}{(n+1)^2(n-1)^2}\left( -4n^2 +8n+12\right) =B(n)\label{h6.72.1d}\\
\frac{-4(n-3)}{(n+1)(n-1)^2} =B(n) .\label{h6.73.1d}
\end{eqnarray}
Thus there is not just one simple solution of the form (\ref{h6.62d}), there is an infinite continuum of solutions given by (\ref{h6.74.1d}) with only values of $n$, where $-1\le n \le +3$ excluded, where $B(n)\ge 0$ and so would make $A(n)$ either negative or complex. However, within this set of values excluded form the continuous values there is another relevant infinite set of discrete values $n=n_d$, say, at which $A(n)$ is real and positive, denoted by $-1<n_d\le 3$. The detailed location of the values for $n_d$ will be discussed later.
\begin{eqnarray}
\varphi_n (\xi)&=&  A(n) \xi^{p(n)} \label{h6.74.1d}\\
p(n)&=& \frac{n^2 -n-4}{n^2-1}\label{h6.75d}\\
A(n)&=&(-B(n))^{\frac{1}{n-1}} = \left(\frac{4(n-3)}{(n+1)(n-1)^2}\right)^{\frac{1}{n-1}} \label{h6.76.1d}\\
-\infty <n <-1&,&-1<n_d< 3\quad ,\quad 3<n< +\infty . \label{h6.77.1d}
\end{eqnarray}
Solution to the modified Lane Emden equation implied by this result is given by formula (\ref{h6.79.1d}) below after using (\ref{h6.74.1d}), (\ref{h6.75d}) and (\ref{h6.76.1d})
\begin{eqnarray}
\vartheta (\xi) &=& \xi^{- \frac{n+2}{n+1}}\varphi (\xi) \label{h6.78.1d}\\
\vartheta (\xi) &=& A(n) \xi^{ - \frac{n+2}{n+1}} \xi^{ \frac{n^2 -n-4}{n^2-1}} = A(n)\xi^q(n),\ say \label{h6.79.1d}\\
q(n)&=&\frac{n^2 -n-4}{n^2-1}- \frac{n+2}{n+1}= \frac{2}{1-n}.\label{h6.80.1d}
\end{eqnarray}
Returning to the starting point of this calculation at (\ref{h6.62d}) we can express the original density function, $\varrho_\prime (r)$, after using (\ref{h6.78.1d}) etc, three lines below,  as at (\ref{h6.62d13})
\begin{eqnarray}
\varrho_{\prime,n} (r) &\rightarrow& \varrho_{\prime,b} \vartheta_n ^n(\xi)\label{h1b}\\
\vartheta_n (\xi) &=& \xi^{- \frac{n+2}{n+1}}\varphi_n (\xi) \label{h6.62d11}\\
\varphi_n (\xi)&=&  A(n) \xi^{p(n)} \label{h6.62d12}\\
\varrho_{\prime,n} (r) &\rightarrow&\varrho_{\prime,b} (\xi^{- \frac{n+2}{n+1}} A(n) \xi^{p(n)})^n\nonumber\\
\varrho_{\prime,n} (r) &\rightarrow& \varrho_{\prime,b} A^n(n) \xi^{nq(n)}=B_{b,n}\xi^{s(n)}, \ say \label{h6.62d13}\\
B_{b,n}&=& \varrho_{\prime,b} \left(\frac{4(n-3)}{(n+1)(n-1)^2}\right)^{\frac{n}{n-1}} \label{h6.62d133}\\
(B_{b,n})_{max\forall n}&\approx & 0.0450819 \varrho_{\prime,b} \label{h6.62d1333}\\
s(n)&=&nq(n)=\frac{2n}{1-n}\label{h6.62d13333}\\
\varrho_{\prime,b} &\leftarrow&\varrho_{\prime}  (r_b)\label{h6.8111x}\\
\vartheta_n (\xi_b)&=&1= \xi_b^q A(n) \label{h6.8211}\\
\xi_b &=& A(n)^{\frac{n-1}{2}}= \left(\frac{4(n-3)}{(n+1)(n-1)^2}\right)^{\frac{1}{2}}.\label{h6.82113}\\
\alpha&=& \left(\frac{(n+1) K\varrho_{\prime,b}^{(1-n)/n} }{4\pi G}\right)^{\frac{1}{2}}. \label{h6.12511}\\
r_b&=& \alpha\xi_b=\left(\frac{(n+1) K\varrho_{\prime,b}^{(1-n)/n} }{4\pi G}\right)^{\frac{1}{2}} \left(\frac{4(n-3)}{(n+1)(n-1)^2}\right)^{\frac{1}{2}}.\nonumber\\
 \label{h6.125111}\\
r_b&=&\left(\frac{ (n-3) K\varrho_{\prime,b}^{(1-n)/n} }{ (n-1)^2\pi G}\right) ^{\frac{1}{2}}. \label{h6.125112}\\
\varrho_{\prime,b} &=&\varrho_\prime (r_b)=\left(\frac{(n-1)^2\pi G}{(n-3) K }r_b^2\right)^{\frac{n}{1-n}}.\label{h6.125114}
\end{eqnarray}
The emergence of the index $S(n)$ at (\ref{h6.62d13}) is important in that its value determines boundary conditions at both $\xi=0$ and $\xi=\infty$ which in modelling galactic structures would need to be finite or positive infinite density at $\xi=0$ and zero density at $\xi=\infty$ for realistic physics. 
Thus the condition needed is
\begin{eqnarray}
-1<S(n)<0\implies -1<n<0\label{111}
\end{eqnarray}
which puts it within the $n_d$, discrete solutions, range, (\ref{h6.77.1d}) $(-1\rightarrow 3)$, at the  left hand side $(-1\rightarrow 0)$.

Let us now return to the problem of locating the discrete solution set denoted by $n_d$ associated with the final formula for $\varrho$ density solutions at (\ref{h6.62d13}). To obtain physical mass density solutions the quantity $ B_{b,n}$, (\ref{h6.62d133}), has to appear with a real positive value. However, in the range of values $(1<n_d\le 3)$ this valued quantity with power $n/(n-1)$ will have many continua sub ranges of complex values and negative values interspersed with an infinite discrete set of the wanted {\it real positive\/} values. This arises because the quantity, $B(n_d)$, under the index is negative over the range $(1<n_d\le 3)$. We can locate the wanted values by using the complex trigonometric representation of $-1= \cos (\pi) +i\sin (\pi)=\exp (i\pi)$ in the formula for $B_{b,n_d }$
\begin{eqnarray}
B_{b,n_d }&=& \varrho_{\prime,b} \left(\frac{4(n_d -3)}{(n_d +1)(n_d-1)^2}\right)^{\frac{n_d }{n_d -1}} \label{h6.125115}\\
&=& \varrho_{\prime,b} \left(\frac{4(3-n_d) \exp (i\pi)}{(n_d +1)(n_d -1)^2}\right)^{\frac{n_d }{n_d -1}} \label{h6.125116}\\
&=& \varrho_{\prime,b} \left(\frac{4(3-n_d) }{(n_d +1)(n_d -1)^2}\right)^{\frac{n_d }{n_d -1}} \cos \left(\frac{ \pi n_d }{n_d -1}\right) \nonumber\\
&+& \varrho_{\prime,b} \left(\frac{4(3-n_d) }{(n_d +1)(n_d -1)^2}\right)^{\frac{n_d }{n_d -1}}i\sin \left(\frac{ \pi n_d }{n_d -1}\right). \label{h6.125118}
\end{eqnarray}
It is immediately obvious from this last equation that real positive values for $B_{b,n_d }$ occur where the two conditions below hold
\begin{eqnarray}
\sin \left( \frac{ \pi n_d }{n_d -1}\right)&=&0\implies \frac{ \pi n_d }{n_d -1}=\pi k\label{h6.125119x}\\
\cos \left(\pi k\right)&>&0\implies k=2l. \label{h6.Z}
\end{eqnarray}
Thus
\begin{eqnarray}
n_d=\frac{2l}{2l-1}, \label{h6.ZA}
\end{eqnarray}
where $l$ is an integer. As $n_d$ needs to be in the range,
\begin{eqnarray}
-1<n_d< 0\  or\  +1<n_d< +3,\label{h6.ZB}
\end{eqnarray} 
it follows from text below (\ref{111}) that it is any rational number in the range
\begin{eqnarray}
-1< \frac{2l}{2l-1}<0,\label{h6.ZC}
\end{eqnarray}
determined by the integer $l$ and of the form displayed at (\ref{h6.ZA}).
 
The rather complicated structure described above, (\ref{h1b}) down to (\ref{h6.Z}), from the modified Lane Emden equation theory does not occur in the standard Lane Emden equation theory because the later  equation theory does not have the very large number of non linear solutions that are available from the former  equation theory. However, both version involve an almost identical function variable transformation, (\ref{h66.8}) to (\ref{h66.10}) or (\ref{h1}) to (\ref{h3}). The difference lies in the choice of a boundary condition for the two versions. In both cases, it is assumed that there is a definite  {\it known\/} mass density isothermal gravitational  equilibrium state described by a function pair available for transforming, $\rho_\prime(r)$, for the old theory and, $\varrho_\prime (r)$, for the new theory. The objective in both cases is to transform to a new description of the mass density using a new function variable pair $\theta (\xi)$. In both cases, it can only be assumed {\it initially\/} that the n index is an arbitrary real number and as described above this and other parameters determines the value of $\alpha$ needed to get the simplest possible Lane Emden form of equation. However, it is  a value for the {\it known\/} function $\rho_\prime (r)$ or $\varrho_\prime (r)$, assumed known in terms of $r$, for some specific value of $r$, $r_c$ or $r_b$ that will depend on $n$ that converts the dimensionless quantity $\theta ^n(\xi)$ into  a mass density by multiplication with $\rho_{\prime,c}$ or $\varrho_{\prime,b}$ that does not depend on $n$ and so determines what can be called a boundary condition that renders the defining  equation, (\ref{h1}),  consistent as $\rho_\prime (r_c)=\rho_{\prime ,c}$ for the value of $\xi_c$ that makes $\theta (\xi_c)=1$ and similarly for the $b$ subscripted case when $\varrho_\prime (r_b)= \varrho_{\prime,b}$. Thus for the list of relations above $\varrho_{\prime,b}$ does not depend on $n$. It only depends on the value, $r_b$ determined by $n$ a dependence not shown explicitly that sets the boundary condition at $\vartheta_n (\xi_{b(n)})=1$, where the $n$ dependence is shown, for all solutions. This rather confusing situation can best be understood by referring to the relation (\ref{h6.125114}) from which it can be seen that if $\varrho_{\prime,b}$ is to be a fixed valued quantity for all the non linear solutions then $r_b$ has to depend on $n$ via that formula. An important result of this situation is the existence of a {\it globally maximally\/} compact solution given by equation (\ref{h6.62d1333}) for a specific $n$ at a value $n\approx 4.753$.  
\section{Appendix 3 Conclusions}
It is argued here that the usual version of the Lane Emden equation is not adequate to supply solutions for use in modelling galactic structures, particularly in the dark matter context. This is based on an earlier discovery of an isothermal gas gravity equilibrium equation, differing from the standard equation, that implies that the usual Lane Emden equation of isotropic theory should itself be modified to remedy its limited type and number of physical solutions problem. The usual system is known to have only three analytic solutions with one of these in the non-linear regime. The limited number of solutions of the usual Lane Emden theory is obvious. However, it is also a disaster in the realm of galactic halo modelling as such halos would be expected to evolve against epoch time through either a continuum of quasi equilibrium conditions or evolve more quantum like with jumps through a dense set of discrete sets. It is thus necessary that a {\it continuum\/} and or a discrete set of equilibrium states should be available within any theoretical structure that can be used to realistically describe astrophysical halos. The modified Lane Emden equation introduced in this paper is shown to present both types of state structure of equilibrium states that can take into account all levels of none linearity measured by the polytropic index $n$ which, in the new system, can range over most of the real numbers. This state structure is worked out in mathematical detail and analogues of the usual $n=0,1$ solutions  of the original Lane Emden equation are discussed in detail. The whole new system is described by a general formula giving all available mass density distributions that can be achieved under isothermal gravitational equilibrium.  
\section{Appendix 4}
\centerline{\Large {\bf Physical Applicability of Self Gravitating}}
\centerline{\Large {\bf Isothermal Sphere Equilibrium Theory III}}
\centerline{\Large {\bf Dark Matter Galactic Halo Mass Densities}}
\centerline{\Large {\bf Quantized Polytropic index, $n(l)$, $l$ a positive integer}}
\centerline{\Large {\bf Galactic Rotation Curves}}
\vskip 0.35cm  
\centerline{May 15, 2011}
\vskip 0.75cm  
\section{Appendix 4 Abstract}
A model for the dark matter mass distribution within a galaxy is constructed using the many available states that have been found in a new isothermal gas gravity equilibrium equation which describes the spherical isothermal situation. However, as most of the mass in galaxies in general assumes a spherical halo location, the model has considerable relevance for the dark matter problem in general. The model involves a core of high mass density of definite radius and mass and also a definite mass distribution outside the core radius for each available solution from an infinite discrete set of solutions. Detailed formulae for galactic rotation curves and  their radial space derivatives are obtained so that the issue of flatness or otherwise everywhere can be easily resolved for all solutions. Three typical rotation curves are illustrated in one diagram and their three radial space derivatives are shown in a second diagram.    
\vskip 0.2cm 
\centerline{Keywords: Cosmology, Dust Universe, Dark Energy, Dark matter}
\centerline{Newton's Gravitation Constant, Galactic Halo, Rotation Curves}
\centerline{Isothermal Gravitational Equilibrium, Lane Emden Equation Modification}
\vskip 0.3cm

\centerline{PACS Nos.: 98.80.-k, 98.80.Es, 98.80.Jk, 98.80.Qc}
\section{ Appendix 4 Introduction}
\setcounter{equation}{0}
\label{sec-intro0}
This paper is a follow up of papers, \cite{71:gil} and \cite{72:gil}, of similar titles on the problem of formulating the equation that describes the equilibrium of a gaseous material in a self gravitational equilibrium  condition in the galaxy modelling context \cite{70:wei}, see also, appendix 2 of (\cite{58:gil}). 

From the Friedman equation below for local acceleration in terms of radial distance from origin, we can infer the general relativity  form taken by the formula for galactic rotation curves
\begin{eqnarray}
\frac{\ddot r (t) }{ r (t)}  = \frac{\Lambda c^2 }{3} - \frac{4\pi G}{3} \left( \rho (t) + \frac{3 P (t)}{c^2} \right). \label{i00}
\end{eqnarray}
Notably when the pressure term is present and under steady state conditions we conclude that the system will involve an additional mass density given by $3 P (r)/c^2$ when the pressure and density $\rho (r)$ will vary with distance $r$ from radial centre.
Thus the general relativistic generalisation of the rotation curve formula becomes
\begin{eqnarray}
v^2_2(r) &=&  \frac{G}{r} \left(M^+( r)  + M_P(r)\right) -\frac{G M_\Lambda(r)}{r}   \label{i01}\\
M^+(r)&=&\int _0^r \rho (r^\prime)dr^\prime\label{i01b}\\
M_P(r)&=& \int_0^r\frac{3 P(r^\prime)dr^\prime}{c^2} \label{i02}\\
M_\Lambda(r) &=& \int_0^r\rho_\Lambda (r^\prime) dr^\prime=V(r) \rho_\Lambda^\dagger. \label{i03}
\end{eqnarray}
Equation (\ref{i01}) is the general relativity rotation curve for transverse velocity in terms of distance from the origin $r$. The three following masses are the total mass due to the local density distribution, the total Einstein additional mass due to local pressure  and the total dark energy mass all added up by integration within the spheres of radius $r$ about the $r$ origin. The last two of these are masses additional to usual mass density, within the region concerned that help determine the form of the function $v_2(r) $. The last mass is due to dark energy mass within the region and because this has a universal constant density distribution, $\rho_\Lambda$, the integral can be replaced with a multiplied volume for the whole spherical region. The minus sign for $G$ goes with this mass because it acts as a negative gravity source.
The Galactic rotation rate formula is derived from equation (\ref{i01}) by replacing the Masses on the right side with definite functions $M(r)$ dependent on the radius variable $r$ and which give the total amount of that mass within a sphere of radius $r$ about its $r=0$ centre. The contribution from the dark energy mass term, $\rho_\Lambda$, is relatively small unless the galaxy is very large so that mass can usually be neglected. However, the pressure term is a vital part from the general relativity point of view. The normal positively gravitating mass density term, $ M^+( r)$, and the general relativity  pressure term $M_P(r)$ both make major contributions to the way the elements of the galaxy structure rotates in relation to the element's distance from the mass distribution origin and the Newtonian attraction it experience from the mass density  distribution which it can be regarded as travelling through. It is now possible to give explicit forms for these two mass density contributions using a new theory of gravitational isotropic equilibrium having a differential equation description  (\ref{i09}). This differential equation is a modification of the usual equation that was thought to describe gravitational isothermal equilibrium with the advantage that it has many solutions in contrast with its precursor. In the paper preceding this one I obtained formulae for those many solutions in terms of the Lane Emden variables $(\vartheta , \xi))$. For application to the Galaxy halo mass density structure problem it is better to work with the more physically orientated variables $(\varrho , r)$. Getting the solution set in terms of the physical variable does take a little effort and this will be carried through next. The objective  of this paper is to solve the problem of finding suitable mass density solutions that can {\it realistically\/}  be used to form the total masses $ M^+( r)$ and $M_P(r)$  so that the general relativistic rotation curve structures for galaxies can be found.
\section{Solutions in Terms of Physical Variables}
\setcounter{equation}{0}
\label{sec-sitpv}
The solutions of equation, (\ref{i09}), as function of the isotropic index $n$ and as involved with and derived through the mathematically powerful but obscure Lane Emden variables, $(\vartheta , \xi))$, with details are displayed from (\ref{i04}) to (\ref{i08}),  
\begin{eqnarray}
\varrho_{\prime,n} (r) &=& B_{b,n}\xi^{s(n)}\label{i04}\\
B_{b,n}&=& \varrho_\prime (r_b) \left(\frac{4(n-3)}{(n+1)(n-1)^2}\right)^{\frac{n}{n-1}} \label{i05}
\end{eqnarray}
\begin{eqnarray}
s(n)&=&\frac{2n}{1-n}\label{i06}\\
r_b&=&\left(\frac{ (n-3) K\varrho_{\prime,b}^{(1-n)/n} }{ (n-1)^2\pi G}\right)^{1/2} \label{i07}\\
\varrho_{\prime,b} &\leftarrow&\varrho_\prime (r_b)=\left(\frac{(n-1)^2\pi G}{(n-3) K }r_b^2\right)^{\frac{n}{1-n}}\label{i08}\\
\frac{\partial}{\partial r}\left(\frac{\partial (r^2 P(r))}{ \varrho_\prime (r)\partial r}\right)&=& \frac{\partial}{\partial r}\left(\frac{\partial (r^2 K\varrho_\prime^\gamma (r))}{ \varrho_\prime (r)\partial r}\right)= -4\pi r^2 G\varrho_\prime(r) \label{i09}\\
P(r)&=&K\varrho ^\gamma (r).\label{i09a}
\end{eqnarray}
The system of solutions displayed above has been derived by taking a rather involved route through defining a modified Lane Emden equation and making much use of the Lane Emden $\theta, \xi$ variables. This course of action relied essentially on these mathematically powerful variables to crack the non-linear barrier involved with the thermal equilibrium equation to get the solution at $\ref{i04}$. However, it occurred to me that the formula at (\ref{i08}) might give a direct route to a solution of the isothermal gravity equilibrium equation, (\ref{i09}), without the full Lane Emden complications. This calculation is carried through in the next few pages. 
Because $r=\alpha\xi$ the modified Lane Emden type solution at (\ref{i04}) with (\ref{i06}) suggests that we can assume a form  $\varrho_\prime(r,n)$ of two {\it independent\/} variables $r$ and $n$ is a solution of the new isothermal equilibrium equation as at (\ref{ia10}),
\begin{eqnarray}
\varrho_\prime (r,n)&=& X(n) r^{\frac{2n}{1-n}},\ say,\label{ia10}\\
\frac{\partial \varrho_\prime (r,n)}{\partial r}&=&  X(n) \frac{2n}{1-n} r^{\frac{2n}{1-n}-1}= X(n) \frac{2n}{1-n} r^{\frac{3n-1}{1-n}}\label{i11}\\
\frac{\partial^2 \varrho_\prime (r,n)}{\partial r^2}&=& X(n) \frac{2n(3n-1)}{(1-n)^2} r^{ \frac{3n-1}{1-n}-1}= X(n) \frac{2n(3n-1)}{(1-n)^2} r^{ \frac{4n-2}{1-n}}\label{i12}
\end{eqnarray}
The two previous lines are the first and second derivatives of the function $\rho_\prime (r,n)$ with respect to $r$. If we now expand the basic differential equation (\ref{i09}) or (\ref{i13}) as below we get (\ref{i15}).
\begin{eqnarray}
\frac{\partial}{\partial r}\left(\frac{\partial (r^2 K\varrho_\prime^\gamma (r))}{ \varrho_\prime (r)\partial r}\right)+ 4\pi r^2 G\varrho_\prime(r) &=& 0 \label{i13}\\
\frac{\partial}{\partial r}\left(2r \varrho_\prime^{\gamma -1} (r)+ \frac{ r^2 \gamma\varrho_\prime^{\gamma -2} (r) \partial \varrho_\prime (r)}{\partial r} \right)+ 4\pi r^2 \frac{G}{K}\varrho_\prime(r) &=& 0\label{i14}\\
2 \varrho_\prime^{\gamma -1} (r)+ 2r (\gamma -1) \varrho_\prime^{\gamma -2} (r) \frac{\partial\varrho_\prime (r) }{\partial r}&+&\nonumber\\
\frac{ 2r \gamma\varrho_\prime^{\gamma -2} (r) \partial \varrho_\prime (r)}{\partial r}+ r^2 (\gamma -2)\gamma\varrho_\prime^{ \gamma -3} (r) \left(\frac{ \partial \varrho_\prime (r)}{\partial r}\right)^2 &+&\nonumber\\
 \frac{ r^2 \gamma\varrho_\prime^{\gamma -2} (r) \partial^2 \varrho_\prime (r)}{\partial r^2}+
 4\pi r^2 \frac{G}{K}\varrho_\prime(r)&=&0 \label{i15}\\
2 (\varrho_\prime)^{\gamma -1}+ 2r (\gamma -1) (\varrho_\prime)^{\gamma -2} \frac{\partial\varrho_\prime }{\partial r}&+&\nonumber\\
2r \gamma (\varrho_\prime)^{\gamma -2} \frac{ \partial \varrho_\prime }{\partial r}+ r^2 (\gamma -2)\gamma(\varrho_\prime)^{ \gamma -3} \left(\frac{ \partial \varrho_\prime}{\partial r}\right)^2 &+&\nonumber\\
 r^2 \gamma(\varrho_\prime)^{\gamma -2} \frac{ \partial^2 \varrho_\prime }{\partial r^2}+
 4\pi r^2 \frac{G}{K}\varrho_\prime&=&0\nonumber\\
 \label{i16}
\end{eqnarray}
\begin{eqnarray}
2 (X(n) r^{\frac{2n}{1-n}})^{\gamma -1}+ 2r (\gamma) (X(n) r^{\frac{2n}{1-n}})^{\gamma -2} X(n) \frac{2n}{1-n} r^{\frac{3n-1}{1-n}}&+&\nonumber\\
r^2 (\gamma -2)\gamma(X(n) r^{\frac{2n}{1-n}})^{ \gamma -3} \left(X(n) \frac{2n}{1-n} r^{\frac{3n-1}{1-n}}\right)^2 &+&\nonumber\\
 r^2 \gamma(X(n) r^{\frac{2n}{1-n}})^{\gamma -2} X(n) \frac{2n(3n-1)}{(1-n)^2} r^{ \frac{4n-2}{1-n}}+
 4\pi r^2 \frac{G}{K} X(n) r^{\frac{2n}{1-n}}&=&0\nonumber\\
 \label{i17}\\
2 ( r^{\frac{2n}{1-n}})^{1/n} + 2r (1+1/n) ( r^{\frac{2n}{1-n}})^{1/n-1}  \frac{2n}{1-n} r^{\frac{3n-1}{1-n}}&+&\nonumber\\
r^2 (1/n-1)(1+1/n)( r^{\frac{2n}{1-n}})^{1/n-2} \left( \frac{2n}{1-n} r^{\frac{3n-1}{1-n}}\right)^2 &+&\nonumber\\
 r^2 (1+1/n)( r^{\frac{2n}{1-n}})^{1/n-1}  \frac{2n(3n-1)}{(1-n)^2} r^{ \frac{4n-2}{1-n}}+
 4\pi r^2 \frac{G}{K} X^{\frac{n-1}{n}}(n) r^{\frac{2n}{1-n}}&=&0\nonumber\\
 \label{i18}
\end{eqnarray}
\begin{eqnarray}
2  + 2 (1+1/n) \frac{2n}{1-n} &+&\nonumber\\
 (1/n-1) (1+1/n) \left( \frac{2n}{1-n} \right)^2 &+&\nonumber\\
 (1+1/n) \frac{2n(3n-1)}{(1-n)^2} +
 4\pi \frac{G}{K} X^{\frac{n-1}{n}}(n)&=&0\nonumber\\
 \label{i19}\\
2  + \frac{8(1+n)}{1-n} +\frac{2(1+n)(3n-1)}{(1-n)^2} +
 \frac{4\pi G}{K} X^{\frac{n-1}{n}}(n) &=&0\nonumber\\
 \label{i20}
\end{eqnarray}
\begin{eqnarray}
\frac{2(1-n)^2}{(1-n)^2}  + \frac{8(1-n^2)}{(1-n)^2} +\frac{2(1+n)(3n-1)}{(1-n)^2} +\frac{4\pi G}{K} X^{\frac{n-1}{n}} &=&0\nonumber\\
\label{i21}
\end{eqnarray}
\begin{eqnarray}
\frac{8}{(1-n)^2}  &+&  \frac{4\pi G}{K} X^{\frac{n-1}{n}}(n)= 0\nonumber\\
\label{i22}\\
X(n)&=& \left( \frac{-8 K }{4\pi G (1-n)^2}\right)^{\frac{n}{n-1}} \label{i23}\\
\end{eqnarray}
However, the expression that we had thought $X(n)$ was representing at (\ref{i04}) etc, was 
\begin{eqnarray}
\left(\frac{(n-1)^2\pi G}{(n-3) K }\right)^{\frac{n}{1-n}}. \label{i24}
\end{eqnarray}
It seems that these two different expressions cannot reasonable be equal without getting some nonsense result. However, there is some subtlety in the definition of the quantity $\varrho_{\prime,b}$ and its use in the Lane Emden definition of the transformed density. This problem can be resolved by making use of the extra mathematical machinery available in the revised form of the Lane Emden Equation as explained in the next section.
\section{Lane Emden Revised Structure}
\setcounter{equation}{0}
\label{sec-lers}
\begin{eqnarray}
\varrho_{\prime,n} (r) &\rightarrow& \varrho_{\prime,b} \vartheta_n ^n(\xi)\label{i25}\\
\vartheta_n (\xi) &=& \xi^{- \frac{n+2}{n+1}}\varphi_n (\xi) \label{i26}\\
\varphi_n (\xi)&=&  A(n) \xi^{p(n)} \label{i27}\\
\varrho_{\prime,n} (r) &\rightarrow&\varrho_{\prime,b} (\xi^{- \frac{n+2}{n+1}} A(n) \xi^{p(n)})^n\nonumber\\
\varrho_{\prime,n} (r) &\rightarrow& \varrho_{\prime,b} A^n(n) \xi^{nq(n)}=B_{b,n}\xi^{s(n)}, \ say \label{i28}\\
B_{b,n}&=& \varrho_{\prime,b} \left(\frac{4(n-3)}{(n+1)(n-1)^2}\right)^{\frac{n}{n-1}} \label{i29}\\
(B_{b,n})_{max\forall n}&\approx & 0.0450819 \varrho_{\prime,b} \label{i30}\\
s(n)&=&nq(n)=\frac{2n}{1-n}\label{i31}\\
\varrho_{\prime,b} &\leftarrow&\varrho_{\prime}  (r_b)\label{h6.8111}\\
\vartheta_n (\xi_b)&=&1= \xi_b^q A(n) \label{i32}\\
\xi_b &=& A(n)^{\frac{n-1}{2}}= \left(\frac{4(n-3)}{(n+1)(n-1)^2}\right)^{\frac{1}{2}}.\label{i33}\\
\alpha&=& \left(\frac{(n+1) K\varrho_{\prime,b}^{(1-n)/n} }{4\pi G}\right)^{\frac{1}{2}}. \label{i34}\\
r_b&=& \alpha\xi_b=\left(\frac{(n+1) K\varrho_{\prime,b}^{(1-n)/n} }{4\pi G}\right)^{\frac{1}{2}} \left(\frac{4(n-3)}{(n+1)(n-1)^2}\right)^{\frac{1}{2}}.\nonumber\\
 \label{i35}\\
r_b&=&\left(\frac{ (n-3) K\varrho_{\prime,b}^{(1-n)/n} }{ (n-1)^2\pi G}\right) ^{\frac{1}{2}}. \label{i36}\\
\varrho_{\prime,b} &=&\varrho_\prime (r_b)=\left(\frac{(n-1)^2\pi G}{(n-3) K }r_b^2\right)^{\frac{n}{1-n}}.\label{i37}
\end{eqnarray}
\begin{eqnarray}
\frac{1}{\xi^2}\frac{\partial }{\partial \xi }\left(\frac{\xi^2 \partial \theta(\xi)}{\partial \xi }\right)&=&- \theta^n(\xi) \label{i38}\\
\frac{\partial}{\xi^2\partial \xi }\left(\frac{2\xi\vartheta(\xi)}{n+1}+ \frac{\xi^2 \partial \vartheta(\xi)}{\partial \xi }\right)&=& - \vartheta^n(\xi)\label{i39}
\end{eqnarray}
However, if we look more closely at the result at (\ref{i29}) that seems to contradict the form assumed at (\ref{ia10}) we see that a more rational expectation is that after using the relation $r=\alpha\xi$ we might expect the transformation from the isothermal equilibrium equation variables  to the Lane Emden type equation variables to cause a transformation as indicated at (\ref{i40}) rather than precise equality holding.
\begin{eqnarray}
\varrho (r,n)=X(n) r^{\frac{2n}{1-n}}&=&\left( \frac{-8 K }{4\pi G (1-n)^2\alpha^2}\right)^{\frac{n}{n-1}} \xi^{\frac{2n}{1-n}} \nonumber\\
\left( \frac{-8 K }{4\pi G (1-n)^2\alpha^2}\right)^{\frac{n}{n-1}} \xi^{\frac{2n}{1-n}} &\rightarrow& \varrho_\prime (r_b) \left(\frac{4(n-3)}{(n+1)(n-1)^2}\right)^{\frac{n}{n-1}}\xi^{\frac{2n}{1-n}} \nonumber\\
\label{i40}
\end{eqnarray}
or
\begin{eqnarray}
 \frac{-8 K }{4\pi G (1-n)^2\alpha^2} \rightarrow \varrho_\prime (r_b)^{\frac{n-1}{n}}\frac{4(n-3)}{(n+1)(n-1)^2}. \label{i41}
\end{eqnarray}
It follows that 
\begin{eqnarray}
\varrho_\prime (r_b) \rightarrow\left(\frac{ K (n+1)}{2\pi G \alpha^2  (3-n)}\right)^{\frac{n}{n-1}}, \label{i42}
\end{eqnarray}
but according to (\ref{i34})
\begin{eqnarray}
\varrho_{\prime,b}  &=& \left(\frac{ 4\pi\alpha^2 G }{ (n+1)K }\right)^{\frac{n}{1-n}}=\left(\frac{ (n+1)K }{ 4\pi\alpha^2 G }\right)^{\frac{n}{n-1}}. \label{i43}
\end{eqnarray}
It follows that $\varrho_{\prime,b}$ only equals  $\varrho_\prime (r_b)$ in the limiting case when $n=1$. Otherwise, $\varrho_{\prime,b}$ should be taken to be given by (\ref{i43}).  It is also the case that $\varrho_{\prime,b}$ only sets the function value at which input functions of $r$ are equal to output functions $\xi$ and this is a matter of personal choice anyway. However, this is different from the situation with the normal Lane Emden equation where only one input function is involved and so $\rho_{\prime,c}$ can be given an unambiguous value. In the next section I shall confine the work to using the thermal equilibrium type of solutions, (\ref{ia10}), rather than the modified Lane Emden version of the solutions, (\ref{i28}). The former solution seems simpler than the latter and are physically more immediate in that they involve the physical variable $r$ rather than the esoteric  mathematical variable, $\xi$, which of course is an advantage. However, the former was found originally by solving its modified Lane Emden transform. As shown above, the thermal equilibrium  collection of solutions is give by 
\begin{eqnarray}
\varrho_\prime (r,n)&=& X(n) r^{\frac{2n}{1-n}} \label{i45B}\\
X(n)&=& \left( \frac{-8 K }{4\pi G (1-n)^2}\right)^{\frac{n}{n-1}}\label{i45C}\\
\varrho_{sc}(n)&=& X(n)= \left( \frac{-8 K }{4\pi G (1-n)^2}\right)^{\frac{n}{n-1}}\label{i45E}\\
\varrho_\prime (r,n)&=& \varrho_{sc}(n)r^{ \frac{2n}{1-n}} \label{i45F}
\end{eqnarray}
We can identify the function $X(n)$ as a scale factor, $ \varrho_{sc}(n)$, representing {\it size\/} associated with the isotropic index $n$ and the index of $r $, $s(n)=2n/(1-n)$, as giving a more intrinsic indication of the solutions shape associated with the isotropic index $n$. Consequently when we observe a galaxy its apparent size will depend on its distance so scale cannot be used to determine the mass distribution within is boundary. However, $s(n)$ will determine its state and that does not depend on its distance from the observer. The point I am making is that if we wish to compare different solutions given by different $n$ in diagrams or discussion, we can take $\varrho_{sc}(n)=X(n)=1$ as a temporary simplification. In diagrams where this temporary simplification is used all solutions will coincide in value when $r=1$. The scale also has to be real and positive for all physical solutions and an inspection of the formulae (\ref{i45E}) and  (\ref{i45F}) indicates that this reality condition implies that the isotropic index for reality can only involve $n$ where $(-1)^{ \frac{n}{1-n}}$ is real and positive. The values of $n$ conforming to this condition is an infinite discrete set. This set of values was worked out in paper (\cite{72:gil}) with reference to a subset of recognisably solutions. Here it has become clear that this condition for reality extends to all solutions of the isothermal sphere equilibrium equation.
\newpage
Thus real positive values occur when
\begin{eqnarray} 
(-1)^{ \frac{n}{1-n}}=\cos \left(\frac{n\pi}{1-n}\right) +i\sin \left(\frac{n\pi}{1-n}\right)=+1.\label{222}
\end{eqnarray}
 That is the two conditions below hold
\begin{eqnarray}
\sin \left( \frac{ \pi n }{n -1}\right)&=&0\implies \frac{ \pi n }{n -1}=\pi k\label{h6.125119}\\
\cos \left(\pi k\right)&>&0\implies k=2l, \label{h6.125120}
\end{eqnarray}
where $k$ is an integer.
Thus
\begin{eqnarray} 
n=\frac{2l}{2l-1}, \label{h6.125121}
\end{eqnarray}
where $l$ is an integer. However, if the mass density distributions are to converge to zero at infinity it is necessary that  
\begin{eqnarray}
s(n)= \frac{2n}{1-n}<0\label{h6.125122}
\end{eqnarray} 
or $n$ remains {\it outside\/} the range
\begin{eqnarray}
0<n<1.\label{h6.125123}
\end{eqnarray}
In other words, convergence to zero at $r=\infty$ is the case outside the relatively small range of values for $n$, $0<n<1$. Thus outside this range any rational number value for $n$ of the form (\ref{h6.125121}), where $l$ is an integer will do. In order to use solutions such as (\ref{i45B}) of the isothermal equilibrium equations in galactic modelling we have to confront the problem that all such solutions chosen using values of $n$ that ensure convergence to zero when $r$ goes to infinity, will certainly diverge when $r$ goes to zero. The implication of this is that these {\it raw\/} density solution involved all diverge towards $r=0$. However, generally physical galactic densities do generally become large at their mass centre but this is not the same as becoming infinite. However, here we are building mathematical models to physical specifications and so the mathematics has to be adapted to give realistic physics. Fortunately this accommodation between mathematics and physics is cleanly and convincingly achieved by introducing the idea of a spherical galactic core. This core can be taken to be of uniform mass density from its outer radius $r_\epsilon$ all the way to the radial origin. This uniform density distribution can also be taken to have the same value as the raw solution density at $r=r_\epsilon$. Thus the model character is such that the finite core mass replaces the offending mathematical infinite mass within spheres centred on and with surfaces near the origin. I shall call the mass of the core $M_\epsilon$ and take its radius to be $r_\epsilon$ and regard it as an unavoidable model construct. In the next section, the core concept will be employed to derive the general relativistic rotation curves and their spatial derivatives with respect to radial distance from the origin, $r$. It will also be assumed that the totality of general relativity contributed mass, $ M_{GR}(r)$, within any spherical region about the radial origin $r=0$ of radius $r$ is just the sum of $ \varrho (r^\prime)$ contributed density outside the core radius $r_\epsilon$ together with the core mass within the spherical core region as at (\ref{i46B}). 
\section{Isothermal Equilibrium  Galactic Rotation Curves}
\setcounter{equation}{0}
\label{sec-grgrc}
\begin{eqnarray}
v^2_2(r) &=&  \frac{G}{r} \left(M^+( r)  + M_P(r)\right) -\frac{G M_\Lambda(r)}{r}   \label{i46}\\
M_{GR}(r)&=& M^+(r)+ M_P(r)- M_\Lambda(r)= \int_{r_\epsilon}^r \varrho_g (r^\prime)dr^\prime +M_\epsilon \label{i46B}\\
v^2_2(r) &=& \frac{G}{r} M_{GR}(r) =\frac{G}{r} \left(\int_{r_\epsilon}^r \varrho _g(r^\prime)dr^\prime + M_\epsilon\right) \label{i46C}\\
\frac{\partial v^2_2(r) }{\partial r}&=& \frac{-G}{r^2} \left(\int_{r_\epsilon}^r \varrho_g (r^\prime)dr^\prime+M_\epsilon\right) +\frac{G}{r}\varrho_g (r) \label{i46D}\\
&=& \frac{-G}{r^2} \left(\int_{r_\epsilon}^r {4\pi\varrho_g(n) r^{\prime 2}r^\prime}^{\frac{2n}{1-n}}
dr^\prime + M_\epsilon \right)    +\frac{G}{r}4\pi\varrho_g(n) r^2  r^{\frac{2n}{1-n}}\nonumber\\
\label{i46E}
\end{eqnarray}
\begin{eqnarray}
&=& \frac{-G}{r^2} \left(\int_{r_\epsilon}^r {4\pi\varrho_g(n) r^\prime}^{\frac{2}{1-n}}
dr^\prime + M_\epsilon \right)    +G4\pi \varrho_g(n)  r^{\frac{1+n}{1-n}}\label{i46F}\\
&=& \frac{(n-1)G4\pi\varrho_g(n)}{(3-n)r^2}(r^{\frac{3-n}{1-n}}- r_\epsilon^{\frac{3-n}{1-n}})
- \frac{G M_\epsilon}{r^2} +   G4\pi\varrho_g(n) r^{\frac{1+n}{1-n}}\nonumber\\
\label{i46FF}\\
&=& \frac{G4\pi\varrho_g(n) (n-1)}{ 3-n}  {r}^{\frac{1+n}{1-n}}\nonumber\\
&-&\frac{ G 4\pi\varrho_g(n)}{ r^2}\left( \frac{(n-1) }{(3-n) }r_\epsilon^{\frac{3-n}{1-n}}
 +\frac{M_\epsilon}{4\pi\varrho_g(n) }\right)
 +G4\pi \varrho_g(n)r^{\frac{1+n}{1-n}}.\label{i46G}
\end{eqnarray}
Let us denote the quantity in the large round brackets above at (\ref{i46G}) multiplied by $(3-n)r_0^\frac{3-n}{n-1}$ as a function, called, $two(r_\epsilon,n)$, of the parameters $r_\epsilon $ and $n$,
\begin{eqnarray}
two(r_\epsilon,n)= r_0^{\frac{3-n}{n-1}}\left( (n-1)r_\epsilon^{\frac{3-n}{1-n}} 
 +\frac{(3-n)M_\epsilon}{4\pi\varrho_g(n)} \right), \label{i46Ga}
\end{eqnarray}
we see that this quantity is to be different according to which isotropic index value $n$ is used but otherwise it depends on what we chose the relation between the core mass, $M_\epsilon$, and the core mass radius $r_\epsilon$ to be. This implies that the function {\it two\/} is a model type dependent quantity and we can therefore select all state $n$ to be of the model type most convenient for simplicity. If we look at the equation at (\ref{i46G}), we see that very substantial simplicity is obtained if the function {\it two\/} is taken to have the fixed value $2$ as its choice of name implies. Thus we get the simplified equation for local rotation curve  gradient with respect to $r$ at (\ref{i49A}). 
\begin{eqnarray}
\frac{\partial v^2_2(r) }{\partial r}&=& -\frac{4 G \pi\varrho_g(n)r_0^{\frac{n-3}{n-1}}}{ r^2(3-n)} two(r_\epsilon,n) \nonumber\\
&+&G4\pi \varrho_g(n)r^{\frac{1+n}{1-n}}+\frac{G4\pi\varrho_g(n) (n-1)}{ 3-n}  {r}^{\frac{1+n}{1-n}}\label{i46Gb}\\
 &=& -\frac{8 G \pi\varrho_g(n)r^{\frac{1+n}{1-n}}}{(3-n)}\left(\frac{r}{r_0}\right)^{\frac{n-3}{1-n}}\nonumber\\
&+&G4\pi\varrho_g(n) r^{\frac{1+n}{1-n}}+\frac{G4\pi\varrho_g(n) (n-1)}{ 3-n}  {r}^{\frac{1+n}{1-n}}\label{i46Gc}\\
&=&G4\pi \varrho_g(n){r}^{\frac{1+n}{1-n}}\left(\frac{ (n-1)}{ 3-n}  -\frac{ 2}{(3-n) } \left(\frac{r}{r_0}\right)^{\frac{n-3}{1-n}} +1 \right) \label{i46Gd}\\
&=&\frac{G8\pi \varrho_g(n)}{n -3}{r}^{\frac{1+n}{1-n}}\left(-1  + \left(\frac{r}{r_0}\right)^{\frac{n-3}{1-n}}\right). \label{i49A}
\end{eqnarray}
This last equation which gives us the radial spatial gradient of any solution everywhere  determined by the isotropic index, $n$, is a key equation which enables us to determine precisely how flat or otherwise any specific given rotation curve solution is.
\newpage
\section{The Galactic Core}
\setcounter{equation}{0}
\label{sec-tgc}
let us now consider the function $two(r_\epsilon,n)$ at (\ref{i46Ga}) and use this to find out how the core radius, $r_\epsilon$, depends on the isotropic index $n$. We have
$$ two(r_\epsilon,n)= r_0^{\frac{3-n}{n-1}}\left( (n-1)r_\epsilon^{\frac{3-n}{1-n}} +\frac{(3-n)M_\epsilon}{4\pi\varrho_g(n)} \right)$$
\begin{eqnarray}
two(r_\epsilon,n)&=&2= r_0^{\frac{3-n}{n-1}}\left( (n-1)r_\epsilon^{\frac{3-n}{1-n}} +\frac{(3-n)M_\epsilon}{4\pi\varrho_g(n)} \right)\label{i49B}\\
M_\epsilon (n,r_\epsilon)&=& 4 \pi r_\epsilon^3 \varrho_g (r_\epsilon,n)/3 =4 \pi \varrho_g(n)r_\epsilon^{\frac{n+3}{1-n}}/3 \label{i49C}
\end{eqnarray}
\begin{eqnarray}
2&=& r_0^{\frac{3-n}{n-1}} \left((n-1)r_\epsilon^{\frac{3-n}{1-n}}
 +(3-n)r_\epsilon^{\frac{3-n}{1-n}}/3 \right)\label{i49D}\\
&=& \left((n-1) \left(\frac{r_\epsilon}{r_0}\right)^{\frac{3-n}{1-n}}
 +(3-n) \left(\frac{r_\epsilon}{r_0}\right)^{\frac{3-n}{1-n}}/3 \right)\label{i49DE}\\
\varrho_g (r_\epsilon,n)&=& \varrho_g(n)r_\epsilon^{\frac{2n}{1-n}}\label{i49E}\\
\left(\frac{r_\epsilon}{r_0}\right) &=& \left(\frac{3}{n}\right)^{\frac{1-n}{3-n}}.\label{i49F}
\end{eqnarray}
Thus the  core radius $r_\epsilon$ decreases  from large values near $n=0+$ to zero as $n$ approaches $+\infty$.
\begin{eqnarray}
v^2_2(r) &=& \frac{ M_{GR}(r)}{r} =\frac{G}{r} \left(\int_{r_\epsilon}^r \varrho_g (r^\prime)dr^\prime + M_\epsilon\right) \label{i50}\\
&=&\frac{G}{r}\left( \varrho_g(n)\int_{r_\epsilon}^r {4\pi r^{\prime 2}r^\prime}^{\frac{2n}{1-n}}
dr^\prime + M_\epsilon \right) \label{i51}\\
&=& \frac{G4\pi\varrho_g(n) (1-n)}{ 3-n}  {r}^{\frac{2}{1-n}}-\frac{ G 4\pi\varrho_g(n)}{ r}\left( \frac{(1-n) }{(3-n) }r_\epsilon^{\frac{3-n}{1-n}}
 -\frac{M_\epsilon}{4\pi\varrho_g(n) }\right)\nonumber\\
 \label{i52}\\
&=& \frac{G4\pi\varrho_g(n) (1-n)}{ 3-n}  {r}^{\frac{2}{1-n}}+\frac{ G 4\pi\varrho_g(n)}{ r}\left( \frac{2 r_0^{\frac{3-n}{1-n}}}{3-n} \right) \label{i53}\\
&=&\frac{G4\pi\varrho_g(n){r}^{\frac{2}{1-n}}}{ 3-n}\left(  1 + 2 \left(\frac{r}{r_0}\right)^{\frac{n-3}{1-n}}-n  \right) .\label{i54}
\end{eqnarray}
Expression (\ref{i54}) is the final form for the isotropic equilibrium galactic rotation curves formula with any specific curve determined by an appropriate value  of the isotropic index $n$. A selection of one curve, green, $index,\  n=1.333$, from the bundle of $25$ curves that are bounded by the $index$ values\  $n= 2,\ red$ and  $n=1.02041,\ blue$ is shown in top diagram that can be found on page $17$ at (\href{http://www.maths.qmul.ac.uk/~jgg/gil135.pdf} 
{QMUL, 2011}). In the lower diagram the spatial derivative curves of the same three curves shows that they are all relatively flat out beyond radial position $r=15$. Even before this position, $10<r<15$, their gradients are between and $0$ and $-0.23,-0.15,-0.13$ respectively, as can be found by direct calculation.

We can find a definite formula for the total mass of the external to  core mass distribution, $ M_{extc}(r)$, up to radius $r$ as a function of the isotropic index $n$ using the definition from (\ref{i50}) as below
\begin{eqnarray}
M_{extc}(r)&=&M_{GR}(r)- M_\epsilon =\int_{r_\epsilon}^r \varrho_g (r^\prime)dr^\prime \label{i55}\\
&=&\varrho_g(n)\int_{r_\epsilon}^r {4\pi r^{\prime 2}r^\prime}^{\frac{2n}{1-n}}dr^\prime \label{i56}\\
&=& \frac{\varrho_g(n)4\pi (1-n)}{ 3-n} \left( {r}^{\frac{3-n}{1-n}}-r_\epsilon^{\frac{3-n}{1-n}}\right)\label{i57}\\
M_\epsilon (n,r_\epsilon)&=& 4 \pi r_\epsilon^3 \varrho_g (r_\epsilon,n)/3 =4 \pi \varrho_g(n)r_\epsilon^{\frac{n+3}{1-n}}/3 .\label{i58}
\end{eqnarray}
with the last entry coming from \ref{i49C}. It follows that the ratio of extra to core mass to core mass ratio, $R(r,r_\epsilon,n)$ up to radius $r$, depends on state index number $n$ as
\begin{eqnarray}
R(r,r_\epsilon,n) &=& \frac{3 (1-n) }{ (3-n) } r_\epsilon^{\frac{n+3}{n-1}} \left( {r}^{\frac{3-n}{1-n}}-r_\epsilon^{\frac{3-n}{1-n}}\right)\label{i59}\\
R(\infty,r_\epsilon,n) &=& \frac{3 (1-n) }{ (3-n) } r_\epsilon^{\frac{n+3}{n-1}} \left(-r_\epsilon^{\frac{3-n}{1-n}}\right)\label{i60}\\
&=& \frac{3 (n-1) }{ (3-n) } r_\epsilon^{\frac{2n}{n-1}}\label{i61}\\
&=& \frac{3 (n-1) }{ (3-n) } \left(\left(\frac{3}{n}\right)^{\frac{1-n}{3-n}}\right)^{\frac{2n}{n-1}} \label{i62}\\
&=& \frac{3 (n-1) }{ (3-n) } \left(\frac{3}{n}\right)^{\frac{2n}{n-3}}. \label{i63}
\end{eqnarray}
Thus we have above the relation between the ratio of total external to core mass to core mass and the isotropic index $n$. This ratio is quite complicated. I note here that it increases from zero at $n=1$ to large positive values between $1$ towards $3$. At the value $n=2$ its value is approximately $0.593$.
 \section{A bundle of solutions}
\setcounter{equation}{0}
\label{sec-abos}
To complete this paper, I shall briefly discuss a specific bundle of {\it physical\/}  solutions. A subset of this bundle was mentioned earlier just below equation (\ref{i54}) and is represented on diagram $1$.  The issue of the existence of other solutions of the same form as this bundle or of different forms will be examined in a future paper. The full bundle discussed here is located in the isotropic index range $2\ge n >1$ and the sub bundle, $2\ge n\ge 1.02041$,  mentioned earlier has three elements, blue, green and red,  shown in the diagram. In terms of the integer parameter $l$ the full bundle is given by all the positive integers, $l$; $(1,2,3,4,5...\infty )$, whilst the sub bundle in the diagram is represented by the integer $l$ in the range, $(1\le l\le 25)$. The limiting boundary solution, $n=1$ or equivalently $l=\infty$, of all the non-linear solutions in the bundles is the {\it linear\/} solution for $n=1$ discussed in detail in \cite{72:gil}, equation (4.35) etc. In any small parameter region greater than and including $n=1$ or $l=\infty$ there is an infinite discrete set of solutions determined by the integral parameter $l$. The existence of the integer parameter $l$ as a descriptor and identifier of these many non-linear solutions is my reason for referring to these special {\it physical gravitational \/} isothermal equilibrium states as {\it quantized\/}. The diagram can be found on page $17$ at (\href{http://www.maths.qmul.ac.uk/~jgg/gil135.pdf} 
{QMUL, 2011}).
\section{ Appendix 4 Conclusions}
The new version of the gravitational isothermal equilibrium equation has very many solutions. Among these solutions there is a smaller discrete infinite collection of physical solutions designated by a parameter $l$ that can take on integral values. A physical model for galaxies which involves a massive core is introduced in order to eliminate a {\it mathematical\/} divergence at the radial origin which would otherwise inevitably occur if there is to be convergence to zero at infinite radius. Detailed formulae are obtained for the physical galactic  rotations curves  and also their radial derivatives as a function of $r$ and also as a function of their isotropic index designation $n$. Thus the gradients of these many curves can be found explicitly and their flatness or otherwise  everywhere can be ascertained. An explicit formula for the core radius and core mass is obtained for all solutions of given  polytropic index $n$. Also an explicit formula for the ratio of mass outside the core radius to the core mass is obtained for all solutions of given  polytropic index.
\section{Appendix 5}
\large
\centerline{\Large { \bf Physical Applicability of Self Gravitating}}
\centerline{\Large {\bf Isothermal Sphere Equilibrium Theory IV}}
\centerline{\Large {\bf Gravitational Polytropic Schr\"odinger Equation}}
\centerline{\Large {\bf Clumping Bridge between Cosmology and Quantum Theories}}
\centerline{\Large {\bf A Cohesive Force for Dark Matter Galactic Halos}}
\vskip0.5cm
\centerline{July 10, 2011}
\vskip 0.75cm  
\section{Abstract}
A cosmological Schr\"odinger equation that has steady state amplitude solutions that give mass densities  coincident with those of a new isothermal gravitational equilibrium equation is used to discuss possible galactic models for dark matter galactic halos.  With the use of an identification of the essence of cosmological gravitational mass clumping, this Schr\"odinger equation is shown to imply a definite molecular cohesive force that has to exist to hold the thermal equilibrium mass densities of the halos in approximate steady state conditions. A formula for the ratio of the numerical value of this new cohesive force between two particles  at separation $r$ to the numerical value of the Newtonian force between two particles at the same separation in terms of the isotropic index $n(l)$ of the system state is found. This ratio is used to discuss the viability of galactic models in terms of the masses involved and the state quantum number $l$. Atomic argon is shown to be one possible particle representation of dark matter under the isothermal equilibrium density constraints. 
\vskip 0.2cm 
\centerline{Keywords: Cosmology, Dust Universe, Dark Energy, Dark matter}
\centerline{Newton's Gravitation Constant, Galactic Halo, Rotation Curves}
\centerline{Isothermal Gravitational Equilibrium, Lane Emden Equation Modification}
\centerline{Schr\"odinger Equation, Clumping}
\vskip 0.3cm
\centerline{PACS Nos.: 98.80.-k, 98.80.Es, 98.80.Jk, 98.80.Qc}
\section{Introduction}
\setcounter{equation}{0}
\label{sec-intro}
This paper is a follow up of papers, \cite{71:gil}, \cite{72:gil} and \cite{73:gil} of similar titles on the problem of formulating the equation that describes the equilibrium of a gaseous material in a self gravitational equilibrium  condition in the galaxy modelling context, \cite{70:wei}, see also, appendix 2 of (\cite{58:gil}). Here I shall relate isothermal gravitational equilibrium theory with a cosmological Schr\"odinger equation obtained earlier, (\cite{64:gil}) and thus demonstrate  a theoretic bridge between quantum theory and cosmology. The structure involved in this cosmological Schr\"odinger equation will, with more detail than originally, firstly be repeated in the next subsection. 
\subsection{Cosmological Schr\"odinger Equation}
\label{sec-cse}
In reference (\cite{64:gil}), I showed that the whole theory for the dust universe model can be obtained from a Schr\"odinger equation (\ref{j0}) or,  (\ref{jj0}) in terms of a quantum density $\rho_{nl}(t)$ from the standard general Schr\"odinger equation (\ref{j3}) with the condition $\nabla^{2} \Psi_{nl,\rho} ({\bf r},t) \equiv 0$ and the external potential $ V({\bf r,t})$ replaced with the feed back term $V_C (t)$ given at (\ref{j1}). The last four equations below fill in all the details. $\rho (t)$ is the spatially uniform positively gravitating substratum mass density from the dust universe model and $\Lambda$ is Einstein's lambda.
\begin{eqnarray}
\frac{i\hbar\partial \Psi_{nl,\rho} (t) }{\partial t} &=& -\frac{\hbar ^2}{2m} \nabla^{2} \Psi_{nl,\rho} ({\bf r},t) +V_C (t) \Psi_{nl,\rho }(t)
\label{j0}\\
\frac{i\hbar\partial \Psi_{nl,\rho} (t) }{\partial t} &=& V_C (t) \Psi_{nl,\rho }(t)
\label{jj0}\\
V_C (t)&=& -(3i\hbar/2)H (t)\label{j1}\\
H(t)&=& (c/R_\Lambda ) \coth(\pm 3ct/(2R_\Lambda)) \label{j2}\\
i\hbar \frac {\partial \Psi ({\bf r},t)}{\partial t} &=& -\frac{\hbar ^2}{2m} \nabla^{2} \Psi({\bf r},t) +V({\bf r,t}) \Psi({\bf r},t). \label{j3}\\
\rho (t) & = & (3/(8\pi G))(c/(R_\Lambda)^2\sinh ^{-2} (3ct/(2 R_\Lambda ))\label{jj4}\\
\Psi_{nl,\rho} (t)&=&\rho ^{1/2}(t) =  A^{1/2}\sinh ^{-1} (3ct/(2 R_\Lambda )) \label{jj5}\\
A & = &(3/(8\pi G))(c/R_\Lambda)^2\label{jj6}\\
R_\Lambda & = & (3/\Lambda)^{1/2}.\label{jj7}
\end{eqnarray}
$H(t)$ above is the Hubble function from the dust universe theory. Here I shall generalise equation (\ref{j0}) by instead of replacing $V({\bf r,t})$ with $V_C (t)$ in (\ref{j3}) , I shall add it and drop all the subscripts to give
\begin{eqnarray}
\frac{i\hbar\partial \Psi ({\bf r,t}) }{\partial t} &=& -\frac{\hbar ^2}{2m} \nabla^{2} \Psi ({\bf r},t) + V({\bf r,t}) \Psi ({\bf r},t) +V_C (t) \Psi ({\bf r},t),\label{j4}
\end{eqnarray}
while not now assuming that $\nabla^{2} \Psi ({\bf r},t)\equiv 0 $. 
I now claim that equation (\ref{j4}) is a general cosmological Schr\"odinger equation. Clearly it differs from the normal general Schr\"odinger equation at (\ref{j3}) in that it has a special external potential of the form $ V_S({\bf r,t}) =V({\bf r,t}) +V_C (t)$ and is only non-general in the very weak sense that it has an additional time only dependent part. Let us now consider solutions to the cosmological Schr\"odinger equation (\ref{j4}) with the special form
\begin{eqnarray}
\Psi ({\bf r},t)= \Psi_1 ({\bf r},t) \Psi_{nl,\rho} (t).\label{j5} 
\end{eqnarray}
Substituting this form into (\ref{j4}), we get 
\begin{eqnarray}
\frac{i\hbar\partial \Psi ({\bf r,t}) }{\partial t} &=&\quad\frac{i\hbar\partial\Psi_1 ({\bf r},t) }{\partial t} \Psi_{nl,\rho} (t) + \Psi_1 ({\bf r},t) \frac{i\hbar\partial\Psi_{nl,\rho}(t)}{\partial t} \nonumber\\
 &=&-\frac{\hbar ^2\nabla^{2} \Psi_1 ({\bf r},t) }{2m} \Psi_{nl,\rho}(t)\nonumber\\
 &+& V({\bf r,t}) \Psi_1 ({\bf r},t) \Psi_{nl,\rho} (t) +V_C (t) \Psi_1 ({\bf r},t) \Psi_{nl,\rho} (t)\nonumber\\
\label{jjj4}
\end{eqnarray}
and using the equation for $\Psi_{nl,\rho} (t)$ at (\ref{jj0}) and the material above {\it after\/} the first equality, we see that the two far right vertically aligned terms above cancel and after dividing through with $\Psi_{nl,\rho} (t)$, we get the result,
\begin{eqnarray}
\frac{i\hbar\partial \Psi_1 ({\bf r,t}) }{\partial t} &=& -\frac{\hbar ^2}{2m} \nabla^{2} \Psi_1 ({\bf r},t) + V({\bf r,t}) \Psi_1 ({\bf r},t).\label{j6} 
\end{eqnarray}
Thus we have recovered the original general Schr\"odinger equation with arbitrary external potential, $V({\bf r,t})$, for the factor $\Psi_1 ({\bf r},t)$ in equation (\ref{j5}). Consequently this factor can be any solution of that equation. Clearly, the  solutions of the cosmological Schr\"odinger equation with its feed back term $V_C (t)\Psi_{nl,\rho} (t)$ embraces all  possible quantum theory solutions to the usual general Schr\"odinger equation (\ref{j6}). Thus this cosmological Schr\"odinger equation can be used as a cosmological platform, via the dust universe model, for any system that can be described by the usual Schr\"odinger equation. The cosmological Schr\"odinger equation has the usual Hermitian scalar product solutions that goes along with its amplitude solutions of the form
\begin{eqnarray}
\rho_S ({\bf r},t)= \Psi ({\bf r},t) \Psi ^* ({\bf r},t)= \Psi_1 ({\bf r},t) \Psi_1 ^* ({\bf r},t) \Psi_{nl.\rho }^2(t).\label{j7} 
\end{eqnarray}
If we choose these solutions to have the character of cosmological mass per unit volume then the $ \Psi_1 ({\bf r},t) \Psi_1 ^* ({\bf r},t)$ will have to be taken as dimensionless as the factor $\Psi_{nl.\rho }^2$ already has the dimensions of mass per unit volume. A direct consequence of this is that the amplitude $\Psi_1 ({\bf r},t)$ of the Schr\"odinger equation (\ref{j6}) will have also to be chosen dimensionless.   
$\rho_S ({\bf r},t)$ is a cosmological mass density with position variability of great generality which can involve all known solutions to the standard Schr\"odinger equation  and any other solutions yet to be found. So that on this basis models of the universe can be found involving individually described galaxies of any known quantum internal structure bundled together to describe the whole universe in great detail. I derived a very simple detailed quantum model for the universe in reference, \cite{64:gil}, using the first version of the cosmological Schr\"odinger equations with the $\nabla^{2} \Psi ({\bf r},t)\equiv 0 $ condition holding. Under this condition there is a multiplicity infinity number of spatially {\it inhomogeneous\/} possible cosmology models. This contrasts sharply with the very limiting homogeneous character of the cosmological models derived directly from general relativity. In the next section, I shall show how the isothermal gravitational equilibrium solutions of galactic structure theory, (\cite{73:gil}),  can be used via the cosmological Schr\"odinger equation to build quantum general relativistic galaxy models. 
\section{Isothermal Cosmological Schr\"odinger Equation}
\setcounter{equation}{0}
\label{sec-icse}The, mass per unit volume, density solutions of the new isothermal gravitational equilibrium equation as functions of the usual dimensioned radius parameter, $r$, and of the isotropic index, $n$, can be written as,
\begin{eqnarray}
\varrho_\prime(r,n)&=& \varrho_{sc}(n)r^{ \frac{2n}{1-n}}, \label{j8}\\
\varrho_{sc}(n)&=&\left( \frac{-8 K }{4\pi G (1-n)^2}\right)^{\frac{n}{n-1}}
.\label{j9}
\end{eqnarray}
The function $\varrho_{sc}(n)$ is a dimensioned scale factor. It is easy to show that the function (\ref{j8}) is a solution to a specific Schr\"odinger equation like (\ref{j6}) by the following steps. Firstly, we note that the kinetic energy term of the Schr\"odinger equation (\ref{j6}) has the resultant form, (\ref{j10}), when $\Psi_1 ({\bf r},t)$ is assumed to have the form,
\begin{eqnarray}
\Psi_1 ({\bf r},t)= \exp\left(-\frac{E(n)it}{\hbar}\right)\varrho_\prime^{1/2}(r,n)= \exp\left(-\frac{E(n)it}{\hbar}\right)\varrho_{sc}^{1/2}(n)r^{ \frac{n}{1-n}}\label{j9.1}
\end{eqnarray}
\begin{eqnarray}
-\frac{\hbar ^2}{2m} \nabla^{2} \Psi_1 ({\bf r},t)&=& -\frac{\hbar ^2}{2m} \frac{\partial(r^2\partial\Psi_1 ({\bf r},t))}{r^2\partial r^2}=- \frac{\hbar^2n\varrho_\prime^{1/2}(r,n) \exp(-\frac{E(n)it}{\hbar })}{2mr^2(1-n)^2}\nonumber\\
\label{j10}\\
i\hbar\frac{\partial}{\partial t} \Psi_1 ({\bf r},t)&=&E(n) \Psi_1 ({\bf r},t)=  E(n) \exp(-E(n)it/\hbar)\varrho_\prime^{1/2}(r,n)\nonumber\\
\label{j10.1}
\end{eqnarray}
the far right hand side of equation, (\ref{j10}), being all that is left from the Laplace operator action on a function that only involves the radial length parameter $r$ and the unitary exponential time factor. The second equation above gives the result of the quantum energy operator acting on $\Psi_1 ({\bf r},t)$. Thus if we denote and define an external potential $V({\bf r})$ by 
\begin{eqnarray}
V({\bf r}) \Psi_1 ({\bf r},t) &=& E(n) \Psi_1 ({\bf r},t) +\frac{\hbar ^2}{2m} \nabla^{2} \Psi_1 ({\bf r},t) \label{j10.2}\\
&=& \left(E(n)+ \frac{\hbar^2n }{2mr^2(1-n)^2}\right)\varrho_\prime^{1/2}(r,n) \exp(-E(n)it/\hbar)\nonumber\\
&=& \left(E(n)+ \frac{\hbar^2n }{2mr^2(1-n)^2}\right) \Psi_1 ({\bf r},t)\label{j10.3}\\
V({\bf r}) &=& E(n)+ \frac{\hbar^2n }{2mr^2(1-n)^2}\label{j10.4}\\
E(n) &=& V({\bf r}) - \frac{\hbar^2n }{2mr^2(1-n)^2}.\label{j10.5}
\end{eqnarray}
 we find looking at equations  (\ref{j10}) $\rightarrow$ (\ref{j10.5}) that the solutions of the new isothermal equilibrium equation are also  solutions of the Schr\"odinger equation,
\begin{eqnarray}
i\hbar\frac{\partial}{\partial t} \Psi_1 ({\bf r},t)= -\frac{\hbar ^2}{2m} \nabla^{2} \Psi_1 ({\bf r},t) + V({\bf r}) \Psi_1 ({\bf r},t),\label{j12}
\end{eqnarray}
provided the an external potential contribution is defined by (\ref{j10.2}) or (\ref{j10.4}).
From (\ref{j12}) and the preceding discussion it follows that the cosmological Schr\"odinger equation, (\ref{j4}), has quantum amplitude solutions that would give the density solutions, (\ref{j7}), involving a factor from isothermal gravitational theory for {\it condensed\/} galactic structure's set within the dark matter expanding substratum, just in fact, as it is thought is the case in the physical world. However, it seems from the quantum Schr\"odinger point of view the stability of galaxies in this model involves an additional inverse square law isothermal potential that is {\it induced\/} by positive gravity as it apparently comes from isothermal theory but it is not quite positive gravity as usually understood because the $G$ coupling constant is not involved in the potential form. The Schr\"odinger equation (\ref{j12}) that I am claiming involves amplitude solutions equivalent to the density solutions of the isothermal gravitational equilibrium equation has been derived by a rather simplistic mathematical argument so that one is entitled to ask to what extent is there a physical basis for  connecting the isothermal route density solutions to the quantum route apparently identical density solutions and what is the external potential that apparently does not involve gravity all about? I discuss this issue in more detail in the next subsection.
\subsection{Cosmological Cohesion}  
\label{sec-cc}
It seems that viewing the isothermal equilibrium equation solutions structure from the schr\"odinger quantum equation point of view conjures up a quantum cohesive force somehow emulating gravity but not quit being gravity. Gravity in the isothermal sphere context involves a gravitational source origin at the centre of the sphere with gravitation potential contributions at any given radius coming from the whole sphere region within that radius and it is this structure that holds the spherical mass together. Because the same radius vector with the same origin as the gravitational vector appears in the  induced quantum external potential, it looks as though the contained gravity of the isothermal sphere situation naturally induces an equivalent local quantum cohesion potential. This perhaps would not have been noticed before because it has been generally recognised that for gravitational clumping to occur, although gravity can be seen as bringing dispersed mass together, other forces are needed such as the electromagnetic or strong nuclear etc to maintain a stable clumped situation. The situation with regard to clumping is so very complicated that a universal gravity induced clumping force that could subsume all the others may not have been considered or even thought necessary. We can compare the quantum force $F_U$ at radius, $r$, derivable from (\ref{j10.4}) with the usual gravitation force $F_G$ at radius $r$ to get the ratio, $R_{U,G}$, of the two force magnitudes at equation (\ref{j15}),
\begin{eqnarray}
F_U(r)&=&+ \frac{\partial V({\bf r})}{\partial r}= \frac{\partial}{\partial r}\frac{+\hbar^2n }{2mr^2(1-n)^2}\nonumber\\
&=& -\frac{ \hbar^2n }{mr^3(1-n)^2}\equiv kgms^{-2}\label{j13.1}\\
F_G(r)&=&+\frac{\partial V_G({\bf r})}{\partial r}=-\frac{mMG}{r^2}\equiv kgms^{-2} \label{j14}
\end{eqnarray}
\begin{eqnarray}
R_{U,G} (M,m,n,r)&=&\frac{F_U(r)}{F_G(r)}= \frac{\hbar^2n}{Mm^2Gr(1-n)^2 }\nonumber\\
&\approx& \frac{1.6667\times 10^{-58}n}{Mm^2r(1-n)^2 }.\label{j15}
\end{eqnarray}
At radii $r_t$ at which the value of this ratio is unity, the cohesive force $F_U(r_t)$ is exactly equal to the gravitational force $F_G(r_t)$ and for larger values of $r>r_t$ the cohesive force becomes less than the gravitational force. The subscript $t$ is an abbreviation for {\it takes over}.
From the way the dimensionless ratio $R_{U,G}$ has been constructed, it seems to me that, a good interpretation of its form and value can be stated as follows. The value of $R_{U,G}(M,m,n,r)$ is a measure of the strength of the new quantum cohesive force exerted on the mass $m$ by the mass $M$ when separated by the distance $r$ compared with the  strength of the Newtonian gravitation force exerted on the mass $m$ by the mass $M$ when separated by the same distance $r$ in a Schr\"odinger quantum mass thermal equilibrium assembly state, $n(l)$.  This is a very clear meaningful statement except for the last nine words starting with, {\it in a Schr\"odinger\/}, this point will be returned to shortly.  I shall confine discussion to the infinite set of discrete states determined by the positive integers, $l=1,2,3\dots \infty$ determined in the previous paper,  \cite{73:gil}.   

Inspecting the formula (\ref{j15}), it is immediately obvious given the very small numerical factor $1.6667\times 10^{-58}$ that relative to the Newtonian force the new cohesion force is negligible if the parameter product, $Mm^2r$, only involves macroscopic values and n is not very close to one. If on the other hand, $r$ is of the order of the very small length, $1.6667\times 10^{-58}\ meters$, the strength of the cohesive force can dominate over the Newtonian gravitation force even if the rest of the parameters are of macroscopic size.
If we try out in this formula astronomical masses such as $M=M_g=10^{11}m_S$, an approximate galactic mass and $m=m_S= 1.98892\times 10^{30}\ kg$ the approximate mass of the sun and $n$ {\it not near one\/}, we get,  
\begin{eqnarray}
R_{U,G} (M_g,m_S,n,r)&\approx& \frac{1.6667\times 10^{-159}n}{(1.98892)^2r(1-n)^2 }\label{j16}
\end{eqnarray}
and if this is to have any cohesion relevance then it looks as though we must go down to ridiculous particle separations of the order of $10^{-159}\ meters$ for the two forces to become comparable. Now this may not be impossible, but at first thoughts, it does seem that using this formula with macroscopic mass values is likely to be very difficult. However, if we lift the restriction {\it n not near one\/}, the complexion of the problem changes dramatically because isothermal cohesive force for any state is determined by the function, $n(l)/(1-n(l))^2$, as shown above and this can contribute a very large numerical value when $n(l)$ is near to one to cancel the numerical factor $10^{-159}$. With $l$ an integer just  $> 10^{1/2}\times 10^{79}$ for example, we get a multiplier of value just greater than $10^{159}$. $l$ can therefore be chosen so that the value of the  parameter length $r$ will be determined by the state implied by the value of $l$ chosen.  Thus the physical viability of the formula is validated if we take into account the great range, $1\le l\le\infty$, of isothermal equilibrium states available within this theory. The implication is that simple models for  galaxies can be set up involving a mass $M$ and secondary masses $m^\prime$ separated from the first mass at radial distances $r^\prime$ and held in  position by gravitational thermal equilibrium force.

The material  above is exactly the course I followed when first thinking about the formula (\ref{j15}). Following this, I tried out the  microscopic case $M=m_p$, the mass of the proton with $m=m_e$, the mass of the electron and was very surprised, though perhaps I shouldn't have been, to find the value of the mass of the proton is $m_p \approx 1.6726\times 10^{-27}\ kg$ whilst the mass of the electron is $m_e \approx  9.1094 \times 10^{-31}\ kg$ and their product would just cancel the $10^{-58}$ factor 	. However, after using
\begin{eqnarray} 
R_{U,G} (M,m,n,r)&=&\frac{F_U(r)}{F_G(r)}= \frac{\hbar^2n}{Mm^2Gr(1-n)^2} \label{A}\\ 
G&=&6.67259\times  10^{-11}\label{B}\\
\hbar&=& 1.054571596\times  10^{-34}\label{C}\\
\hbar^2/G&=& \frac{(1.054571596)^2 10^{-68}}{6.67259\times 10^{-11}}\approx 1.6667\times 10^{-58}\label{D}
\end{eqnarray}
and as as the electron mass is squared, the ratio turns out in this case  to be 
\begin{eqnarray}
R_{U,G} (M,m,n,r)&=&\frac{ 1.6667\times 10^{-58}n}{Mm^2r(1-n)^2}\label{zzz}\\ 
R_{U,G} (M_p,m_e,n,r) &\approx& \frac{1.6667\times 10^{31} n}{1.6726\times (9.1094)^2 r(1-n)^2 }\label{j17}
\end{eqnarray}
and for the quantum cohesive  force to take over from Newtonian gravity in an $n=2$ state the radial separation of the particles involved has to be of order approximately $r\approx 2\times 10^{29}\ meters$. This great distance does strongly suggest, as one might expect, the cohesive force is not involved at the atomic level as all the chemical elements involve microscopic electron proton separations  and further no correction is available from the {\it positive\/} isothermal equilibrium  state  factor $n(l)/(1-n(l))^2$. Let us try a more extreme pair of masses, $(M,m)$, one macroscopic and one microscopic, in this formula. The approximate macroscopic mass of a galaxy $ M=M_g=1.989\times 10^{41}$ with the approximate mass of an argon gas atom, $m=m_{Ar}=6.6\times 10^{-26}$.
\begin{eqnarray}
R_{U,G} (M,m,n,r)&=&\frac{ 1.6667\times 10^{-58}n}{Mm^2r(1-n)^2}\label{zzzz}\\ 
R_{U,G} (M_g,m_{Ar},n,r)&=&\frac{ 1.6667\times 10^{-58}n}{(1.989\times 10^{41})(6.6\times 10^{-26})^2r(1-n)^2}\label{j17z}\\
&=&\frac{ 1.6667\times 10^{-58}n}{(1.989\times 10^{41})(6.6^2\times 10^{-52})r(1-n)^2}\label{j17zz}\\
&=&\frac{ 1.6667\times 10^{-47}n}{(1.989)(6.6^2)r(1-n)^2}.\label{j17zzz}
\end{eqnarray}
Here again, we can correct for the very small factor $10^{-47}$ by making use of the versatile {\it positive\/} state factor $n(l)/(1-n(l))^2$. Thus there are clearly a substantial range of possible models for galactic dark matter structures using the many state system from the new isothermal gravitational equilibrium equation and the last case above, argon gas, identifies a definite possible gaseous atomic material for dark matter. 
In the next subsection, I discuss the cosmological cohesive force as a possible contributory factor of the cosmological {\it clumping process\/} otherwise described as the  condensation of uniformly spread expanding substratum mass to become separated massive  galactic bodies of relatively stable volume.
\section{The Essence of Clumping}
\setcounter{equation}{0}
\label{sec-teoc}
There is a general tendency or perhaps consensus among workers in astronomical physics to regard clumping as the mechanism which takes spherical regions of the primitive expanding mass substratum, reverses its expansion by gravitation attraction towards the spherical centre ultimately to produce the non-expanding isolated galactic mass structures we see today. This, of course, is a very simple and indeed persuasive idea. However, when we begin to think about clumping as a definite physically describable process, we soon recognise a very great deficiency in what the mechanism described above actually is on any precise and detailed level. For a start it cannot be said that we understand what the physical cosmological mass substratum is. Is it mass or energy? Is it continuous or is it particulate? If it is particulate, how varied in size and indeed how massive are objects of which it is composed and is the mass built from fundamental particles that we presently know about? Then there is the question of how it changes over the vast lengths of time that are thought to be involved. All we know about this change with time is that it is overwhelmingly complicated with mass accumulations changing their character, forms, motions and interactions with all aspects of physics involved from gravity down to fundamental particle interactions under quantum mechanics. After all these changes have occurred, can it be said that we know when there is a clumped state of the cosmos even locally? I suggest not, because we do not know if there is such a thing as maximally clumped packet of mass or even what such a physical condition would mean. I think the remarks above greatly {\it understate\/} the difficulties of finding a definite theoretical structure to describe clumping. However, I also think that the cosmological Schr\"odinger equation described earlier offers a small step towards beginning to understand what the overall process of clumping is all about, or expressed otherwise the Schr\"odinger view of the clumping process can help us find the {\it essence\/} of clumping and this will be explained next.

At the human common experience level we all accept that the solid surfaces that we walk on or come in contact with in the course of every day life is clumped mass usually occupying some closed spatial region. However, although indirectly over time such mass regions may have come together as a result of gravitational attraction it is not gravity that holds such closed regions of mass together. Rather a loose rock just lying on the earth's surface is held together by internal and surfaces forces that are certainly quantum electronic and nuclear in character. At such a surface the force involved is the Van der Waal force between molecules. On the other hand, gravity apparently does keep the rock in contact with the earth's surface. Gravity also acts between the clumped mass moon and the clumped mass earth to produce the clumped mass earth-moon system involving the moon's velocity to overcome the gravitational centripetal force and maintain the total clumped mass system in an approximate steady state that is not expanding. Clearly we can build up the idea of what a clumped mass system is as far even to the galactic scale regarded as a clumped mass but when we look at the details of such a large system we can always find surfaces separating component mass configuration themselves clumped in which gravity and other atomic forces such as the electronic and nuclear conspire to generate some form of steady state equilibrium locally with the whole large system being in an approximate steady state and not expanding. Of course, clumped mass is also often present within fluid boundaries, the sun for example, or even gaseous surfaces under which conditions Van de Waal forces are operative in holding massive material together, see for example \cite{74:mf}. Thus it seems that basic to clumping there is always a complex diverse systems of forces complementary to the gravity force that helps maintains a local steady state equilibrium that may originally have been generated by gravitational attraction between mass elements of some sort or other. Thus as first thoughts it appears that the clumping idea is as complicated as is the cosmos its self. However from these considerations there does seem to emerge some abstract idea of what a  clumped cosmological system is which can be expressed as follows. A clumped cosmological system involves an assembly of mass packets in relative motions under gravity and individually held together by self {\it non-gravitational\/}  forces so that the whole assembly is in an approximate steady state condition. Thus far our definition of complex clumped systems is defined in terms of sub clumped systems and therein the {\it essence\/} of clumping can be seen to emerge as follows. A clumped system is a mass packet in an approximate steady state condition constrained by internal non gravitational and or surface forces  so that it is not expanding with the cosmological substratum. Thus clumping as just defined is very close to the somewhat consensus view mentioned at the start of this discussion. However, this definition does mention the very important internal forces involved.

The new isothermal gravitational equilibrium solutions were obtained by consideration of the end condition of a spherical gas like distributed mass density settling into self equilibrium under its internal gravity. This is a very simple example of mass actually clumping under gravity to form a steady state system that is not expanding. Putting such solutions into the cosmological Schr\"odinger equation and finding the complementary force to gravity and its potential that is necessary to maintain this equilibrium is possible because the solutions I am using are so simple.  The derivation of a single simple force to describe the gas gravity clumped situation is only possible because the end point states used here are vastly simpler than the general complicated real world physics states described above that occur in nature. As a result of such considerations, I suggest that the force (\ref{j13.1}) is an idealised smooth quantum microscopic, local and very regular force. It arises in this model of a mass clumped gas situation state and takes over from gravity to give the steady state conditions implied by the Schr\"odinger equation. I say smooth because the mass distributions, the isothermal gravity equilibrium states, being employed are smooth very symmetric and do not have the gaps between chunky parts as I have been describing earlier. On the contrary, each state is a single system in a steady state configuration and so one would expect the force needed to hold the state together in a steady state configuration would reflect this simplicity characteristic. The recognition that there is such a force from this work is simply the result of looking at the local galactic mass stability situation from the Schr\"odinger equation quantum theory point of view which locally describes a possible quantum end point of clumping for a gas cloud condensed from the expanding substratum by gravity. This new force is essentially the Van de Waal force associated with dark matter as represented by an isothermal gravitational gas.
\section{Conclusions}
\setcounter{equation}{0}
\label{sec-con}
The work described above brings together two distinct features arising from the dust universe cosmological model studies. The two features are a cosmological non-linear Schr\"odinger equation and the many solutions of a new isothermal gravitational  equilibrium equation. The cosmological Schr\"odinger equation is constructed so that it has steady state amplitude solutions that are such that by forming their Hermitian scalar products, densities are produced that are exactly the mass densities that arise in the new isothermal equilibrium equation set in a background uniform time dependent density that arises in the general relativity dust universe model describing the expanding substratum. The solutions of the isothermal equilibrium equation give mass densities that imply for orbits, not too near the mass centroid of the galaxy, flat galactic rotation curves just as are presently observed. Thus the solutions of the cosmological schr\"odinger equation give, as eigen-functions  dependent on an integer quantisation number $l$ such that the isotropic index, $n(l)$, depends on $l$ in a definite way, $n(l)=2l/(2l-1)$, representations of galaxies embedded in the expanding substratum background.

Brief considerations of the complexity of physical cosmological mass clumping in the formation of galaxies over epoch time leads to a succinct identification of the essence of clumping. A view of the clumping process that is not far from the usual consensual understanding but does additionally put emphasis on the possible end point of the clumping process. The equilibrium arising from gravitational assembling mass has to terminate in an approximate steady state condition maintained in equilibrium by local quantum level molecular forces. This situation is very successful described by the cosmological Schr\"odinger equation by supplying the external potential that is necessary generate the force field needed hold the density solutions of the isothermal equilibrium equation in their steady state equilibrium condition. Thus the cosmological Schr\"odinger equation bridges the great conceptual gulf between the cosmological substratum and the world described by quantum physics. The quantum force operative at this terminal point of clumping in the isothermal gravitational gas situation is found to be an inverse cube law force just depending on the isotropic index of the condensed state being considered.   
\vskip 0.5cm
\section{Appendix 6}
\vskip 0.4cm
\large
\centerline{\Large {\bf Physical Applicability of Self Gravitating}}
\centerline{\Large {\bf Isothermal Sphere Equilibrium Theory V}}
\centerline{\Large {\bf Quantized Dark Matter Mass Densities}}
\centerline{\Large {\bf Gravitational Quantization from $\Lambda$}}
\centerline{\Large {\bf Why Einstein's Cosmological Constant is Essential}}
\centerline{\Large {\bf Steady State Stabilised Galactic Halos}}
\vskip0.2cm
\centerline{April 12, 2012}
\vskip 0.75cm  
\section{Abstract Appendix 6}
Using a new isothermal gravity equilibrium theory, the dust universe model together with a cosmological Schr\"odinger equation are applied to solving the problem of generating mass spectra. The masses generated can range from sub fundamental particle rest masses to masses greater than that of the universe. The ranges all depend on a quantum integer number $l$, related to the isotropic index $n$,  which can lie between unity and infinity. One such mass obtained is given by $l=8$ and  can represent a small galaxy. The rotation curves for stars, in  motion, within this galaxy are examined for flatness and found to have gradients of approximately, $-10^{-23}$. Examination of the Newtonian gravitation potential associated with these mass quanta  reveals that it is, consistent with the dust universe model, based on Einstein's cosmological constant,  $\Lambda$, rather than on Newton's gravitational constant, $G$, as this last constant disappears by fractional cancellation within the theory structure. Thus this quantization of gravity is based on the cosmological constant. There is found within this theory structure a simulation of negative mass from suitably geometrically orientated positive mass. It is suggested that this feature could supply an explanation for the character of {\it dark energy\/} mass as being due to suitably orientated positive mass. However, this last point needs further study. This paper is a corrected version involving an added section (\ref{sec-eoc}) explaining the corrections.
\newpage    
\centerline{Keywords: Cosmology, Dust Universe, Dark Energy, Dark matter,}
\centerline{Newton's Gravitation Constant, Galactic Halo,}
\centerline{Quantized Isothermal Gravitational Equilibrium,}
\centerline{Cosmological Constant, Schr\"odinger Equation}
\vskip 0.3cm
\centerline{PACS Nos.: 98.80.-k, 98.80.Es, 98.80.Jk, 98.80.Qc}
\vskip 0.3cm
\section{Introduction Appendix 6}
\setcounter{equation}{0}
\label{sec-introod}
This paper is a follow up of papers, \cite{71:gil}, \cite{72:gil}, \cite{73:gil} and \cite{75:gil} of similar titles on the problem of formulating the equation that describes the equilibrium of a gaseous material in a self gravitational equilibrium  condition in the galaxy modelling context, \cite{70:wei}, see also, appendix 2 of (\cite{58:gil}). Here I shall examine, in more detail, the quantum set of dark matter density distributions $\varrho (r,n)$ which depend on the radial distance $r$ and the isotropic quantum state index $n(l)=2l/(2l-1)$ rational number which itself depends on the quantum integer numbers $l:1.2.3...\infty$.
The, mass per unit volume, density solutions of the new isothermal gravitational equilibrium equation as functions of the usual dimensioned radius parameter, $r$, and of the isotropic index, $n$, can be written as,
\begin{eqnarray}
\varrho_\prime(r,n)= \left( \frac{-2 K }{\pi G (1-n)^2}\right)^{\frac{n}{n-1}}r^{ \frac{2n}{1-n}}, \label{k8}
\end{eqnarray}
It is easy to show that the function (\ref{k8}) can be used as the space dependent part of a steady state solution to a specific Schr\"odinger equation by the following steps. Firstly, given a function, $E(n)$, for a set of energies dependent on the parameter $n$, we note that the kinetic energy term of the Schr\"odinger equation (\ref{k12}) will have the form, (\ref{k10}), when $\Psi_1 ({\bf r},t)$ is assumed to have the product form,
\begin{eqnarray}
\Psi_1 ({\bf r},t)&=& e^{-\frac{E(n)it}{\hbar}} \varrho^{1/2}_\prime(r,n) \nonumber\\
&=& e^{-\frac{E(n)it}{\hbar}} \left(\frac{-2 K}{\pi G (1-n)^2}\right)^{\frac{n}{2(n-1)}}r^{\frac{n}{1-n}}\label{k9.1}
\end{eqnarray}
\begin{eqnarray}
-\frac{\hbar ^2}{2m} \nabla^{2} \Psi_1 ({\bf r},t)&=& -\frac{\hbar ^2}{2m} \frac{\partial(r^2\partial\Psi_1 ({\bf r},t))}{r^2\partial r^2}=- \frac{\hbar^2n\varrho_\prime^{1/2}(r,n) \exp(-\frac{E(n)it}{\hbar })}{2mr^2(1-n)^2}\nonumber\\
\label{k10}\\
i\hbar\frac{\partial}{\partial t} \Psi_1 ({\bf r},t)&=&E(n) \Psi_1 ({\bf r},t)=  E(n) \exp(-E(n)it/\hbar)\varrho_\prime^{1/2}(r,n)\nonumber\\
\label{k10.1}
\end{eqnarray}
The second equation above gives the result of the quantum energy operator acting on $\Psi_1 ({\bf r},t)$. Thus if we denote and define an external potential $V({\bf r})$ by 
\begin{eqnarray}
V({\bf r}) \Psi_1 ({\bf r},t) &=& E(n) \Psi_1 ({\bf r},t) +\frac{\hbar ^2}{2m} \nabla^{2} \Psi_1 ({\bf r},t) \label{k10.2}\\
&=& \left(E(n)+ \frac{\hbar^2n }{2mr^2(1-n)^2}\right) \Psi_1 ({\bf r},t)\label{k10.3}\\
V({\bf r}) &=& E(n)+ \frac{\hbar^2n }{2mr^2(1-n)^2},\label{k10.4}
\end{eqnarray}
we find looking at equations  (\ref{k10}) $\rightarrow$ (\ref{k10.4}) that the square roots of solutions of the new isothermal equilibrium equation are also  solutions of the Schr\"odinger equation,
\begin{eqnarray}
i\hbar\frac{\partial}{\partial t} \Psi_1 ({\bf r},t)= -\frac{\hbar ^2}{2m} \nabla^{2} \Psi_1 ({\bf r},t) + V({\bf r}) \Psi_1 ({\bf r},t),\label{k12}
\end{eqnarray}
provided the an external potential contribution is defined by (\ref{k10.4}). It follows from this that the mass densities of the new isothermal equilibrium equations, apart from a multiplicative dimensioned constant, coincide with the probability densities of the Schr\"odinger equation.  The formula at lines (\ref{k10.2}) and (\ref{k10.3}) has a well known significance in the quantum regime. It represents as shown at (\ref{k14}) the statement that $ \Psi_1 ({\bf r},t)$ is an eigen-function of the operator version of the external potential, 
\begin{eqnarray}
\hat V({\bf r})&=& i\hbar\frac{\partial}{\partial t} +\frac{\hbar ^2}{2m} \nabla^{2}\label{k13}\\
\hat V({\bf r})\Psi_1 ({\bf r},t))&=& V({\bf r})\Psi_1 ({\bf r},t)\label{k14}\\
V({\bf r}) &=& E(n)+ \frac{\hbar^2n }{2mr^2(1-n)^2}, \label{k15}
\end{eqnarray}
and in this case the eigen-values of this operator are given by the function $ V({\bf r})$ at (\ref{k15}), these actual values being determined by whatever the appropriate value of the isotropic index $n$ happens to be. Thus more appropriately, we should make the notation changes at (\ref{k16}) and (\ref{k17}) with the consequent change in the schr\"odinger equation (\ref{k12}) recorded at (\ref{k18}).
\begin{eqnarray}
\Psi_1 ({\bf r},t) &\rightarrow& \Psi_1 ({\bf r},t,n) \label{k16}\\
V({\bf r})&\rightarrow& V({\bf r},n)\label{k17}\\
i\hbar\frac{\partial}{\partial t} \Psi_1 ({\bf r},t,n)&=& -\frac{\hbar ^2}{2m} \nabla^{2} \Psi_1 ({\bf r},t,n) + V({\bf r},n) \Psi_1 ({\bf r},t,n). \label{k18}
\end{eqnarray}
Thus we have, perhaps, the  unusual quantum situation, that what might be called an augmented Laplace operator $\hat V({\bf r})$, (\ref{k13}), has steady state eigen-functions which are solutions of an {\it eigen}-Schr\"odinger equation (\ref{k18}). The solutions $\Psi_1 ({\bf r},t,n)$  of the Schr\"odinger equation (\ref{k18}) can be used to give a space variable character to the spatially constant but epoch time variable solutions of the basic quantum solution, $\rho ^{1/2}(t) $, of the dust universe model just by multiplication as below
\begin{eqnarray}
\Psi ({\bf r},t)= \Psi_1 ({\bf r},t,n) \Psi_{nl,\rho} (t),\label{k19} 
\end{eqnarray}
where
\begin{eqnarray}
\Psi_{nl,\rho} (t)&=& \rho ^{1/2}(t) =  A^{1/2}\sinh ^{-1} (3ct/(2 R_\Lambda )) \label{k20}\\
A & = &(3/(8\pi G))(c/R_\Lambda)^2\label{k21}\\
R_\Lambda & = & (3/\Lambda)^{1/2}.\label{k22}
\end{eqnarray}
and
\begin{eqnarray}
\frac{i\hbar\partial \Psi_{nl,\rho} (t) }{\partial t} &=& V_C (t) \Psi_{nl,\rho }(t)\label{k23}\\
V_C (t)&=& -(3i\hbar/2)H (t)\label{k24}\\
H(t)&=& (c/R_\Lambda ) \coth(3ct/(2R_\Lambda)) \label{k25}\\
\rho (t) & = & (3/(8\pi G))(c/(R_\Lambda)^2\sinh ^{-2} (3ct/(2 R_\Lambda )).\label{k26}
\end{eqnarray}
$H(t)$ above is the epoch time variable Hubble {\it constant\/} and $\rho (t)$ is the epoch time variable substratum density both from the dust universe model.
The objective of this work so far is to establish that the wave function $\Psi({\bf r},t)$ defined at (\ref{k19}) is the solution of a {\it cosmological Schr\"odinger\/} equation which might be described as a hybrid structure giving a theoretical mixture of general relativity and quantum theory from a new isothermal gas gravity self equilibrium theory. The cosmological Schr\"odinger equation takes the form,
\begin{eqnarray}
\frac{i\hbar\partial \Psi ({\bf r,t}) }{\partial t} &=& -\frac{\hbar ^2}{2m} \nabla^{2} \Psi ({\bf r},t) + V({\bf r,t}) \Psi ({\bf r},t) +V_C (t) \Psi ({\bf r},t),\label{k27}
\end{eqnarray}
where $V_C (t)$, the feed back potential, is given by equation (\ref{k24}).
There is a freedom to choose the numerical multiplier both in magnitude and dimensionality that goes along with the solutions of the Schr\"odinger equation (\ref{k18}) because of its linearity. This multiplier will be determined by the way the solutions are to be used. The intention here is to use these solutions to modulate with space variability the otherwise time only dependent substratum quantum solutions from general relativity. Because these wave functions are not complex the mass density solutions from the quantum Hermitian product is just the square of the wave function for the substratum from general relativity, $\rho (t)=\Psi_{nl,\rho} (t) \Psi_{nl,\rho}^* (t)\rightarrow \Psi_{nl,\rho}^2 (t)$. This squared quantity has the  built in dimensionality of mass per unit volume. Thus if the density solutions of the cosmological Schrodinger equation is to have the dimensions of mass per unit volume then the wave function  $\Psi_1 ({\bf r},t,n)$ used as a multiplier at (\ref{k19}) needs to be taken,  initially at least, as dimensionless. Recalling the definition of this wave function at (\ref{k9.1})
\begin{eqnarray}
\Psi_1 ({\bf r},t)= e^{-\frac{E(n)it}{\hbar}} \left(\frac{-2 K}{\pi G (1-n)^2}\right)^{\frac{n}{2(n-1)}}r^{\frac{n}{1-n}}\label{k28}
\end{eqnarray}
it can be seen that a dimensionless version of this is easily obtained if it is written in the form 
\begin{eqnarray}
\Psi_1 ({\bf r},t)= e^{-\frac{E(n)it}{\hbar}} \left(\frac{-2 a}{\pi (1-n)^2}\right)^{\frac{n}{2(n-1)}}(r/r_0)^{\frac{n}{1-n}},\label{k29}
\end{eqnarray}
where $a$ is a dimensionless real number numerically equal to $K/G$ is used to   replaces the dimensioned quantity, $K/G$, and $r_0$ is a dimensioned length both determined by the physical context of application.
Thus finally we can write out in full the solution for the cosmological Schr\"odinger equation (\ref{k27}) associated with the isotropic index $n$ as at (\ref{k31}) etc 
\begin{eqnarray}
\Psi ({\bf r},t,n)&=& \Psi_1 ({\bf r},t,n) \Psi_{nl,\rho} (t)\label{k30}\\
&=& e^{-\frac{E(n)it}{\hbar}} \left(\frac{-2 a}{\pi (1-n)^2}\right)^{\frac{n}{2(n-1)}}(r/r_0)^{\frac{n}{1-n}}\Psi_{nl,\rho} (t)\nonumber\\
\label{k31}\\
\Psi_{nl,\rho} (t)&=& \rho ^{1/2}(t) =  A^{1/2}\sinh ^{-1} (3ct/(2 R_\Lambda )) \label{k32}\\
A &=& \left(\frac{3}{8\pi G}\right)\left(\frac{c}{R_\Lambda}\right)^2\label{k32.1}\\
R_\Lambda & = & (3/\Lambda)^{1/2}.\label{k33}
\end{eqnarray}
We should note that the mass density per unit volume solutions of the cosmological schr\"odinger equation are given by the usual Hermitian scalar product, 
\begin{eqnarray}
\rho_S ({\bf r},t,n)&=&\Psi ({\bf r},t,n) \Psi^\dagger ({\bf r},t,n)\label{k34}\\
&=& \left(\frac{-2 a}{\pi (1-n)^2}\right)^{\frac{n}{n-1}}(r/r_0)^{\frac{2n}{1-n}} \Psi^2_{nl,\rho} (t).\label{k35}
\end{eqnarray}
From here on in this paper, the work will be carried through in terms of the integer quantization parameter, $l$, rather than in terms of the isotropic index, $n=2l/(2l-1)$. The quantum number $l$ will be placed as a subscript so that we have
\begin{eqnarray}
\rho_{S,l}({\bf r},t)&=& \rho_S ({\bf r},t,n(l)),\quad n(l)=\frac{2l}{2l-1},\quad \Psi^2_{nl,\rho} (t)=\rho(t)\label{k36}\\
\rho_{S,l}({\bf r},t)&=& \left(\frac{2 a(2l-1)^2}{\pi }\right)^{2l} (r/r_0)^{-4l}\Psi^2_{nl,\rho} (t) \label{k37}\\
&=& \rho_{1,l}({\bf r})\rho(t)= \rho_{b,l}({\bf r})( \rho(t) / \rho(t_b))\label{k37.1}\\
\rho_{1,l}({\bf r}) &=& \left(\frac{2 a(2l-1)^2}{\pi }\right)^{2l} (r/r_0)^{-4l} \label{k37.2}\\
\rho_{b,l}({\bf r}) &=& \rho(t_b)\left(\frac{2 a(2l-1)^2}{\pi }\right)^{2l} (r/r_0)^{-4l} \label{k37.21}\\
&=& \sigma_l (r_0)r^{-4l},\ say,\ with \label{k37.3}\\
\sigma _l(r_0)&=& \rho(t_b)\left(\frac{2 a(2l-1)^2}{\pi }\right)^{2l} r_0^{4l}.\label{k37.4}
\end{eqnarray}
The formula at (\ref{k37}) above gives the density solutions, $\rho_{S,l}({\bf r},t)$, of the cosmological schr\"odinger in terms of the {\it integer\/} quantization parameter $l$ which is now placed as a subscript on the density.

The introduction of the constant $\rho(t_b)$ at that line is self cancelling so that the solution of the cosmological Schr\"odinger is not changed but in effect the two distinct differential equations are renormalized, if from now on, $\rho_{b,l}({\bf r}) $ is taken to be a density solution of the differential equation for $\Psi_1 ({\bf r},t,n)$. The introduction of the self cancelling function at line (\ref{k37.1}) is important for the physical-philosophical interpretation of the solutions of the cosmological Schr\"odinger equation. By construction the solutions of this equation take the product form (\ref{k37.1}), one factor of this product is pure quantum mechanics and the other is pure cosmology from the dust universe model. However, as we have seen, without the $\rho(t_b)$, the solutions of the quantum part $\Psi_{l,1}(\bf{r})$ have to be dimensionless and so the Hermitian product form  cannot represent a mass density. It can however, be regarded as a spatial modulation of the cosmological factor. My suggestion is that the product with or without the $\rho (t_b)$  factor represents two points of observational view. With the $\rho (t_b)$ the density solutions, $\rho_{b,l}({\bf r})$, of the $\Psi_1 ({\bf r},t,n)$ equation represent the view of an observer within and part of its quantum system with cosmology somewhat sidelined. Without the $\rho (t_b)$ , the cosmological Schr\"odinger equation solutions represent the view of a general observer not particularly interested in any specific galaxy but being conscious that regions within galactic domains are spatially different from the substratum. With this philosophical slant on the meaning of the product solutions of the cosmological Schr\"odinger equation, the with $\rho (t_b)$ can be explained as follows. It is usually assumed that galaxies have been around for a very long time. Often it is suggested that the milky way is nearly as old as the universe itself. This seems to be a very reasonable idea and along with this idea it seems likely that a galaxy is a large amount of mass conserved within a not expanding volume. Thus galaxies seem to be objects of almost constant mass density over very large epoch times. Obviously a region of astro-space which accommodates a galaxy is greatly distinguished by its mass density from the substratum mass density in which it swims. Now although mass density may be conserved, if this mass density is mean or average mass density in the region of occupation, great changes or evolution of the local distribution of this mass within the galactic region is not precluded from taking place over time. We can envisage the beginning of a new galaxy as a birth process taking place at a definite epoch time, $t_b$, by a {\it quantum\/} process in which a spherical region of the substratum at the time $t_b$, when the substratum density is $\rho (t_b)$, stops expanding with the substratum by a spatially extended  change of state. A region fractures from the substratum at time $t_b$ to retain the mass and volume at its birth to follow its own evolution under the cyclic steady state factor (\ref{k31}). Thus for all following times the actual internal mass mean density will retain its birth value $\rho(t_b)$ whilst the environment substratum mass density  outside the galactic region will at time $t$ have assumed the much reduced evolved value $\rho (t)$ for $t>t_b$. Thus the birth of a galaxy can be regarded as a random centred and time determined quantum change of state process that effects spherical volumes of the substratum which then evolves scale wise independently of their environment except for their mass centroids which will move with the environment. This is what the wave function $\Psi_1 ({\bf r},t,n)$ describes. It is convenient at this point to introduce a useful conceptual radius associated with the birth of a galaxy. The structure is such that we know two physical characteristics involved with a galactic birth. Its mass $ M_l$ can be found from the theory given its quantum state $l$ and its uniform mass density $ \rho (t_b)$ equal to the substratum mass density at the moment of birth given by the assumed time of birth $t_b$. Thus we can define a conceptual spherical volume $V_b=\frac{ M_l }{\rho (t_b)}=\frac{4\pi r_b^3}{3}$ and the conceptual radius $r_b$ associated with the birth process. I shall interpret this {\it conceptual\/} radius as the radius of a sphere of visible material that suddenly appears at time $t_b$ although it is not likely that there will have been any observers to see the creation event. However, this is not mass creation from nothing, it is a visible change of state of the pre-existing substratum mass. Thus I shall call $r_b$ the {\it visibility\/} radius of the galaxy and as previously discussed this is a feature that stays with the galaxy for very many following years. This radius is theoretically important because the object that it represents is mathematically an infinitely radially extended material sphere.
This can be the recognition of the recently substantiated conclusion that with galaxies what you see is only part of the story.             

The $n$ in the subscript $nl$ above at (\ref{k36}) which is short for non-linear should not be confused with the isotropic index $n$. The version at (\ref{k37.1}) gives the density  solution of the cosmological schr\"odinger in terms of the corresponding density solution of the related schr\"odinger equation (\ref{k18}) with (\ref{k37.3}) and (\ref{k37.4}) giving  a convenient  abbreviation for this function.
So far in this paper, the argument has been developed on the premiss that eigen-values for steady state energies for the mass density distributions (\ref{k8}) are given in the form of the  function $E(n)$ of the isotropic index $n$. In the next section, I shall show that there is a natural function within the {\it quantized\/} isothermal theory for the dark matter galaxy halos that fits this bill.
\section{Steady State Dark Matter Energies, E(n)}
\setcounter{equation}{0}
\label{sec-sdme}
In reference, \cite{73:gil}, I showed that in general relativity the total gravitationally {\it effective\/} mass within a sphere of radius $r$ for a spherically extended source in an isotropic equilibrium state can be written as, 
\begin{eqnarray}
M_{GR}(r)&=& M^+(r)+ M_P(r)- M_\Lambda(r)= \int_{r_\epsilon}^r \varrho_g (r^\prime)dr^\prime +M_\epsilon . \label{k38}
\end{eqnarray}
The actual mass as opposed to effective mass within the same sphere is
\begin{eqnarray}
M_{gr}(r)&=& M^+(r)+ M_P(r)+ M_\Lambda(r) \label{k39}
\end{eqnarray}
because all masses and mass densities are to be taken as positive.
To avoid confusion, I am using the lower case subscript $gr$ for actual mass.
The effective mass $M_{GR}(r)$ expression above is a convenient abbreviation for
\begin{eqnarray}
 M_{GR}(r)&=& (G_+M^+(r)+ G_+M_P(r)+G_- M_\Lambda)/G\label{k40}\\
G_+&=& +G\label{k41}\\
G_-&=& -G .\label{k42}
\end{eqnarray}
The Newtonian gravitational potential at radius r from the centre of a distribution such as (\ref{k37}) above is given by 
\begin{eqnarray}
V_G(r)=\frac{M_{gr}(r)G}{r}. \label{k43}
\end{eqnarray}
Here, as indicated by lower case subscript $gr$, the mass should be the actual mass. In earlier versions of this paper, I mistakenly used the effective mass. This change has the consequence that further changes have been made in the following text. This includes the addition of a section (\ref{sec-eoc}) in this version of this paper entitled {\it Explanation of Corrections\/} in which my mistake is explained and is implications are discussed.
If we are to use the potential (\ref{k43}) then $M^+(r)$ for example needs to be calculated from the formula,
\begin{eqnarray}
M^+(r)&=& \int_0^r\rho_{b,l}({\bf r},t)4\pi r^2dr. \label{k44}\\
&=& \sigma_l (r_0)\int_0^r r^{-4l} 4\pi r^2dr \label{k44.1}\\
&=& \sigma_l (r_0)\int_0^r r^{2-4l} 4\pi dr \label{k45}\\
&=& 4\pi \sigma_l (r_0)\left[\frac{r^{3-4l}}{3-4l}\right]^r_0 .\label{k46}
\end{eqnarray}
The $4\pi r^2$ factor in the first two integrals above converts the mass density per unit volume to mass per unit radius. The integer quantization parameter $l$ can have the numerical values, $1,2,3,4\dots\dots\infty$. It follows that $3-4l$ is always negative.
\begin{eqnarray}
3-4l <0\ \forall l . \label{k47}
\end{eqnarray}
Thus the upper value for r in (\ref{k46}) can be $\infty$ when $r^{3-4l}\rightarrow 0$ but at the lower limit when $r\rightarrow 0$, the lower value of $r^{3-4l}$ diverges to $\infty$. It follows that the raw density functions cannot {\it comfortably\/} be used in calculations. In fact, nature comes to the rescue here with the factual existence of galactic cores. What seems to me to be the simplest assumption is to replace the densities $\rho_{b,l}({\bf r})$ with a more physical realistic densities, $\rho_{b,l,\epsilon}({\bf r})$, defined as follows
\begin{eqnarray}
\rho_{b,l}({\bf r})&\rightarrow& \rho_{b,l,\epsilon}({\bf r})= \rho_{b,l}({\bf r})\quad r\ge r_\epsilon\label{k48}\\
\rho_{b,l}({\bf r})&\rightarrow& \rho_{b,l,\epsilon}({\bf r})= \rho_{b,l}({\bf r}_\epsilon )\quad r < r_\epsilon \label{k49}\\
\lim_{r_\epsilon\rightarrow 0} \rho_{b,l,\epsilon}({\bf r})&=& \rho_{b,l}({\bf r})\quad\forall l \label{k49.1}
\end{eqnarray}
with the region within the radius $r_\epsilon$ being regarded as the galactic core and having the constant density $\rho_{b,l}({\bf r}_\epsilon)$. The last equation above shows that this modification is reversible by taking the limit $r_\epsilon \rightarrow 0$. Thus for practical calculational purposes we can work with the always finite densities $\rho_{1,l,\epsilon}({\bf r})$ and if needs be take the limit $r_\epsilon \rightarrow 0$ afterwards. However, I shall usually drop the $\epsilon\/$ subscript on these densities and only restore it, if it is really needed in context. Let us now return to calculating the effective mass within a sphere of radius $r$ using the finite everywhere densities. Firstly consider the positively gravitating mass excluding the pressure generated part $M_P(r)$ calculated above,
\begin{eqnarray}
M^+(r)&=& \int_0^r\rho_{b,l,\epsilon}({\bf r},t)4\pi r^2dr \label{k50}\\
&=& \sigma_l (r_0)\int_0^{r_\epsilon} {r_\epsilon}^{-4l} 4\pi r^2dr+\sigma_l (r_0)\int_{r_\epsilon}^r r^{-4l} 4\pi r^2dr  \label{k51}\\
&=& \sigma_l (r_0)\int_0^{r_\epsilon} r_\epsilon^{-4l}r^2 4\pi dr+\sigma _l(r_0)\int_{r_\epsilon}^r r^{-4l} 4\pi r^2dr  \label{k52}\\
&=&\sigma_l (r_0)r_\epsilon^{-4l}\left[\frac{r^3}{3}\right]_0^{r_\epsilon} 4\pi+ 4\pi \sigma_l (r_0)\left[\frac{r^{3-4l}}{3-4l}\right]^r_{r_\epsilon} \label{k53}\\
&=&4\pi \sigma_l (r_0)\left(r_\epsilon^{-4l}\frac{r_\epsilon^3}{3} + \frac{r^{3-4l}}{3-4l}-\frac{r_\epsilon^{3-4l}}{3-4l}\right) \label{k54}\\
&=&4\pi r_0^{4l}\rho(t_b)\left(\frac{2 a(2l-1)^2}{\pi }\right)^{2l} \left(r_\epsilon^{-4l}\frac{r_\epsilon^3}{3} + \frac{r^{3-4l}}{3-4l}-\frac{r_\epsilon^{3-4l}}{3-4l}\right)\nonumber\\
&=&4\pi r_0^{4l}\rho(t_b)\left(\frac{2 a(2l-1)^2}{\pi }\right)^{2l} \left(\frac{r^{3-4l}}{3-4l}-\frac{4lr_\epsilon^{3-4l}}{3(3-4l)}\right) \label{k56}\\
&=&A_l\left(\frac{4lr_\epsilon^{3-4l}}{3}-r^{3-4l}\right),\ say \label{k561}\\
A_l &=&\frac{4\pi r_0^{4l}\rho(t_b)}{4l-3}\left(\frac{2 a(2l-1)^2}{\pi }\right)^{2l} .\label{k562}
\end{eqnarray}
According to construction here the mass of the core $M_\epsilon$ should be given by $r=r_\epsilon$ in equation (\ref{k50}) and inspection of (\ref{k54}) shows that the core mass is
\begin{eqnarray}
M_\epsilon=M^+(r_\epsilon)=4\pi \sigma_l (r_0) \frac{r_\epsilon^{3-4l}}{3}= A_l\frac{r_\epsilon^{3-4l}}{3}. \label{k57}
\end{eqnarray}
Using $s(t)$ to denote the function $\sinh^{-2} ( 3ct/(2 R_\Lambda))$, the total mass of this type is
\begin{eqnarray}
M^+(\infty)&=& 4\pi r_0^{4l}\rho (t_b)\left(\frac{ 2a(2l-1)^2}{\pi }\right)^{2l} \left(\frac{4lr_\epsilon^{3-4l}}{3(4l-3)}\right)=A_l \left(\frac{4lr_\epsilon^{3-4l}}{3}\right)\nonumber\\
\label{k58}\\
&=&\frac{12\pi r_0^{4l}s(t_b)}{8\pi G}\left(\frac{c}{R_\Lambda}\right)^2
\left(\frac{2 a(2l-1)^2}{\pi }\right)^{2l} \left(\frac{4lr_\epsilon^{3-4l}}{3(4l-3)}\right). \label{k59}
\end{eqnarray}
Thus the ratio of total mass of this type to core mass is 
\begin{eqnarray}
\frac{M^+(\infty)}{M^+(r_\epsilon)}&=&\frac{4l}{4l-3}.\label{k60}
\end{eqnarray}
Formula (\ref{k59}) can be regarded a giving the total mass $M_g(r_\epsilon,r_0,l)= M^+(\infty)$ of this type of a galaxy represented as having values for its parameters given by $(r_\epsilon,r_0)$, if additionally it is in the gravitational equilibrium quantum state given by the integer, $l$. Other parameters used in this formula have approximate known numerical values. The parameter $a$ can be taken to be just the non {\it dimensioned numerical\/} value of the dimensioned ratio $R/G$ from isotropic gravitation theory. Let us now consider the positively gravitation mass arising from Einstein's pressure term
\begin{eqnarray}
M_P(r) &=& \int_0^r \frac{3 P (r^\prime) 4\pi r^{\prime 2}}{c^2}dr^\prime= \int_0^r 3  K_\prime\varrho ^{\frac{4l-1}{2l}} (r^\prime) 4\pi r^{\prime 2}dr^\prime \label{k42.2}\\
&=& \int_0^r 3 K_\prime(\rho_{1,l}({\bf r}^\prime))^{\frac{4l-1}{2l}} 4\pi r^{\prime 2}dr^\prime\nonumber\\
&=& \int_0^r3 K_\prime\left( \left(\frac{2 a(2l-1)^2}{\pi }\right)^{2l}\left(\frac{r^\prime}{r_0}\right)^{-4l}    \right)^{\frac{4l-1}{2l}} 4\pi r^{\prime 2}dr^\prime \label{k42.3}\\
&=&\frac{12\pi K_\prime}{r_0^{2-8l}}\left(\frac{2 a(2l-1)^2}{\pi}\right)^{4l-1} \int_0^r \left(r^\prime\right)^{4-8l} dr^\prime \label{k42.3b}\\
&=& \frac{12\pi K_\prime}{r_0^{2-8l}}\left(\frac{2 a(2l-1)^2}{\pi}\right)^{4l-1} \left(\frac{r_\epsilon ^{2-8l}r_\epsilon^3 }{3}+ \frac{r^{5-8l}}{5-8l}-\frac{r_\epsilon^{5-8l}}{5-8l}\right)\nonumber\\
&=&\frac{12\pi K_\prime}{r_0^{2-8l} (8l-5)}\left(\frac{2 a(2l-1)^2}{\pi}\right)^{4l-1} \left(\frac{r_\epsilon ^{5-8l}(8l-2)}{3}- r^{5-8l}\right)\nonumber\\
&=& B_l \left(\frac{r_\epsilon ^{5-8l}(8l-2)}{3}- r^{5-8l}\right),\ say \label{k42.3c}\\
B_l&=& \frac{12\pi K_\prime r_0^{8l-2}}{ (8l-5)}\left(\frac{2 a(2l-1)^2}{\pi }\right)^{4l-1} \label{k42.32}\\
P(r)&=&c^2 K_\prime \rho_{1,l,\epsilon}({\bf r})^{\frac{4l-1}{2l}} .\label{k42.4}
\end{eqnarray}
The last equation above is the Lane-Emden type polytropic gas equation used above in a form most suitable for use with this work in terms of  the quantum number $l$.  $K_\prime $ is a constant with dimensions of mass per unit volume and $\rho_{1,l,\epsilon}({\bf r})^{\frac{4l-1}{2l}}$ is the dimensionless mass density defined at (\ref{k37.1}) and (\ref{k37.2}) and core modified, see (\ref{k48}) etc.  
The negatively gravitating mass $M_\Lambda (r)$ within a sphere of radius $r$ and volume $4 \pi r^3/3$ is the easiest term to obtain. It is
\begin{eqnarray}
M_\Lambda (r)&=& \frac{4 \pi r^3}{3}(3/(4\pi G))(c/R_\Lambda)^2= \frac{c^2\Lambda r^3}{3G} \label{k42.7}\\
&=&C_lr^3,\ say\label{k42.71}\\
C_l&=& \frac{c^2\Lambda}{3G}.\label{k42.72}
\end{eqnarray}
That is to say $M_\Lambda (r)$  is the volume times twice Einstein's dark energy density term,
\begin{eqnarray}
(3/(8\pi G))(c/R_\Lambda)^2= \frac{c^2\Lambda}{8\pi G}. \label{k42.8}
\end{eqnarray}
Thus the total gravitationally {\it effective\/} mass within a spherical volume is given by the sum of the three components  $M^+$, $M_P$ and $-M_\Lambda$,
\begin{eqnarray}
M_{GR}(r)&=& M^+(r)+ M_P(r)- M_\Lambda(r) \label{k42.9}\\
&=& A_l\left(\frac{4lr_\epsilon^{3-4l}}{3}-r^{3-4l}\right)+ B_l \left(\frac{r_\epsilon ^{5-8l}(8l-2)}{3}- r^{5-8l}\right) - C_lr^3\nonumber\\
\label{k42.91}\\
A_l(r_0) &=& \frac{4\pi r_0^{4l}\rho (t_b)}{4l-3}\left(\frac{2 a(2l-1)^2}{\pi }\right)^{2l}=
\frac{r_0^{4l}s(t_b) c^2\Lambda}{2G(4l-3)}\left(\frac{2 a(2l-1)^2}{\pi }\right)^{2l} \nonumber\\
\label{k42.92}\\
B_l(r_0) &=& \frac{12\pi K_\prime r_0^{8l-2}}{ (8l-5)}\left(\frac{2 a(2l-1)^2}{\pi }\right)^{4l-1}\nonumber\\
&=&\frac{3 r_0^{8l-2}s(t_b)c^2\Lambda}{2 G(8l-5)} \left(\frac{2 a(2l-1)^2}{\pi }\right)^{4l-1}
\label{k42.93}\\
C_l&=& \frac{c^2\Lambda}{3G}. \label{k42.94}
\end{eqnarray}
From (\ref{k42.91}), we can find the total core mass is given by,
\begin{eqnarray}
M_{GR}(r_\epsilon) &=& A_l\left(\frac{4lr_\epsilon^{3-4l}}{3}-r_\epsilon^{3-4l}\right)+ B_l \left(\frac{r_\epsilon ^{5-8l}(8l-2)}{3}- r_\epsilon^{5-8l}\right) - C_lr_\epsilon^3\nonumber\\
 &=& A_l\left(\frac{4l }{3}-1 \right) r_\epsilon^{3-4l}+ B_l \left(\frac{ (8l-2)}{3}-1 \right) r_\epsilon^{5-8l} - C_lr_\epsilon^3 \label{k42.95}\\
&=& A_l\left(\frac{4l -3}{3}\right) r_\epsilon^{3-4l}+ B_l \left(\frac{ (8l-5)}{3}\right) r_\epsilon^{5-8l} - C_lr_\epsilon^3 .\label{k42.96}
\end{eqnarray}
It follows that
\begin{eqnarray}
 M_{GR}(r)-M_{GR}(r_\epsilon) &=& A_l\left(r_\epsilon^{3-4l}  -r^{3-4l}\right)+ B_l \left(r_\epsilon^{5-8l} -r^{5-8l}\right)\nonumber\\
& &\quad - C_l(r^3- r_\epsilon^3)\label{k42.97}\\
&=& M_{GR,\epsilon}+ A_l\left(  -r^{3-4l}\right)+ B_l \left( -r^{5-8l}\right)- C_l(r^3)\nonumber\\
\label{k42.98}\\
M_{GR,\epsilon}&=& A_l\left(r_\epsilon^{3-4l}\right)+ B_l \left(r_\epsilon^{5-8l}\right)- C_l(- r_\epsilon^3) \label{k42.99b}\\
M_{GR}(r) &=& M_{GR,\epsilon^\prime}+ A_l\left(  -r^{3-4l}\right)+ B_l \left( -r^{5-8l}\right) - C_l(r^3)\nonumber\\
 \label{k42.990}\\
M_{GR}^+(r) &=& M_{GR,\epsilon^\prime}+ A_l\left(  -r^{3-4l}\right)+ B_l \left( -r^{5-8l}\right)\label{k42.991}\\
M_{GR,\epsilon^\prime}&=& M_{GR,\epsilon}+ M_{GR}(r_\epsilon)\label{k42.992}\\
M_{GR}^+(r) &=& M_{GR}(r) + C_l(r^3).\label{k42.992b} 
\end{eqnarray}
Formula (\ref{k42.991}) is a key result for this section, in a suitably simplified form, which can be used to find, the quantum steady state energy values $E_l$ and  which I also intend to try out for the special case quantum state $l=8$ as a generator of galactic rotation curves. The reason for this choice of special case will be explained later. This involves evaluating the coefficients having given the free parameters specific values. This will be carried through in the next subsection after firstly dealing with the steady state energies issue. From formula (\ref{k42.991}) we can obtain the total {\it positively\/} gravitating mass $M_l$ associated with each quantum state $l$ by taking the limit $r\rightarrow\infty$ with the result
\begin{eqnarray}
M_l=M_{GR}^+(\infty) &=& M_{GR,\epsilon^\prime}= M_{GR,\epsilon}+ M_{GR}(r_\epsilon) \label{k42.993} 
\end{eqnarray}
as both $3-4l $ and $5-8l $ are negative.
To obtain this result the {\it negatively\/} gravitating dark energy mass involved in the term $- C_l(r^3)$ has to be excluded as at (\ref{k42.991}). In more detail 
\begin{eqnarray}
M_l=M_{GR}^+(\infty) &=& A_l\left(r_\epsilon^{3-4l}\right)+ B_l \left(r_\epsilon^{5-8l}\right)- C_l(- r_\epsilon^3)+\nonumber\\
& &A_l\left(\frac{4lr_\epsilon^{3-4l}}{3}-r_\epsilon^{3-4l}\right)+\nonumber\\
& & B_l \left(\frac{r_\epsilon ^{5-8l}(8l-2)}{3}- r_\epsilon^{5-8l}\right) - C_lr_\epsilon^3\label{k42.994}\\
&=&\frac{r_0^{4l}s(t_b) c^2\Lambda}{2G(4l-3)}\left(\frac{2 a(2l-1)^2}{\pi }\right)^{2l} \left(\frac{4lr_\epsilon^{3-4l}}{3}\right)+\nonumber\\
& & \frac{3 r_0^{8l-2}s(t_b)c^2\Lambda}{2 G(8l-5)} \left(\frac{2 a(2l-1)^2}{\pi }\right)^{4l-1}\left(\frac{r_\epsilon ^{5-8l}(8l-2)}{3}\right)\nonumber\\ \label{k42.994b}\\
&=& \frac{r_0^{4l}s(t_b) M_G 2lr_\epsilon^{3-4l}}{R_\Lambda^3(4l-3)}\left(\frac{2 a(2l-1)^2}{\pi }\right)^{2l}+\nonumber\\
& & \frac{3 r_0^{8l-2}s(t_b)M_G r_\epsilon ^{5-8l}(4l-1)}{ R_\Lambda^3(8l-5)} \left(\frac{2 a(2l-1)^2}{\pi }\right)^{4l-1}.\label{k42.997} 
\end{eqnarray}
It is interesting to consider the meaning of this last formula under the factored  dimensioned decomposition of the gravitational constant, $G$, as in the last two lines above
\begin{eqnarray}
G&=&M_G^{-1} R_\Lambda^3(R_\Lambda/c)^{-2}, \label{k42.996c}\\
&\rightarrow& \frac{c^2\Lambda}{G}=\frac{3 M_G}{R_\Lambda^3}\label{k42.996bb}
\end{eqnarray}
 where $R_\Lambda$ is the {\it de Sitter\/} radius, and which essentially defines a mass $M_G$ \cite{76:gil} and if we represent the total mass associated with a galaxy in a quantum state $l$ and as defined by the parametric values $r_\epsilon$ and $r_0$ as  $M_{g,l}(r_\epsilon,r_0)$, then 
\begin{eqnarray}
M_{GR}(\infty)& \rightarrow & M_{g,l}(r_\epsilon,r_0)\nonumber\\
M_{g,l}(r_\epsilon,r_0)&= & \frac{r_0^{4l}s(t_b) M_G2lr_\epsilon^{3-4l}}{ R_\Lambda^3(4l-3)}\left(\frac{2 a(2l-1)^2}{\pi }\right)^{2l}+\nonumber\\
& & \frac{3 r_0^{8l-2}s(t_b) M_G r_\epsilon ^{5-8l}(4l-1)}{ R_\Lambda^3 (8l-5)} \left(\frac{2 a(2l-1)^2}{\pi }\right)^{4l-1}\label{k42.995}\\ 
N_l^{-1}(r_0,r_\epsilon) &=&\frac{ M_{g,l}(r_\epsilon,r_0) }{ M_G }\nonumber\\
&= & \frac{r_0^{4l}s(t_b)2lr_\epsilon^{3-4l}}{ R_\Lambda^3(4l-3)}\left(\frac{2 a(2l-1)^2}{\pi }\right)^{2l}+\nonumber\\
& & \frac{3 r_0^{8l-2}s(t_b) r_\epsilon ^{5-8l}(4l-1)}{ R_\Lambda^3 (8l-5)} \left(\frac{2 a(2l-1)^2}{\pi }\right)^{4l-1},\label{k42.995b} 
\end{eqnarray}
where $ N_l(r_0,r_\epsilon)$ is the number of galaxies in quantum state $l $ that would be needed to form a universe of total mass $M_G$.

The objective of this section was to find the steady state energies
\begin{eqnarray}
E_l =E(n(l))\label{k42.996b}
\end{eqnarray}
to be associated with the sub-factor density solutions of the cosmological Sch\"odinger equation represented as functions of the quantization integer $l$. From the above discussion, after taking into account the more detailed specification of the solutions by $r_\epsilon$ and $r_0$, a good choice seems to be 
\begin{eqnarray}
E_l(r_\epsilon,r_0)= M_{g,l}(r_\epsilon,r_0)c^2 .\label{k63}
\end{eqnarray}
The mass $M_G$ from the decomposition of the gravitational constant has an approximate value $2.00789 \times 10^{53} kg$ which is close to estimates of the total mass of the universe that have been made in recent years. This actual {\it theoretical\/} value is a possible candidate for an {\it exact\/} value for the mass of the universe. Thus the formula (\ref{k42.995}) gives a quantized relation between a possible mass for the universe $M_G$ and how that as a total mass can be additively built from a number $N_l$, (\ref{k42.995b}), of galactic masses of specific type, quantum number $l$ and of parametric form determined by the values given to  $r_0,r_\epsilon$. I shall examine this rather unexpected relation between the possible large mass of the universe and galactic sub-masses in the next section.
\section{Galactic Masses relation to Universe Mass}
\setcounter{equation}{0}
\label{sec-gmrtu}
I have shown in reference (\cite{76:gil}) that, if it is assumed that the total mass of the universe is given by $M_G$, then the de Sitter radius $R_\Lambda$ is the radius of the universe at time, $t_c$, when the acceleration of the expansion of the universe was exactly zero. The epoch time $t_c$ is much in the past and very roughly about half the age of the universe now. The radius of the universe {\it now\/} is also very roughly twice the de Sitter radius $R_\Lambda$. It is obvious that the radii of the galactic cores will be many orders of magnitude less than the radius of the universe now and therefore also many orders of magnitude less than $R_\Lambda$. In practice, appropriate values for the adjustable constants $(r_0,r_\epsilon, K)$ and $a$ may be obtained from the physical context. $G$, of course, is well known and tabulated by CODATA.
  
In the last two sections a relation between total mass of the universe, if taken to be $M_G$,  and a possible set of constituent quantum number described galactic masses is given by (\ref{k42.995b}). From this relation the number $N_l$ of such constituent masses involved, if all in the same quantum state, is given by (\ref{k42.995b}). Of course, the type of galaxy involved in the actual universe from the usual or quantum point of view ranges over many different forms or quantum states. However, using this theory formulation we can raise the idea of spatially uniform cosmology to a new superior level of galactic identity uniformity, a collection of galaxies all with the same quantum number $l$. In the next section I shall examine an {\it internal to dark matter\/} quantized version of Newton's law of gravitation. 
\section{Quantized Newtonian Law of Gravitation}
\setcounter{equation}{0}
\label{sec-qnlg}
We note the decomposition of $M_l(r)$ and $ M_l^\prime (r)$ into positive and negative parts,  
\begin{eqnarray}
M_l(r) &=& M_{l+}+ A_l\left(-r^{3-4l}\right)+ B_l \left( -r^{5-8l}\right)- C_lr^3\label{k86c}\\
&=& M_{l+}+ M_{l-}(r),\ say,\label{k86d}\\
M_{l-}(r)&=& A_l\left(  -r^{3-4l}\right)+ B_l \left( -r^{5-8l}\right)-C_l(r^3)\label{k86e}\\
M_l^\prime(r) &=& M_{l+} + A_l\left(-r^{3-4l}\right)+ B_l \left( -r^{5-8l}\right)+ C_lr^3 \label{k86cc}\\
&=& M_{l+}^\prime (r)  + M_{l-}^\prime (r),\ say,\label{k86dd}\\
M_{l-}^\prime(r)&=& A_l\left(  -r^{3-4l}\right)+ B_l \left( -r^{5-8l}\right) \label{k86ee}\\
M_{l+}^\prime (r) &=& M_{l+}(r)  + C_lr^3 .\label{k86ff} 
\end{eqnarray}
The last formulae with the primes on the $M$s is the actual rather than effective mass version. I am here using primes instead of the equivalent $gr$ subscripts to maintain simplicity of notation.
Given the detailed formula for $M_l(r)$ (\ref{k86c}), the total amount of mass within spheres of radius $r$, the Newtonian gravitational potential felt at radius $r$ can be written down as,
\begin{eqnarray}
V_l(r)=\frac{M_l^\prime (r)G}{r}= \frac{M_{l+}^\prime G}{r}+\frac{M_{l-}^\prime G}{r}. \label{k95}
\end{eqnarray}
The prime on the $M_{l-}$ is here being used to indicate that the actual mass version is being used in this definition rather than the effective mass version.
If we now substitute $M_{l+}^\prime (r)$ from equation (\ref{k86ff}) into the above equation we get 

\begin{eqnarray}
V_l(r)=\frac{M_l^\prime (r)G}{r}&=& \frac{ (M_{l+}(r)  + C_lr^3) G}{r}+\frac{M_{l-}^\prime G}{r} \label{k95.1}\\
&=&\frac{ (M_{l+}(r)  + C_lr^3) G}{r}+\frac{M_{l-}^\prime G}{r}.\label{k95.2} 
\end{eqnarray}

The emergent feature here is that the Newtonian gravitational constant $G$ appears, as usual, here in the numerator with the $M$s but an inspection of lines (\ref{k42.92}), (\ref{k42.93}) and (\ref{k42.94}) show that it also appears in the denominator of the same formulae. Thus it cancels out and makes no contribution to the {\it dark matter\/} gravitational potential. On first encounter, this seems a very startling result. However, in can be explained within the structure of the dust universe model as follows. All the mass functions at line (\ref{k86c}) have an initial coefficient, (\ref{k96}), which becomes, they having been multiplied by $G$, an initial coefficient, (\ref{k97}) and (\ref{k98}), for the gravitational potential terms at line (\ref{k95}). 
\begin{eqnarray}
4\pi r_0^{4l}(3/(8\pi G))(c/R_\Lambda)^2\ & & \label{k96}\\
\rightarrow 4\pi r_0^{4l}(3G/(8\pi G))(c/R_\Lambda)^2&=&4\pi r_0^{4l}(3/(8\pi ))(c/R_\Lambda)^2 \label{k97}\\
&=&r_0^{4l}( c^2\Lambda /2). \label{k98}
\end{eqnarray}
The result (\ref{k98}) follows from the definition of $R_\Lambda$ in terms of $\Lambda$. Thus according to (\ref{k98}) the usual gravitational coupling constant $G$ as a multiplier effectively converts to the cosmological constant $\Lambda$ as the coupling constant for describing the distant gravitational effect of dark matter. Hence the gravitational field for dark matter is not quite the usual Newtonian result in spite of the fact that it looks superficially identical to it. On reflection this is not surprising as in this theory {\it dark matter\/} density, $\rho (t) $, the dominant type of matter in the universe before  dark energy at the present epoch, appears as {\it some sort\/} of time variable {\it disturbance\/}, $\sinh ^{-2} (3ct/(2 R_\Lambda ))$, of the {\it dark energy\/} space time constant density field, $(3/(8\pi G))(c/(R_\Lambda)^2$, by the formula 
\begin{eqnarray}
\rho (t) & = & (3/(8\pi G))(c/(R_\Lambda)^2 \sinh ^{-2} (3ct/(2 R_\Lambda))\label{k99}\\
& = &M_G/V_U(t).\label{k100}
\end{eqnarray}
The last formula being valid if the mass of the universe $M_U =M_G$ and $V_U(t)$ is the volume of the universe at epoch time $t$. The constant density 
\begin{eqnarray}
(3/(8\pi G))(c/(R_\Lambda)^2=(\Lambda c^2/(8\pi G))\label{k101}
 \end{eqnarray}
is Einstein's {\it dark energy\/} density introduced to explain his mathematical cosmological constant $\Lambda$ as being due to an actual {\it physical\/} dark energy mass density but appearing here as the basis of all energy density, particularly {\it dark matter\/} density. From lines (\ref{k99}) and (\ref{k100}),  it is very clear that the time variable disturbance I referred to earlier is no mystery at all in this cosmological model. The disturbance is just a consequence of the fact that a constant amount of dark matter, $M_G$, remains within the expanding with epoch time volume of the universe $V_U(t)$. Consequently, the mass {\it density\/} of the universe decreases with time and it is this process that accounts for the $\sinh ^{-2} (3ct/(2 R_\Lambda))$ factor at (\ref{k99}). However, the point I wish to emphasise is that the constant amount of {\it dark matter\/} within the expanding spherical universe is intimately related by (\ref{k99}) to Einstein's universally constant {\it dark energy density\/}, a constancy and existence for which  that extends to outside the expanding spherical sphere of the universe. Dark energy is hyper-universal. The remarks above are perhaps digressional, but the point I am making about the Newtonian gravitational potential within the dark matter halos discussed above is that it differs {\it fundamentally\/} from the usual gravitational Newtonian potential in that its source is the disturbed density for {\it dark energy\/}. Notably, the gravitational potential for dark matter (\ref{k95}) is, as has been shown above, distinctly not of the usual $G$ coupled  Newtonian type. Further more, it is quantized with a quantum state number $l$ associated with the mass sources involved being constructed from solutions of a cosmological Schr\"odinger equation for dark matter halo wave functions. A {\it quantization\/} of gravity is here coming from the cosmological constant, $\Lambda$. This remark is reinforced by the unexpected result of the decomposition of the {\it total\/} gravitation mass $ M_{l}$ for a galaxy into the two parts $M_{l+}$ and $M_{l-}$ at (\ref{k86d}). The positive part $M_{l+}$ within the local gravitational potential generates the usual attraction to the centre  Newtonian field whilst the negative part generates negative gravity repulsion. This is additional to the repulsive field arising directly from Einstein's cosmological constant. Thus negatively gravitating material is greatly involved with {\it dark matter\/} additionally to its involvement arising from the existence of $\Lambda$. The explanation for the existence of the negative term in the gravitating mass source or its gravitating potential is very clear from the mathematics and the way it has been obtained from this theory. The formula for the gravitating mass $M (r)$ within a sphere of radius $r$ is basic to this work. However, the total object mass involved $M (\infty)$ extends to infinite distance. Thus the mass outside a sphere of radius $r$ centered on a galaxy, $M_{out}(r)$, is 
\begin{eqnarray}
M_{out}(r)=M (\infty) - M (r)= M_{l+}- M_l (r)= -M_{l-}(r)>0,\label{kk20}
\end{eqnarray}
by (\ref{k86d}). Thus the positive mass $-M_{l-}(r)$ outside the sphere of radius $r$ on which the galaxy is centred contributes negatively to the gravitational potential at $r$. Thus clearly in this case the negative mass within the sphere is  fictional and simply is a reflection of actual positive mass outside the sphere. This result at (\ref{kk20}) can be seen to be an explanation for the essential need for a cosmological constant in the Einstein field equations by the following considerations. There are obviously very large numbers of galaxies spread throughout the cosmos. According to this theory each galaxy is an infinitely radially extended structure. Hence if we consider any location distance $r$ from the centre of a galaxy at such a position its gravitation potential will be felt included in which will be the influence from its mass outside  radius $r$. This location will also be usually outside all the other infinitely extended galaxies in the universe. Thus all the rest of the galaxies will make a resultant but usually very small negative gravitation contribution.  This small negative gravity component is supplied by the Einstein additional $\rho_\Lambda$ input density contribution in the form of an extra a component of the stress energy momentum tensor so that in this case this mass within the sphere is actually negatively gravitational and not a reflection from actual positively  gravitating mass outside the sphere. It seems that Einstein's field equations without $\Lambda$ only describe a local object in isolation from the rest of the universe. Thus negatively gravitating material is not quite so weird as it has seemed for the past decade since $\Lambda$ was reinstated. Negative  gravitation is just a collective very small attraction felt at any point towards all the rest of the rest of universe but usually masked by the existence of local positively gravitating material towards that point and a directional uniformity. However, this issue is philosophically deep and complicated and I intend to discuss it in more detail in future publications. 
\subsection{Galactic Rotation Curves}
In order to study the galactic rotation curves as a function of $r$ generated by a mass function of $r$ such as (\ref{k86c}),(\ref{k86d})and (\ref{k86e}) we can look at the simplest case and also use a reasonable set of values for the free parameters $r_0,r_\epsilon$. We need the values,
\begin{eqnarray}
\Lambda&=&1.35\times 10^{-52}\label{k101.5}\\
c&=&299792458.\label{k101.6}
\end{eqnarray}
\begin{eqnarray}
A_l(r_0) &=& \frac{r_0^{4l}s(t_b) c^2\Lambda}{2G(4l-3)}\left(\frac{2 a(2l-1)^2}{\pi }\right)^{2l}= \frac{c^2\Lambda s(t_b)\beta^{2l}(2l-1)^{4l}}{2G(4l-3)}\label{kkk0}\\
B_l(r_0) &=& \frac{3 r_0^{8l-2}s(t_b)c^2\Lambda}{2 G(8l-5)} \left(\frac{2 a(2l-1)^2}{\pi }\right)^{4l-1}= \frac{3c^2\Lambda s(t_b)\beta^{4l-1}(2l-1)^{8l-2}}{2G(8l-5)}\nonumber\\
\label{k103}\\
C_l&=& \frac{c^2\Lambda}{3G}\rightarrow\frac{299792458^2\times 1.35\times 10^{-52}}{3G}= \frac{4.0444\times 10^{-36}}{G}. \label{k104}\\
M_{l+}&=&M_{GR,\epsilon}+M_{GR}(r_\epsilon)=\nonumber\\
& & A_l\left(r_\epsilon^{3-4l}\right)+ B_l \left(r_\epsilon^{5-8l}\right)- C_l(- r_\epsilon^3) +\nonumber\\
& & A_l\left(\frac{4l -3}{3}\right) r_\epsilon^{3-4l}+ B_l \left(\frac{ (8l-5)}{3}\right) r_\epsilon^{5-8l} - C_lr_\epsilon^3.\nonumber\\
\label{k104.0}
\end{eqnarray}
The quantity $\beta =\frac{2ar_0^2}{\pi}$ introduced above at (\ref{kkk0}) has the dimensions length squared, $m^2$, is a useful simplifier as it is arbitrary because $r_0$ is arbitrary it can be given the value unity when convenient. Thus largely we can ignore $r_0$ and sideline $a$. 
All the mass contributions combined for the quantum state $l$ are given at  (\ref{k104u})
\begin{eqnarray}
M_{l}(r)&=& M^+(r)+ M_P(r)- M_\Lambda(r) \label{k42.99}\\
&=& A_l\left(\frac{4lr_\epsilon^{3-4l}}{3}-r^{3-4l}\right)+ B_l \left(\frac{r_\epsilon ^{5-8l}(8l-2)}{3}- r^{5-8l}\right) - C_lr^3.\nonumber\\
\label{k104u}
\end{eqnarray}
In more detail, we have
\begin{eqnarray}
M_l(r)&=& \frac{c^2\Lambda s(t_b)\beta^{2l}(2l-1)^{4l}}{2G(4l-3)}\left(\frac{4lr_\epsilon^{3-4l}}{3}-r^{3-4l}\right)+\nonumber\\
& & \frac{3c^2\Lambda s(t_b)\beta^{4l-1}(2l-1)^{8l-2}}{2G(8l-5)} \left(\frac{r_\epsilon ^{5-8l}(8l-2)}{3}- r^{5-8l}\right) - \frac{c^2\Lambda}{3G} r^3.\nonumber\\
\label{kk0}
\end{eqnarray}
This can be separated into positive, $M_{l+}(r)$, and negative, $M_{l-}(r)$, signed terms as follows
\begin{eqnarray}
M_{l+}(r)&=&\frac{c^2\Lambda s(t_b)\beta^{2l}(2l-1)^{4l}}{2G(4l-3)}\left(\frac{4lr_\epsilon^{3-4l}}{3}\right)+\nonumber\\
& & \frac{3c^2\Lambda s(t_b)\beta^{4l-1}(2l-1)^{8l-2}}{2G(8l-5)} \left(\frac{r_\epsilon ^{5-8l}(8l-2)}{3}\right) = \nonumber\\
M_{l+}&=&\ a\  constant\ with\ respect\  to\ r\ variation.\label{kk1}
\end{eqnarray}
\begin{eqnarray}
M_{l-}(r)&=& \frac{c^2\Lambda s(t_b)\beta^{2l}(2l-1)^{4l}}{2G(4l-3)}\left(-r^{3-4l}\right)+\nonumber\\
& & \frac{3c^2\Lambda s(t_b)\beta^{4l-1}(2l-1)^{8l-2}}{2G(8l-5)} \left(- r^{5-8l}\right) - \frac{c^2\Lambda }{3G} r^3\label{kk2}\\
M_{l-,0}(r)&=& \frac{c^2\Lambda s(t_b)\beta^{2l}(2l-1)^{4l}}{2G(4l-3)}\left(-r^{3-4l}\right)+\nonumber\\
& & \frac{3c^2\Lambda s(t_b)\beta^{4l-1}(2l-1)^{8l-2}}{2G(8l-5)} \left(- r^{5-8l}\right) ,\label{kk2.1}
\end{eqnarray}
the last version above not including Einstein's dark energy term.
Thus we have 
\begin{eqnarray}
M_l(r)&=& M_{l+}+ M_{l-}(r) \label{kk3}\\
M_{l,0}(r)&=& M_{l+}+ M_{l-,0}(r) \label{kk3.1}
\end{eqnarray}
and, using the actual  masses case, the Newtonian gravitational potential at radius $r$ is 
\begin{eqnarray}
V_l(r)&=& \frac{ M_{l+}G}{r}+ \frac{c^2\Lambda s(t_b)\beta^{2l}(2l-1)^{4l}}{2(4l-3)}\left(-r^{2-4l}\right)+\nonumber\\
& & \frac{3c^2\Lambda s(t_b)\beta^{4l-1}(2l-1)^{8l-2}}{2(8l-5)} \left(- r^{4-8l}\right) + \frac{c^2\Lambda}{3} r^2.\label{kk4}
\end{eqnarray}
It follows that the galactic rotation curves given as transverse velocity squared as a function of $r$ have the equation
\begin{eqnarray}
v_l^2(r)&=& \frac{ M_{l+}G}{r}+ \frac{c^2\Lambda s(t_b)\beta^{2l}(2l-1)^{4l}}{2(4l-3)}\left(-r^{2-4l}\right)+\nonumber\\
& & \frac{3c^2\Lambda s(t_b)\beta^{4l-1}(2l-1)^{8l-2}}{2(8l-5)} \left(- r^{4-8l}\right) + \frac{c^2\Lambda}{3} r^2.\label{kk5}
\end{eqnarray}
The gradients of these curves with respect to $r$ are 
\begin{eqnarray}
\frac{\partial v_l^2(r)}{\partial r}&=& -\frac{ M_{l+}G}{r^2}+ \frac{c^2\Lambda s(t_b)\beta^{2l}(2l-1)^{4l}(2-4l)}{2(4l-3)}\left(-r^{1-4l}\right)+\nonumber\\
& & \frac{3c^2\Lambda s(t_b)\beta^{4l-1}(2l-1)^{8l-2}(4-8l)}{2(8l-5)} \left(- r^{3-8l}\right) + \frac{c^2\Lambda 2}{3} r. \label{kk6}
\end{eqnarray}
\subsection{Galactic Curves for a Small Galaxy}
I have decided to check out the galactic curve kinematics that this theory delivers for a small galaxy which will be identified below.  However, it was initially and in fact still remains unclear how to identify galaxies within the quantum set off galaxies derivable from this new theory. Thus to get going with the use of this theory some trial and error was required which I will now briefly explain. The theory can deliver an infinite discrete set of quantized mass values. However, the actual numerical values involved with this set is determined by the free input parameters which are $r_\epsilon$, $\beta$ and $t_b$. It seems to me that these three parameters can take on arbitrary values. However, it is desirable that physically reasonable values are chosen. If one takes the view, among other trial possibilities, that we use a set of quantum states determined by the quantum parameter $l=1,2,3 \dots 9$ associated with some definite value, to be explained later, of $r_\epsilon =1.3213133\ m$, in meters say, $\beta=1\ m^2$ in meters squared, with $t_c$ the approximate epoch time when the universe has zero radial acceleration,
\begin{eqnarray}
t_c&=&\frac{2 R_\lambda\sinh^{-1}(2^{-1/2})}{3 c}  \label{kk12.1}\\
&=& 2.18285\times 10^{17}\ s,\label{kk12.2}
\end{eqnarray}
the function $ M_{l+}$ at (\ref{kk1}) generates in kilograms the  nine values,
\begin{eqnarray}
& &3.4114\times 10^{-25},\ 5.18055\times 10^{-20},\ 2.52104\times 10^{-12},\nonumber\\
& &0.00246163\times 10^{0},\  2.09608\times10^{7},\ 9.76751\times 10^{17},\nonumber\\
& &1.84565\times 10^{29},\  1.14693\times 10^{41},\  2.00789\times 10^{53}.\label{kk12.3}
\end{eqnarray}
I have only taken values for the quantum number up to $l=9$ because the masses generated beyond $9$ with this value for $r_\epsilon$ are substantially greater than the usual ideas of what the mass of the universe is likely to be.
The last mass displayed above for $l=9$ coincides with the value of $M_G$, the value that could be taken to be the mass of the universe. This last value was deliberately achieved by choosing $r_\epsilon =1.3213133$.
The incredibly wide range of mass values generated  for ranges of the integer quantum number $l$ is to my mind very striking. The rest mass of the top quark is $3.11966 \times 10^{-25}$ and the rest mass of the Higgs boson is thought to be  approximately $2.22833\times 10^{-25}$ kilograms so that all the masses in this range are astro-physically interesting. 0bjects in the range $10^{-12}$ to $10^{29}$ above are very common and could include large molecules through planets to objects as heavy as small stars and lastly, the last but one entry above  $10^{41}$, could be a small galaxy in relation to the milky which probably has a mass of about $10^{42}$ kilograms. With a different values of $r_\epsilon$ the usually assumed value of the milky way can be obtained but my choice of $r_\epsilon$ was to include exactly the mass $M_G$ in the hope that the other masses generated would somehow acquire  special significance from its inclusion. Clearly the route forward is uncertain and deserving of much more investigation. I have examined  the galactic rotation curves of the small galaxy above, quantum state $l=8$, with the formula for  velocity gradient with the formula (\ref{kk31}) below that can be obtained in detail using (\ref{kk5}) and (\ref{kk6}).
\begin{eqnarray}
\frac{\partial v(r)}{\partial r}=\frac{\partial v^2(r)}{2 v(r)\partial r}.\label{kk31}
\end{eqnarray}
The following list of values of tangential rotation velocity curve gradients just before, $0.8 r_{SM}$, the visibility boundary at $r=r_c=r_{SM}$ and then extending out further to $1.2 r_{SM}$, the results are shown below,
\begin{eqnarray}
r\ in\ meters & & \ Gradient\ of\ v(r)\nonumber\\
0.8\times 1.24\times 10^{22}& &-5.04\times10^{-23}\nonumber\\
0.9\times 1.24\times 10^{22}& &-4.48\times 10^{-23}\nonumber\\
r_{SM}=1.0\times 1.24\times 10^{22}& &  -4.032\times 10^{-23}\nonumber\\
1.1\times 1.24\times 10^{22}& &-3.67\times 10^{-23}\nonumber\\
1.2\times 1.24\times 10^{22}& & -3.36\times 10^{-23}.\label{kk32}
\end{eqnarray}
From the second list above, these curves are decisively flat.
\section{Stability of Dark Matter Galactic Halos}
\setcounter{equation}{0}
\label{sec-sdmg}
There are some deep and complicated issues about the stability of particle distributions assembled under the mutual Newtonian gravitation attraction of the component particles (\cite{77:phc}).  This problem does not impact on the theory for the dark matter galactic halos discussed in this paper, as I shall now explain. It is widely believed nowadays that the {\it missing\/} matter referred to as {\it dark matter\/} exists within a spherical halo that engulfs the visible parts of a galaxy and usually extends greatly beyond the visible parts. I think there is very little evidence that this assumption is correct but it does seem to be a plausible working assumption. Thus let us assume that this view of the situation is correct then it also seems likely that the dark matter is not necessarily rotating with the galaxy for otherwise it would be flattened and not spherical as is usually its parent visible galaxy. Further, remaining spherical and not having clumped into a flattened form in its evolution suggests its compulsive  dynamics is not usual.  If it is in fact spherical and not rotating then what keeps it in its extended state? The obvious answer to this question is that it is a gaseous structure and in equilibrium caused by an outwards pressure from the gas and an inwards pull from gravitation effectively from its center. The fact that such equilibrium conditions can likely occur at some time, using Newtonian gravitation theory with gas dynamics theory, the forms of possible equilibrium mass densities can be found. However, such solutions are not necessarily time wise stable any more than there is necessarily zero motion when acceleration is zero in general. An absolutely static structure is clearly not  appropriate for the description of a galaxy as complex rotational motion is in fact observed.
The main static aspect is the requirement that there should be no {\it overall radial\/} motion and that the galaxy should not be expanding with the substratum.  Clearly, the idea of {\it steady state\/} motion in the quantum context is just right for galaxy description. In quantum theory, systems with very complex internal motions are successfully described under this tag and usually such systems have quantum state numbers attached to a range of discrete quantum states. However, such quantum systems play out their motion under the influence of some central potential energy, such as the coulomb potential for example. The quantized dark matter densities with integer quantum number $l$ that I found from isothermal gravitational equilibrium theory are just crying out to be Schr\"odinger densities formed from the Hermitian scalar product from  Schr\"odinger wave functions. I have shown above and elsewhere \cite{75:gil} that the Schr\"odnger equation needed in this context involves a non-gravitational potential, $V_1 ({\bf r})$ if it is to play the part of supplying steady state solutions that space wise coincide exactly with the solutions from the new isothermal equilibrium theory. 
However, each solution $\Psi_1 (\bf{r},t)$ has its own Schr\"odinger equation and potential function $V_l({\bf r})$ as given below in terms of the quantum integer $l$ instead of the isotropic index $n(l)$.
\begin{eqnarray}
\Psi_{1,l} ({\bf r},t)&=& e^{-\frac{ E_l(r_\epsilon,r_0)it}{\hbar}} \left(\frac{-2 a(1-2l)^2}{\pi}\right)^l(r/r_0)^{-2l} \label{k105}\\
E_l(r_\epsilon,r_0)&=& M_{g,l}(r_\epsilon,r_0)c^2  \label{k105.1}\\
M_{g,l}(r_\epsilon,r_0)&= & \frac{r_0^{4l}s(t_b) M_G2lr_\epsilon^{3-4l}}{ R_\Lambda^3(4l-3)}\left(\frac{2 a(2l-1)^2}{\pi }\right)^{2l}+\nonumber\\
& & \frac{3 r_0^{8l-2}s(t_b) M_G r_\epsilon ^{5-8l}(4l-1)}{ R_\Lambda^3 (8l-5)} \left(\frac{2 a(2l-1)^2}{\pi }\right)^{4l-1}\label{k105.2}\\ 
V_l({\bf r}) &=& E_l(r_\epsilon,r_0)+ \frac{\hbar^2l(2l-1)}{mr^2}\label{k105.3}\\
i\hbar\frac{\partial}{\partial t} \Psi_{1,l} ({\bf r},t)&=& -\frac{\hbar ^2}{2m} \nabla^{2} \Psi_{1,l} ({\bf r},t) + V_l({\bf r}) \Psi_{1,l} ({\bf r},t). \label{k105.4}
\end{eqnarray}
The equations above summarise the basic results from a {\it quantum view\/} of the type of wave functions and their parametric dependants that needs pertain if the new isothermal gravitation theory solutions are also solutions, $\Psi_{1,l} $, of a Schr\"odinger equation with a potential function $V_l({\bf r})$. Mathematically, they can in fact be regarded as the solution of an {\it unusual\/} classical eigen-value problem expressed as follows. Find the {\it eigen-potentials\/} $V_l({\bf r})$  and steady state energy wave functions $\Psi_{1.l}$ that must be operative if the classical Newtonian energy equation is replaced by what might be called a potential function operator version of Schr\"odinger shape $\hat V(r)$ with eigen-values $V_l({\bf r})$ and eigen-wave functions $\Psi_1 ({\bf r},t)$ as below   
\begin{eqnarray}
\hat V({\bf r})&=& i\hbar\frac{\partial}{\partial t} +\frac{\hbar ^2}{2m} \nabla^{2}\label{k105.6}\\
\hat V({\bf r})\Psi_{1,l} ({\bf r},t))&=& V_l({\bf r}) \Psi_{1.l} ({\bf r},t).\label{k105.7}
\end{eqnarray}
Of course, given only the last two equation, the solutions could not be found from them alone, but they do correctly describe the basis of the problem to be in classical {\it eigen-value theory\/} and, importantly for this papers, emphasise the conclusion that galactic halos exist under a special quantized stabilising internal potential in addition to their actual formative gravitational potential structure which itself is also not usual, but rather is $\Lambda$ orientated as has been shown in section $5$. The Schr\"odinger equation for $\Psi_{1.l} ({\bf r},t)$ does not involve gravity at all. It is a purely quantum structure that represents a discrete infinity of endpoints to cosmological clumping and so the potential involved that conditions steady state motion is a representation of the variety of possible forces involved in clumped mass stability, \cite{75:gil}. Gravitation comes into the picture through the cosmological Schr\"odinger equation and its solutions $\Psi_S ({\bf r},t)$ of which the $\Psi_{1.l} ({\bf r},t)$ solutions are possible quantized modulating factors that supplies the space variability with radial position $r$ and also importantly ensures physical stability via {\it steady state motion\/}. 
\section{Conclusions Appendix 6}
\setcounter{equation}{0}
\label{sec-conc}
The cosmological dust universe model is applied to the problem of galactic  modelling using the quantized mass density solutions of a new theory of gravitational isothermal equilibrium. These solutions depend on a key {\it integer\/} state determining parametric pure number  $l$ related to the isotropic index $n$ and three other physically adjustable parameters. Assuming the adjustable parameters fixed in value, it is shown that all these density solutions are derivable from  {\it amplitude\/} solutions of a  Schr\"odinger equation with a special quantized inverse square law eigen-potential. To adequately define these density solutions and then use them to describe dark matter galaxy halos it is necessary to redefine these initially space origin divergent solutions by replacing a small region at the origin with a constant section of radius $r_\epsilon$, which then becomes one of the input adjustable parameters and the core radius of the galaxy. The density solution are all infinitely extended in space but having finite mass core radii they can be integrated over all space to generate  mass spectra dependent on ranges of the quantum number $l$. One such spectrum is calculated so as to terminate at quantum number $l=9$ giving a theoretical total mass of the universe, $M_G$.   
The last but one value $l=8$ generates a possible small galaxy mass the galactic star rotation curves of which are derived. They are shown to be very flat. The mass spectrum in this case for values near $10^{-25}$ could possibly represent the most fundamental particle of them all, the Higgs boson. In using the quantized mass values to form the usual Newtonian gravitation potential in the study of galactic rotation curves involving the usual $G$, coupling constant, it is found that $G$ becomes replaced with Einstein's cosmological constant $\Lambda$ by fraction cancellation. This implies that consistent with the dust universe model basis on $\Lambda$, the {\it quantization\/} of gravity implied by this model is $\Lambda$ dependent rather than $G$ dependent. The structure of the gravitation potential $V(r)$ reveals that a {\it simulation of dark energy} is involved in its form. For any value of $r$ {\it apparent\/} negative mass nearer to the radial origin than $r$ is {\it actual\/} positive mass at positions further from the radial origin than $r$. This has the effect that the apparent negative mass is actual positive mass outside the radius of reference. Thus suitably orientated positive mass can appear elsewhere to be negative. It is suggested that this rather unexpected structure in this formalism could be used to explain {\it actual\/} dark energy mass in terms of positively gravitating mass. This last point is an idea {\it under construction\/} and will need be examined to see if it reinforces or conflicts with my earlier work on dark energy. However, Einstein's cosmological constant is absolutely essential to all aspects of this physical theory.
\vskip 0.5cm 
\leftline{Added section on corrections, 21st June 2012}
\vskip 0.2cm
\section{Explanation of Corrections Appendix 6}
\setcounter{equation}{0}
\label{sec-eoc}
The mistake I made and have corrected in this paper was real and regrettable.  However, it has turned out to be useful for the understanding of what seems to me to be a rather subtle power associated with using the {\it gravitational potential function\/}. It seems that the process of taking the gradient of that function is mathematically rather subtle. In writing down the gravitational potential to use in deriving the equivalent of Newton's inverse square law formula for force on a particle for the case when dark energy mass was involved as a source of the gravity, I assumed that the mass should be proceeded with a minus sign. Consequently, I chose what I have called in this paper the effective mass version of the total mass for constructing the potential function. This has turned out to be a wrong choice. The process of taking the gradient of a gravitation potential function, operating with $\nabla $, distinguishes between ordinary gravitational mass and negatively gravitating mass on purely analytical geometrical properties of the mass distributions and thus generates the correct sign automatically. Thus by adding the negative sign to the mass simply undid the built in cleverness of the standard $\nabla$ operation on the potential function. The correction to this problem was obviously to start with the actual mass as opposed to the effective mass and let the standard procedure do the work. Thus the corrections involved just changing the use of the effective mass to the actual mass wherever appropriate in the paper. I am writing this explanation of my mistake because I think it reveals something significant about the dark energy concept generally. This dependence of the dark energy concept on geometrical orientation has, as the reader will have seen, has come up strongly in interpreting what the dark energy concept means. Thus my mistake has, I think, had very positive consequences. I can translate these remarks onto a definite mathematical explanation as follows.
Consider the total actual gravitating mass and its potential
\begin{eqnarray}
V_l(r)=\frac{M_l^\prime (r)G}{r}&=& \frac{M_{l+}^\prime G}{r}+\frac{M_{l-}^\prime G}{r}. \label{k105.8}\\
M_{l-}^\prime(r)&=& A_l\left(  -r^{3-4l}\right)+ B_l \left( -r^{5-8l}\right) \label{k105.9}\\
M_{l+}^\prime (r) &=& M_{l+}(r)  + C_lr^3 .\label{k106} 
\end{eqnarray}
The acceleration per unit mass caused by this potential at distance $r$ from the origin is
\begin{eqnarray}
\hat {\bf r}\cdot \nabla V_l(r)&=& \frac{c^2\Lambda s(t_b) \theta^{2 l}r_\epsilon^3
 (2l-1)^{4l}}{2 (4l-3)}\left(-\frac{4l}{3r^2}+\frac{(4l-2)r^{1-4l}}{r_\epsilon^{3-4l}} \right)+\nonumber\\
& & \frac{3c^2\Lambda s(t_b) \theta^{4l-1}r_\epsilon^3
 (2l-1)^{8l-2}}{2 (8l-5)}\left(-\frac{ (8l-2)}{3r^2}+\frac{(8l-4) r^{3-8l}}{r_\epsilon^{5-8l}} \right) +\nonumber\\
& & \frac{c^2\Lambda r_\epsilon^3}{3} \left(\frac{2r}{ r_\epsilon^3}\right),\label{k107}
\end{eqnarray}
where the dimensioned parameter $\beta$ has been replaced by the dimensionless parameter $\theta =\beta /r_\epsilon^2$ to clarify the dimensionality of the various contributions. Thus all the last bracketed quantities become dimensionally inverse square but not all variably inverse square. All the coefficients of the large brackets have dimensions $m^3s^{-1}$. Thus all the terms are accelerations. Notably, Newton's gravitation constant $G$ does not occur. In fact, $G$ is replaced by $\Lambda$. This quantized gravitational expression is clearly a substantial generalisation of Newton's law of gravitation. However, we can identify main inverse square law forms as the first terms in the first two large brackets. Both of these terms have minus signs and so represent the usual Newtonian gravitational law of attraction towards the origin. However, both of the large brackets contain also many possible positive signed terms of inverse form determined by the quantum state parameter $l$. They thus represent repulsions from the origin. Clearly the last positive term above represents the repulsive effect of twice Einstein's dark energy term. The two first large brackets originate in the galactic context, from the galactic mass density and the Einstein pressure term mass density from general relativity respectively. 
The inverse repulsive terms in the first two brackets with their positive signs appear to go along with the negative gravity of the last term. They are the terms which simulate negative mass by contributing repulsion and actually exist outside the reference sphere of radius $r$. I mention one more effect from the correction. The negative gravitating term contributed by Einstein's dark energy, the last term above, was left out when I calculated the rotation curve for the small galaxy on the grounds that for a small galaxy it would only make a negligible contribution on account of the smallness of $\Lambda$. However, if is used in such calculations under the corrected version of this theory it would contribute a small positive addition to the rotation curve gradient formula for large galaxies. For sufficiently large galaxies the rotation curves would eventually curve up from their flat condition at very large distances from the origin. There has been mention of observations to this effect.  

\vskip 0.5cm
\leftline{\bf Acknowledgements}
\vskip 0.5cm
\leftline{I am greatly indebted to Professors Clive Kilmister and} 
\leftline{Wolfgang Rindler for help, encouragement and inspiration}

\end{document}